\documentclass[twoside,11pt]{article}

\usepackage{blindtext}

\usepackage[preprint]{jmlr2e}

\usepackage{amsmath,amsfonts,bm}

\def\eqref#1{equation~\ref{#1}}

\def\1{\bm{1}}

\DeclareMathAlphabet{\mathsfit}{\encodingdefault}{\sfdefault}{m}{sl}
\SetMathAlphabet{\mathsfit}{bold}{\encodingdefault}{\sfdefault}{bx}{n}

\newcommand{\softmax}{\mathrm{softmax}}

\usepackage[colorlinks = true,
            linkcolor = blue,
            urlcolor  = blue,
            citecolor = blue,
            anchorcolor = blue,
            pagebackref = true]{hyperref}

\usepackage[english]{babel}
\usepackage{amsmath}
\usepackage{url}
\usepackage{booktabs}
\usepackage{graphicx,wrapfig}
\usepackage{subfig}
\usepackage{xcolor}
\usepackage{algpseudocode}
\usepackage{siunitx}
\usepackage{multirow}
\usepackage{pgfplotstable}
\usepackage{tcolorbox}

\usepackage{blindtext,tikz}
\usetikzlibrary{calc}

\pgfplotstableset{col sep=comma}
\pgfplotstableset{header=true}
\pgfplotstableset{
every head row/.style={
before row=\toprule,
after row=\midrule,
},
every last row/.style={
after row=\bottomrule,
},}
\usepackage{wrapfig}
\usepackage{pifont}%
\newcommand{\cmark}{\ding{51}}%
\newcommand{\xmark}{}%

\newcommand{\ours}{Moshi\xspace}
\newcommand{\mimi}{Mimi\xspace}
\newcommand{\helium}{Helium\xspace}

\newcommand{\interleaving}{Inner Monologue\xspace}
\newcommand{\proba}[1]{\mathbb{P}\left[#1\right]}

\makeatletter
\newcommand{\ensuremathmode}[1]{%
  \ifmmode
    #1%
  \else
    $#1$%
  \fi
}
\makeatother

\newcommand{\waveform}{\ensuremathmode{x}}

\newcommand{\dims}{D}
\newcommand{\numsamples}{L}
\newcommand{\numtokens}{S}
\newcommand{\numquantizers}{\ensuremathmode{Q}}
\newcommand{\cardaudio}{\ensuremathmode{N_A}}
\newcommand{\pmr}[1]{\scriptsize$\pm$#1}

\definecolor{Sand}{RGB}{194, 178, 128}
\definecolor{Orange}{RGB}{255, 191, 128}
\definecolor{Red}{RGB}{255, 87, 90}
\usepackage{tcolorbox}

\usepackage{lastpage}

\ShortHeadings{Moshi: a speech-text foundation model for real-time dialogue}{Moshi: a speech-text foundation model for real-time dialogue}
\firstpageno{1}

\definecolor{kyutaigreen}{RGB}{57,242,174}

\begin{document}

\title{Moshi: a speech-text foundation model for real-time dialogue}

\author{%
  \name Alexandre D\'efossez$^{*}$ \email alex@kyutai.org \\
  \vspace{-0.6cm}
  \AND
  \name Laurent Mazar\'e$^{*}$ \\
  \vspace{-0.6cm}
  \AND
  \name Manu Orsini \\
  \vspace{-0.6cm}
  \AND
  \name Am\'elie Royer \\
  \vspace{-0.6cm}
  \AND
  \name Patrick P\'erez \\
  \vspace{-0.6cm}
  \AND
  \name Herv\'e J\'egou \\
  \vspace{-0.6cm}
  \AND
  \name Edouard Grave$^{*}$ \\
  \vspace{-0.6cm}
  \AND
  \name Neil Zeghidour$^{*}$ \email neil@kyutai.org \\
  {\addr Kyutai}\\
  {\addr $^*$Equal contribution}\\
}

\maketitle

\begin{abstract}%
  \looseness=-1
  We introduce \ours{}, a speech-text foundation model and full-duplex spoken dialogue framework. Current systems for spoken dialogue rely on pipelines of independent components, namely voice activity detection, speech recognition, textual dialogue and text-to-speech. Such frameworks cannot emulate the experience of real conversations. First, their complexity induces a latency of several seconds between interactions. Second, text being the intermediate modality for dialogue, non-linguistic information that modifies meaning--- such as emotion or non-speech sounds--- is lost in the interaction. Finally, they rely on a segmentation into speaker turns, which does not take into account overlapping speech, interruptions and interjections. \ours{} solves these independent issues altogether by casting spoken dialogue as speech-to-speech generation. Starting from a text language model backbone, \ours{} generates speech as tokens from the residual quantizer of a neural audio codec, while modeling separately its own speech and that of the user into parallel streams. This allows for the removal of explicit speaker turns, and the modeling of arbitrary conversational dynamics. We moreover extend the hierarchical semantic-to-acoustic token generation of previous work to first predict time-aligned text tokens as a prefix to audio tokens. Not only this ``\interleaving'' method significantly improves the linguistic quality of generated speech, but we also illustrate how it can provide streaming speech recognition and text-to-speech. Our resulting model is the first real-time full-duplex spoken large language model, with a theoretical latency of 160ms, 200ms in practice, and is available at \href{https://github.com/kyutai-labs/moshi}{github.com/kyutai-labs/moshi}.
\end{abstract}
\medskip

\newpage

\section{Introduction}

\looseness=-1
Voice has provided a convenient interface to early conversational systems, from Alexa\footnote{\url{https://www.alexa.com}} to Siri\footnote{\url{https://www.apple.com/siri}} and Google Assistant.\footnote{\url{https://assistant.google.com/}} In this context, a ``wake word'' spoken by the user typically triggers an automatic speech recognition (ASR) system which transcribes the subsequent user's request. Then, a natural language understanding (NLU) pipeline converts this query to a structured format used to produce a text answer through natural language generation (NLG). Eventually, a text-to-speech (TTS) system tells the answer back to the user. While this process can handle short, constrained interactions (e.g. triggering an action or retrieving a fact), the rise of large language models (LLMs)~\citep{gpt3,chinchilla,llama1} has called for a consequent extension of voice interfaces to multi-turn, open-ended conversations. A solution to this challenge is handling the NLU and NLG with an LLM, while the ASR and TTS provide the voice interface during the user's and the system's turn respectively~\citep{llama3herd}. This framework supports the current generation of spoken dialogue systems such as Gemini~\citep{team2023gemini} or ChatGPT.\footnote{\url{https://openai.com/index/chatgpt-can-now-see-hear-and-speak/}}%

\looseness=-1
Yet, the experience offered by these interfaces remains far from natural conversations. First, latency compounds along the many components of these pipelines, resulting in a typical global latency of several seconds. This is unlike natural conversations which demonstrate response times of a few hundred milliseconds. Second, as language understanding and generation happens in the textual domain, any non-written information is ignored by the model. This goes from paralinguistic information, such as emotion and accent, to non-speech audio, such as surrounding acoustic events. Finally, these models remain fundamentally turn-based, assuming that dialogue is a sequence of well-defined single-speaker segments. While this paradigm is suited to text dialogue, it falls short in modeling aspects of spoken conversations such as interruptions, overlapping speech--- which amounts for $10$ to $20\%$ of spoken time~\citep{overlap_stats_2} ---and backchanneling (i.e. non-interrupting interjections such as ``OK'' or ``I see'').

\looseness=-1
In this work we introduce \ours{}, a speech-text foundation model and real-time spoken dialogue system that aims at solving the aforementioned limitations: latency, textual information bottleneck and turn-based modeling. \ours{} augments a text LLM backbone with a smaller audio language model~\citep{audiolm,uniaudio} that ingests and predicts discrete audio units. This removes the information bottleneck of text by understanding inputs and generating outputs directly in the audio domain, while benefiting from the knowledge and reasoning abilities of the underlying text LLM. We extend previous work on audio language models and design a streaming, hierarchical architecture, with a theoretical latency of \SI{160}{\ms}---lower than the \SI{230}{\ms} average in natural conversations measured over 10 languages~\citep{Stivers2009UniversalsAC}. We furthermore introduce the first multi-stream audio language model, i.e. a model that explicitly processes the input and output audio streams jointly into two autoregressive token streams. This altogether removes the concept of speaker turn and thus allows training the model on natural conversations with arbitrary dynamics including overlap and interruptions. Our resulting model is the first \textit{full-duplex}--- it always listens and always generates sound, either speech or silence---real-time conversational LLM. We summarize our contributions below:
\begin{itemize}
\item We present \helium, a 7B-parameter text LLM that we pretrain on $2.1$T tokens of public English data. \hyperref[sec:helium]{Section \ref{sec:helium}} describes the architecture and training of the model, while \hyperref[sec:text_dataset]{Section \ref{sec:text_dataset}} provides details on the pretraining data collection and filtering.
\looseness=-1
\item We train \mimi, a neural audio codec~\citep{soundstream,encodec} that converts audio into the discrete tokens predicted by \ours{} and back, using residual vector quantization (RVQ). Audio language models typically combine such \textit{acoustic} tokens with \textit{semantic} tokens from a self-supervised speech model as it is necessary to produce intelligible speech in absence of text conditioning~\citep{audiolm}. We rather extend the approach of~\citet{zhang2024speechtokenizer} by distilling semantic information into the first level of acoustic tokens and introduce improved training tricks. \hyperref[sec:mimi]{Section~\ref{sec:mimi}} describes the architecture and training of \mimi while \hyperref[sec:mimi_eval]{Section~\ref{sec:mimi_eval}} details ablation studies.
\looseness=-1
\item We propose \ours{}, a new architecture for audio language modeling, which combines Helium with a smaller Transformer~\citep{attentionvaswani} model to predict audio tokens in a hierarchical and streaming fashion. We show how challenging it is for such unconditioned audio language models to generate intelligible speech, and we provide solutions that outperform the intelligibility and audio quality of non-streaming models while generating audio in a streaming fashion. We furthermore extend this architecture to model several audio streams in parallel, allowing for a conceptually and practically simple handling of full-duplex dialogues with arbitrary dynamics. \hyperref[sec:modeling]{Section \ref{sec:modeling}} describes this architecture.
\looseness=-1
\item In \hyperref[sec:interleaving]{Section~\ref{sec:interleaving}}, we introduce \interleaving, a new training and inference setup for audio language models that significantly improves the factuality and linguistic quality of generated speech by predicting time-aligned text tokens before audio tokens. \ours{} is a speech-to-speech model as it allows reasoning about non-linguistic information, both from the user audio and from \ours{}'s audio. Yet, this is not incompatible with \ours{} producing text along its speech output. Based on the past observation~\citep{audiolm,zhang2024speechtokenizer} that coarse-to-fine generation (from semantic to acoustic tokens) is critical to generating consistent speech, we extend this hierarchy to using text tokens as a per-timestep prefix to the semantic token. Our experiments show that not only this drastically improves the length and quality of generated speech, but we also show how forcing a delay between text and audio tokens allows deriving streaming ASR and streaming TTS from a \ours{} model.
\item We evaluate all components of \ours{} along several axes, including text understanding, speech intelligibility and consistency, audio quality and spoken question answering. Our experiments, reported in \hyperref[sec:evaluation]{Section~\ref{sec:evaluation}}, show that our model is state of the art among existing speech-text models for speech modeling and spoken question answering while being streaming compatible and able to model several minutes of context (\SI{5}{\minute} in our experiments).
\end{itemize}
We encourage the reader to talk to \ours{} using our web demo.\footnote{\url{https://moshi.chat/}}

\newpage
\section{Related Work}

\paragraph{Audio Language Modeling.}
\looseness=-1
Early developments in speech foundation models have improved speech understanding across many discriminative tasks, from automatic speech recognition (ASR)~\citep{wav2vec2,whisper,google_usm} to speaker verification~\citep{wavlm} and speech classification~\citep{superb}. A key factor in this development is self-supervised learning~\citep{hubert, wav2vec2,wavlm} which allows learning generic, discriminative speech representations. As these speech understanding models build on previous work done on masked language modeling for text~\citep{bert}, generative text pretraining~\citep{gpt} has similarly inspired a large family of speech generation models. In particular,~\citet{gslm} propose quantizing aforementioned self-supervised representations. The resulting discrete \textit{audio tokens} represent a speech segment as a sequence of categorical variables, thus casting speech generation as a language modeling task. AudioLM \citep{audiolm} furthermore combines these \textit{semantic} tokens with \textit{acoustic} tokens from a neural audio codec~\citep{soundstream}, which allows for modeling arbitrary voices, recording conditions and non-speech sounds.  These audio language models have redefined the state of the art in speech generation, from text-to-speech~\citep{valle, spear_tts} to speech-to-speech translation~\citep{audiopalm, gemini_1.5} and speech enhancement~\citep{uniaudio}. Beyond these supervised tasks, a parallel line of work has explored training and scaling unsupervised audio-only models, trained for autoregressive speech generation~\citep{zerospeech21,gslm,audiolm}. The abilities of these models have progressively expanded, from generating short sentences in a single speaker voice~\citep{gslm} to producing meaningful and consistent speech continuations across dozens of seconds in arbitrary voices and conditions~\citep{audiolm}, thanks to a hierarchical modeling of semantic and acoustic tokens. A main challenge is that audio requires the modeling of long sequences, up to a few minutes, to produce meaningful and exploitable outputs. However, latent representations for audio are typically less compact than equivalent representations for text. Thus, discrete representations from neural audio codecs require multiple predictions per timestep when modeled autoregressively. \citep{liu2023audioldm} and \citep{evans2024stable} use latent diffusion~\citep{ho2020denoising} for general audio and music modeling
to alleviate the need for hierarchical discrete tokens. However, these methods cannot be used in a streaming fashion, and it is unclear whether they could generate consistent speech. \citet{musicgen} instead show that the number of auto-regressive steps can be reduced by introducing a delay between the different levels of tokens, and performing parallel prediction over them. Inspired by the RQ-Transformer method by~\citet{rqtransformer} and the hierarchical MegaByte transformer model~\citep{yu2024megabyte},
\citet{uniaudio} and \citet{zhu2024generativepretrainedspeechlanguage} leverage a smaller nested transformer to model the different tokens at a single time step. In this work, we extend these previous works to push the limits of autoregressive speech generation by proposing a scalable hierarchical modeling of audio tokens which can handle several minutes of context while generating audio in real time. Still, while speech-only models learn linguistic structure---lexicon, syntax, semantics--- from raw speech~\citep{zerospeech21}, they typically demonstrate poor-to-nonexistent factual knowledge and reasoning abilities. This has led to the development of speech-text models, intended to combine the knowledge and reasoning abilities of text models with the generative power of audio models.

\paragraph{Speech-text Models.}
\looseness=-1
Such models typically start from a pretrained text language model and either finetune it to predict audio~\citep{twist}, or propose a speech-text finetuning task~\citep{audiopalm,voxtlm,spectron,spiritlm,mitsui2024pslmparallelgenerationtext,zhang2024speechgpt}: For instance, AudioPALM~\citep{audiopalm} starts from a pretrained PALM~\citep{palm} model, and extends its text vocabulary with semantic audio tokens. Then, the model is trained for a mixture of speech-text tasks, including TTS, ASR and speech-to-speech translation. VoxTLM~\citep{voxtlm} adopts a similar approach for TTS and ASR. While these models are trained in a supervised fashion with specific input and output sequences, Spirit-LM~\citep{spiritlm} uses temporal alignment between speech and its transcript to perform modality switch (from speech tokens to text tokens, or conversely) inside a sequence. This allows the model to learn consistent internal representations of language regardless of it being represented as text or speech, as measured through commonsense evaluation. Another approach, adopted by Spectron~\citep{spectron}, SpeechGPT~\citep{speechgpt_emnlp} and PSLM~\citep{mitsui2024pslmparallelgenerationtext}, combines speech and text in a hierarchical manner rather than as interchangeable representations. Similar to how AudioLM~\citep{audiolm} decomposes speech generation into predicting semantic tokens and \textit{then} acoustic tokens, Spectron and SpeechGPT use a ``Chain-of-Modality'' and first produce an utterance as text tokens, subsequently used as a prefix to generate speech. This allows guiding speech generation with the output of an underlying text LLM, however this is fundamentally incompatible with live interactions as the model needs to produce an entire answer as text before it starts speaking. PSLM alleviates this limitation by modeling text and speech tokens in parallel. In this work, we propose \interleaving as a main architectural and training component to combine aligned text and speech data. \interleaving decomposes speech into a chain of text, semantic and acoustic tokens, and predicts this structured sequence in a hierarchical manner. Unlike Spirit-LM, this allows representing all utterances both as text and speech, rather than switching between modalities; In addition, the integration of acoustic tokens into the same generative model enables generating arbitrary voices and conditions, rather than a single speaker. Besides, this hierarchical modeling described in \hyperref[sec:interleaving]{Section~\ref{sec:interleaving}} allows decomposing the generation task without increasing the sequence length of the Transformer~\citep{attentionvaswani} outputs, unlike Chain-of-Modality, while benefiting from producing text a prefix to audio tokens rather than in parallel like PSLM. Moreover, \interleaving decomposes speech on a per-frame basis, which means that each prediction step outputs a speech frame. This is unlike Spectron and SpeechGPT which require generating a complete sequence as text before generating audio tokens, and this makes \ours{} compatible with real-time generation. Moreover, we show in \hyperref[sec:interleaving]{Section~\ref{sec:interleaving}} how \interleaving, when combined with a delay between token types, allows deriving streaming TTS and ASR systems from \ours{}. Finally, while Spectron, SpeechGPT and PSLM model both user and system speech and text tokens into a single stream, which requires properly segmented turns, \ours{} benefits from a novel multi-stream architectures which removes the concept of speaker turns and allows for modeling any type of overlap, interruptions and interjections.

\paragraph{Spoken Dialogue Models.}
\looseness=-1
Spoken dialogue is one of the less explored tasks in speech generation, as it requires addressing several challenges: 1) The model should run in real time and allow for long conversations in full-duplex---the model always listens and can speak at any moment; 2) it should be speech-to-speech to handle paralinguistic communication; 3) it should display knowledge and reasoning abilities that make it amenable to helpful and enjoyable conversations. Spectron benefits from its underlying text LLM (as measured by spoken question answering), however it is not compatible with real-time generation due to Chain-of-Modality. PSLM proposes generating speech and text tokens in parallel to reduce this latency, however it reduces the quality of answers, and the model still relies on ASR, which removes paralinguistic information. More importantly, these models cannot handle full-duplex communication, where there is no boundary between speaker turns, as any side of the conversation can be active at any time. An attempt at modeling these dynamics has been proposed by \citet{fullduplexfsm}, with an ASR system running in parallel to a text generator that feeds into a streaming TTS. While this allows modeling more complex scenarios than previous approaches, it still relies on a cascaded pipeline and models both the user's and the system's speech into a single token stream, which is challenging in presence of significant overlap. The only previous full-duplex dialogue system is dGSLM~\citep{nguyen-etal-2023-generative}, which models user and system speech as separate audio token streams and proposes a Siamese architecture to process both streams jointly. While dGSLM is full-duplex, it remains a proof-of-concept: it does not run in an online fashion, it does not benefit from the knowledge of a text language model, and it does not model acoustic information as it only models semantic tokens. \ours{} addresses these limitations altogether: by modeling two streams of semantic and acoustic tokens hierarchically, \ours{} is full duplex and can exploit all the information from the user (linguistic and non-linguistic) while producing speech in real time. Thanks to text pretraining and \interleaving, \ours{} benefits from the knowledge of its \helium backbone. Finally, as the model produces acoustic tokens along with text and semantic tokens, it can generate an arbitrary range of emotions, voices and acoustic conditions. To the best of our knowledge, \ours{} is the first audio language model that successfully addresses the many aforementioned challenges of spoken dialogue.

\section{Model}

\subsection{Overview}

\begin{figure}[t]
    \centering
    \includegraphics[width=0.85\linewidth]{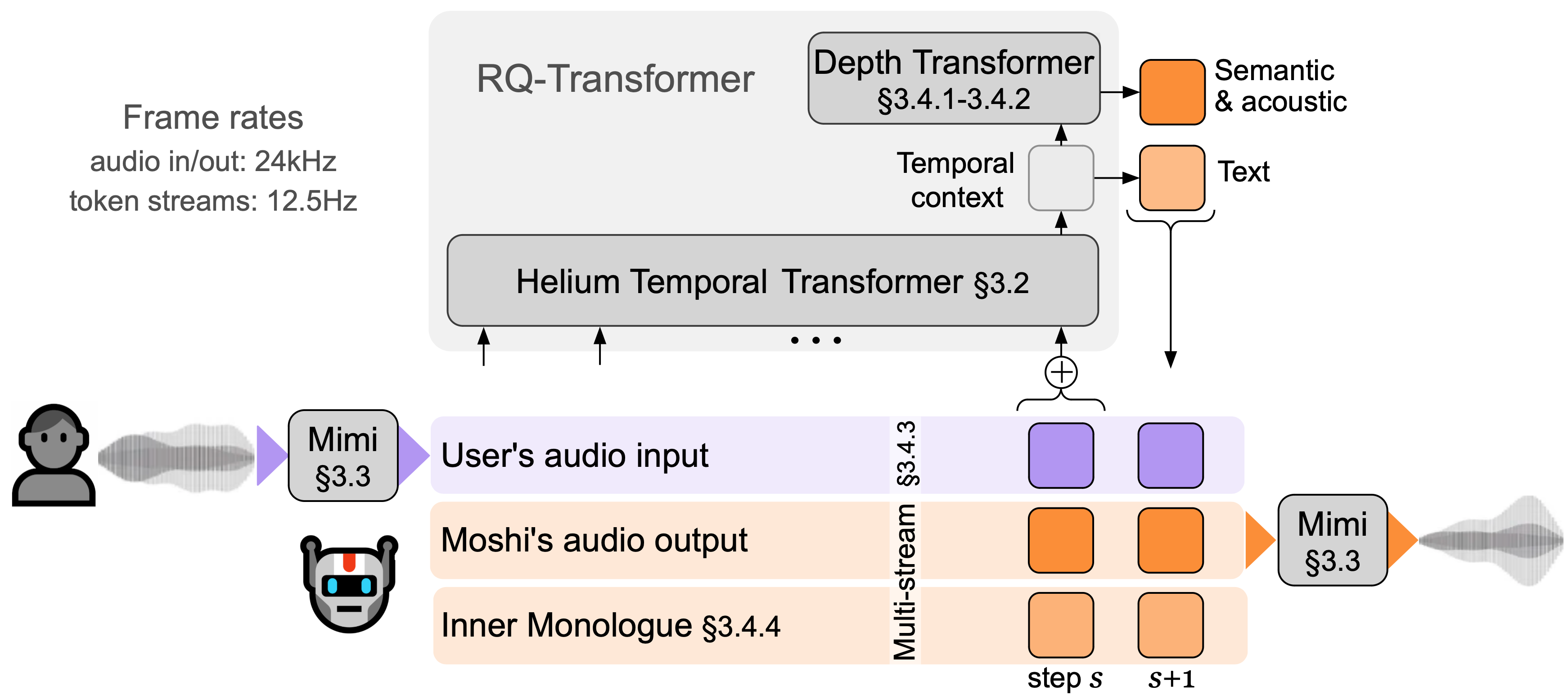}
    \caption{\textbf{Overview of \ours{}}. \ours{} is a speech-text foundation model which enables real-time spoken dialogue. 
    The main components of \ours's architecture are: a bespoke text language model backbone (\helium{}, see \hyperref[sec:helium]{Section \ref{sec:helium})}; a neural audio codec with residual vector quantization and with semantic knowledge distilled from a self-supervised speech model (\mimi{}, \hyperref[sec:mimi]{Section \ref{sec:mimi}}); the streaming, hierarchical generation of semantic and acoustic tokens for both the user and \ours, along with time-aligned text tokens for \ours when using \interleaving{} (\hyperref[sec:modeling]{Section \ref{sec:modeling}}). 
    }
    \label{fig:moshi-overview}
\end{figure}

\ours{} is a multi-stream speech-to-speech Transformer model, which allows for full-duplex spoken dialogue with a user thanks to an innovative architecture summarized in \hyperref[fig:moshi-overview]{Figure \ref{fig:moshi-overview}}.
\ours{} is built on top of Helium, a text LLM which we build from scratch (\hyperref[sec:helium]{Section \ref{sec:helium}}), relying on high-quality text data to provide strong reasoning abilities to the model. 
We also propose \interleaving{} (\hyperref[sec:interleaving]{Section \ref{sec:interleaving}}), a training and inference procedure in which we jointly model text and audio tokens. This allows the model to fully exploit the knowledge imparted from the text modality, while remaining a speech-to-speech system. 
To enable real-time dialogue, we also design \ours{} as a multi-stream architecture from the get-go (\hyperref[sec:multistream]{Section \ref{sec:multistream}}): The model is able to both speak and listen to the user at the same time, and does not need to explicitly model speaker turns. 
In addition, to capture the input user audio and output \ours{}'s voice with high quality and in an efficient manner, we propose Mimi (\hyperref[sec:mimi]{Section \ref{sec:mimi}}), a neural audio codec combining semantic and acoustic information into a single tokenizer by using residual vector quantization and knowledge distillation.   
To jointly model the audio streams from \ours{} and the user, as well as \ours{}'s text tokens, we rely on a Depth Transformer compatible with streaming inference (\hyperref[sec:joint_ar]{Sections \ref{sec:joint_ar}}, \ref{sec:audio_modeling}).

In this section, we further detail each of these components. We then describe the training datasets and the different training phases we used to train \ours{} in \hyperref[sec:maindatasets]{Section \ref{sec:maindatasets}}. 
Finally, in \hyperref[sec:evaluation]{Section \ref{sec:evaluation}}, we report thorough evaluation results on \ours{}'s abilities, both linguistic and acoustic, as well as ablation experiments on its main components, while \hyperref[sec:safety]{Section~\ref{sec:safety}} provides analyses on the safety of our system.

\subsection{The \helium Text Language Model}
\label{sec:helium}
\subsubsection{Architecture}
\helium is an autoregressive language model, based on the Transformer architecture~\citep{attentionvaswani}.
Following previous work in this area, we make the following changes to the original architecture: 
First, we use RMS normalization~\citep{zhang2019root} at the input of the attention blocks, the feed-forward blocks and the output linear layer of the model.
We use rotation positional embeddings~\citep[RoPE]{su2024roformer}, a context length of 4,096 tokens and FlashAttention~\citep{dao2022flashattention} for efficient training.
Finally, we change the architecture of the feed-forward blocks and use Gated Linear Units~\citep{shazeer2020glu}, with the SiLU activation as a gating function~\citep{hendrycks2016gaussian}.
Our tokenizer is based on the unigram model from SentencePiece~\citep{kudo2018sentencepiece}, and contains 32,000 elements mostly targeting English.
We split all numbers into single digits, and use byte-backoff to ensure that our tokenizer does not lose information.
We train the model with the AdamW~\citep{loshchilov2017decoupled} optimizer, with a fixed learning rate followed by a cosine learning rate decay~\citep{loshchilov2016sgdr}.

\begin{table}[t]
  \centering
  \caption{\label{tab:hparams}\textbf{Models' hyper-parameters}. Hyper-parameters for the architecture and training of our 7B-parameter Helium language model and of \ours, our speech-text dialogue model. 
  The training of \ours{} goes through 4 phases: Pre-training on unsupervised data (with Temporal Transformer initialized from \helium); Post-training with simulated multi-stream based on diarization; Fine-tuning on the Fisher dataset~\citep{cieri2004fisher} to gain its fully duplex capabilities; Instruction fine-tuning on a custom dataset built from synthetic interaction scripts. During the pre-training phase, we keep training half of the time on full text batches from the same dataset as used for Helium, using a separate optimizer state.}
  \footnotesize
  \begin{tabular}{l|c|cccc}
    \toprule
      &
    \multicolumn{1}{c|}{Helium training} &
    \multicolumn{4}{c}{Moshi training} \\
    Hyper-parameter & pre-training  & pre-training & post-training & fisher & fine\\
    \midrule
    \multicolumn{6}{c}{\textit{Temporal Transformer}} \\
    \midrule[0.3pt]
    Model dimension    & 4096 & \multicolumn{4}{c}{\multirow{4}{*}{same}}\\
    MLP dimension      & 11264 \\
    Number of heads    & 32 \\
    Number of layers   & 32 \\
    Context size       & 4096 & \multicolumn{4}{c}{3000 steps, e.g. 4 min.} \\
    Learning rate      & $3 \cdot 10^{-4}$ & $3 \cdot 10^{-5}$ & $3 \cdot 10^{-6}$ &  $2 \cdot 10^{-6}$ &  $2 \cdot 10^{-6}$\\
    \midrule
    \multicolumn{6}{c}{\textit{Depth Transformer}} \\
    \midrule[0.3pt]
    Model dimension    & - & \multicolumn{4}{c}{1024}\\
    MLP dimension      & - & \multicolumn{4}{c}{4096} \\
    Number of heads    & - & \multicolumn{4}{c}{16}\\
    Number of layers   & - & \multicolumn{4}{c}{6}\\
    Learning rate      & - & $2 \cdot 10^{-4}$ & $5 \cdot 10^{-5}$ &  $4 \cdot 10^{-6}$ &    $2 \cdot 10^{-6}$\\
    \midrule
    \multicolumn{6}{c}{\textit{Input / Output space}} \\
    \midrule[0.3pt]
    Text cardinality & 32000 & \multicolumn{4}{c}{32000}\\
    Audio cardinality & - & \multicolumn{4}{c}{2048}\\
    Frame rate & - & \multicolumn{4}{c}{12.5 Hz} \\
    \midrule
    \multicolumn{6}{c}{\textit{Common parameters}} \\
    \midrule[0.3pt]
    Batch size (text)    & 4.2M tok. & 1.2M tok. & 1.2M tok. &  - & - \\
    Batch size (audio)   & - & 16h & 8h & 40min & 2.7h\\
    Training steps       & 500k & 1M & 100k & 10k & 30k \\
    LR Schedule          & cosine & cosine & - & - & -\\
    Acoustic delay       & - & 2 & 1 & 1 & 1 \\
    Text delay           & - & $\pm$0.6 & 0 & 0 & 0 \\
    \bottomrule
  \end{tabular}
\end{table}

\subsubsection{Pre-training data filtering}
\label{sec:data_filtering}
Training data is one of the critical ingredients to train LLMs: we now describe our method to obtain a large and high-quality text dataset.
We start from high-quality data sources, such as Wikipedia, Stack Exchange and a large collection of scientific articles.
As the quantity of data from these sources is too small to train a LLM, we also rely on web crawled data, specifically from CommonCrawl, to extend our dataset. See more details on data sources in \hyperref[sec:text_dataset]{Section \ref{sec:text_dataset}}. 
Web data requires extensive processing to obtain a high-quality training set: we perform deduplication, language identification and quality filtering.
In the following, we describe each operation in more details.

\paragraph{Deduplication.}
We start from the WET files, which contain only the text content of web-pages, which was extracted by the CommonCrawl project.
Because this format contains all the text of a page, it includes a lot of boilerplate such as navigation menus.
Thus, the first step of our pipeline is to deduplicate each shard (there is 100 shards per crawl) at the \emph{line level}, to remove this boilerplate.
To do so, we compute the \texttt{FNV-1a}\footnote{\url{http://www.isthe.com/chongo/tech/comp/fnv}} hash of each line, and use a bloom filter to remove duplicates.
We also train a \texttt{fastText}~\citep{joulin2016bag} classifier on duplicates \emph{vs.} non-duplicates, to perform fuzzy deduplication:
here we only remove blocks of at least 3 consecutive lines that are classified as duplicates.

\paragraph{Language identification.}
Once deduplication is performed, we apply a language identifier based on \texttt{fastText} to keep English data only.
Language identification is performed at the document level, and we only keep documents above a certain threshold (0.85).

\paragraph{Quality filtering.}
\looseness=-1
The last step is to filter the remaining data, to keep high-quality webpages only.
To perform this step, we train a \texttt{fastText} classifier on lines from our high quality data sources and from random CommonCrawl webpages.
We obtain a classifier with 9 categories, corresponding to our different high quality sources such as Wikipedia or Wikibooks and to subsets of StackExchange such as STEM or humanities.
The motivation is to obtain a finer control over which documents to keep, not only based on similarity to high quality sources, but also based on their domains.
This classifier is applied at the \emph{line level}, and an aggregated score is obtained by computing the average scores of each line, weighted by their length.
Again, we keep documents corresponding to scores above a certain threshold.

\subsection{Audio Tokenization}
\label{sec:mimi}

\looseness=-1
To discretize waveforms into audio tokens, we introduce Mimi, a neural audio codec~\citep{soundstream,encodec} that operates as an autoencoder with a discrete bottleneck~\citep{vqvae}. In the literature, and following the terminology defined by~\citet{audiolm}, these tokens are referred to as \textit{acoustic} tokens, as they model fine audio details and are optimized for high-quality reconstruction. While these acoustic tokens provide appropriate targets for conditioned text-to-audio models (e.g. text-to-speech~\citep{valle} or text-to-music~\citep{musicgen}), unconditioned speech generation requires combining them with \textit{semantic} tokens extracted from self-supervised speech models~\citep{wav2vec2,hubert,w2vbert}. Unlike their acoustic counterpart, semantic tokens do not allow for reconstructing high-quality audio but correlate strongly with linguistic content. This similarity with language allows generating intelligible and consistent speech, even without text conditioning, by using semantic audio tokens as a prefix to predicting acoustic tokens. Yet, this hybrid tokenization approach is not compatible with real-time generation. Semantic tokens are typically not causal and can thus only be computed in an offline manner. Moreover, generating acoustic and semantic tokens with separate encoders represents a non-negligible computational burden. Consequently, and taking inspiration from previous work on SpeechTokenizer~\citep{zhang2024speechtokenizer}, Mimi uses distillation to transfer non-causal, high-level semantic information into the tokens produced by a causal model, allowing for streaming encoding and decoding of semantic-acoustic tokens.  
\begin{figure}[t]
    \centering
        \includegraphics[width=\textwidth]{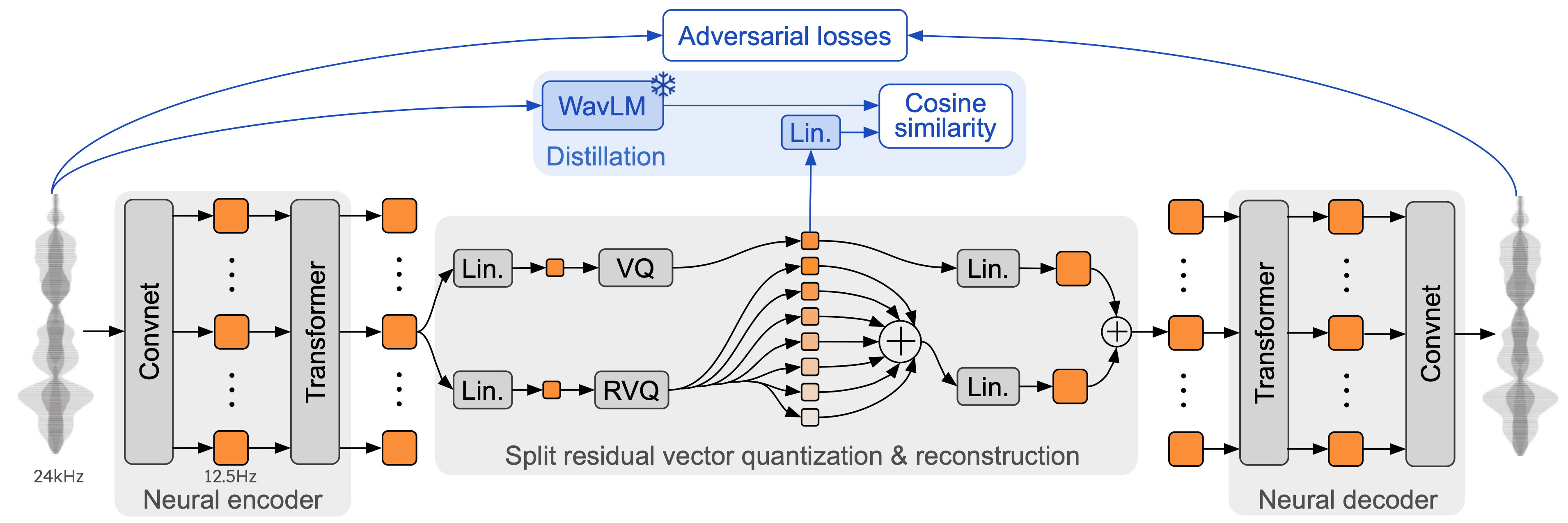}
    \caption{\textbf{Architecture and training of \mimi{}, our neural audio codec, with its split residual vector quantization}. %
    During training (blue part, top), we distill non-causal embeddings from WavLM~\citep{wavlm} into a single vector quantizer which produces semantic tokens, and is combined with separate acoustic tokens for reconstruction.
    }\label{fig:split_quantizer}
\end{figure}

\subsubsection{Architecture}
Our baseline architecture takes inspiration from SoundStream~\citep{soundstream} and Encodec~\citep{encodec} and consists of a SeaNet~\citep{tagliasacchi2020seanet} autoencoder and a Residual Vector Quantizer~\citep{soundstream}. The encoder projects a single-channel waveform $\waveform \in \mathbb{R}^{\numsamples}$ to a latent representation $\mathrm{enc}(x) \in \mathbb{R}^{\numtokens\times\dims}$ by cascading residual convolutional blocks that interleave dilated~\citep{oord2016wavenet} and strided convolutions along with ELU~\citep{elu} non-linearities and Weight Normalization~\citep{weightnorm}. All convolutions are causal, such that this autoencoder can run in a streaming fashion. With 4 convolutional blocks and respective striding factors $(4, 5, 6, 8)$, and a final 1D convolution with stride 2, \mimi's encoder projects a 24kHz waveform to a latent representation of $12.5$ frames per second and dimension $D = 512$. Symmetrically, the decoder adopts a similar structure but with transposed convolutions rather than strided ones, to project the latent representation back to 24kHz audio. We discretize the latent space with a Residual Vector Quantizer~\citep{soundstream}, which iteratively applies vector quantization (VQ) to the residuals of the previous quantizer. With \numquantizers{} quantizers, each with a codebook of \cardaudio{} centroids, the RVQ discretizes the latent space into $\{1,\dots,\cardaudio\}^{\numtokens\times\numquantizers}$. As a baseline, we train this model with a combination of reconstruction and adversarial losses, following the setup of Encodec~\citep{encodec}. We detail below the main changes of \mimi with respect to this default configuration. %
\paragraph{Transformer-based bottleneck.}
\looseness=-1
To improve the ability of \mimi to encode speech into compact representations while reconstructing high-quality audio, we add Transformer modules in the bottleneck, one right before quantization and one after. These Transformers have 8 layers, 8 heads, RoPE position encodings, a finite context of 250 frames (20 seconds), GELU~\citep{gelu} activations, a model dimension of 512 and an MLP dimension of 2048. To stabilize training, we use LayerScale~\citep{layerscale}, with initialization of the diagonal values at $0.01$. Both Transformers use causal masking, which preserves the compatibility of the whole architecture with streaming inference. Both Transformers prove to be useful with regard to perceived audio quality, while Transformer in the encoder also improves the distillation of semantic information described below (see \hyperref[tab:mimi_ablations]{Table~\ref{tab:mimi_ablations}} for ablation studies).

\looseness=-1
\paragraph{Causality and streaming.} With the mentioned hyper-parameters, Mimi is causal and can be used in a streaming fashion, both for encoding and decoding. Both its initial frame size and overall stride correspond to 80ms, meaning that given a first audio frame of 80ms, Mimi outputs a first latent timestep, which can be decoded to 80ms of output audio.

\paragraph{Optimization.}
\looseness=-1
Unlike purely convolutional codecs that use Adam~\citep{adam}, the introduction of Transformers into the architecture requires additional regularization with weight decay along with using the AdamW~\citep{adamw} optimizer. More precisely, we apply weight decay only to the parameters of the Transformers, with a weight of $5\cdot 10^{-2}$. We use a learning rate of $8\cdot 10^{-4}$, a momentum decay of $0.5$ and a decay of the squared gradient of $0.9$, and an exponential moving average of weights with a decay of $0.99$. We train with a batch size of $128$ on random windows of $\SI{12}{\second}$, for 4M steps, while the context of Transformers is limited to 10s (250 frames before the last downsampling layer of the encoder, and symmetrically for the decoder).

\paragraph{Quantization rate.}
\looseness=-1
We use $\numquantizers= 8$ quantizers, each with a codebook size of $\cardaudio = 2048$. At 12.5Hz, this represents a bitrate of 1.1kbps. While the latent dimension is 512, we project embeddings to 256 dimensions before applying the RVQ, and project back to 512 before the decoder. Consistently with previous work, we use quantizer dropout~\citep{soundstream} to provide the codec with bitrate scalability. We moreover follow the observation of ~\citet{dac} that not applying quantization with a certain probability during training improves audio quality. More precisely, we only apply quantization $50\%$ of the time, on a per-sequence basis, during training. Unlike ~\citet{dac}, this means passing unquantized embeddings to the decoder, rather than passing embeddings quantized with all quantizers. \hyperref[tab:mimi_ablations]{Table~\ref{tab:mimi_ablations}} shows that this significantly improves objective quality metrics, while human evaluations are not conclusive. Across our experiments, we make the somehow counter-intuitive observation that this gain gets more significant as we lower the bitrate.

\paragraph{Adversarial-only training.}
\looseness=-1
As a baseline, we train Mimi with the same combination of reconstruction and adversarial losses as ~\citet{encodec}, namely a multi-scale mel-spectrogram reconstruction loss along with a multi-scale STFT discriminator. The exact parameters can be found in the Audiocraft repository.\footnote{\url{https://github.com/facebookresearch/audiocraft/blob/main/config/solver/compression/default.yaml}} While previous neural codecs rely on such combinations of reconstruction and adversarial losses, we experiment with pure adversarial training, where we only keep the feature loss and discriminator loss. We note that this was previously experimented in the context of bandwidth extension by~\citet{tagliasacchi2020seanet} and ~\citet{hauret2023eben}. While removing reconstruction losses majorly degrades objective metrics, we observed during development that the resulting audio sounded much better than expected based on aforementioned metrics. Subjective evaluations reported in \hyperref[tab:mimi_baselines]{Table~\ref{tab:mimi_baselines}} confirm this observation and demonstrate a remarkable boost in audio quality from training with adversarial losses only.

\subsubsection{Learning semantic-acoustic tokens with a split RVQ}
\label{sec:splitrvq}
\looseness=-1
Similarly to SpeechTokenizer~\citep{zhang2024speechtokenizer}, we distill semantic information from a self-supervised model (WavLM\footnote{\url{https://huggingface.co/microsoft/wavlm-large}}~\citep{wavlm} in our case) into the first level of the RVQ. WavLM projects a 16kHz waveform into 1024-dimensional embeddings sampled at 50Hz, while \mimi projects a 24kHz waveform into 512-dimensional at 12.5Hz. During training, we thus produce targets for distillation by downsampling the input waveform to 16kHz before computing WavLM embeddings followed by average pooling with a stride of 4 and a kernel size of 8, to reach 12.5 Hz. Interestingly, we observed that it was critical for performance to perform this average pooling in a non-causal way, which is compatible with streaming inference as these embeddings are only used during training. We apply a linear projection with an output dimension of 1024 to the output of the first RVQ level, parallel to the actual embedding going into the decoder. We then compute a cosine distance between the output of the first quantizer and the transformed WavLM embeddings, to perform distillation. \hyperref[tab:mimi_ablations]{Table~\ref{tab:mimi_ablations}} shows that this distillation loss conflicts with reconstruction and adversarial losses targeting quality. Indeed, while distillation significantly improves the phonetic discriminability of the first quantizer (as measured by ABX~\citep{abx}), it also affects audio quality negatively. We hypothesize that this is due to distilling semantic information into the first level of a single RVQ: As higher-order quantizers operate on the residual of the first one, the latter needs to trade audio quality for phonetic discriminability. We address this issue by proposing a split RVQ. Rather than a single RVQ with 8 levels, we distill semantic information into a plain VQ and apply an RVQ with 7 levels in parallel. We sum their outputs, such that while both can be used for reconstruction, we remove the constraint that acoustic information should be conserved in the residual of the semantic quantizer. \hyperref[fig:split_quantizer]{Figure~\ref{fig:split_quantizer}} illustrates this architecture and \hyperref[tab:mimi_ablations]{Table~\ref{tab:mimi_ablations}} shows that this solution provides a better semantic-acoustic trade-off overall.

\subsection{Generative Audio Modeling}
\label{sec:modeling}
\looseness=-1
We now describe how we extend the base Helium model to support the modeling
of the audio tokens provided by the \mimi{} codec. With our goal of achieving realistic
spoken dialogue interactions, we further show how to model not just a single stream of audio,
but two at the same time, one representing the user, and one the system.
Finally, we detail a novel feature, the \emph{\interleaving}, which consists in a joint
modeling of the textual and audio modalities on the system side, to improve the quality of interactions.

\subsubsection{Hierarchical autoregressive modeling with RQ-Transformer}
\label{sec:joint_ar}

Let $U \in \{1, \ldots, N\}^S$ be a discrete random sequence, with cardinality $N$ and a sequence length $S$. For convenience, we also denote $U_0 = 0$, a deterministic initial token value.
Autoregressive modeling consists in estimating the joint distribution $\proba{U_1, \ldots, U_S}$
through estimating the conditional distributions $\proba{U_s | U_0, \ldots U_{s - 1}}$ for all steps $1 \leq s \leq S$. Text language models, such as GPT~\citep{gpt2} or Helium, fit this paradigm.

\begin{figure}[t]
    \centering
    \includegraphics[width=0.5\textwidth]{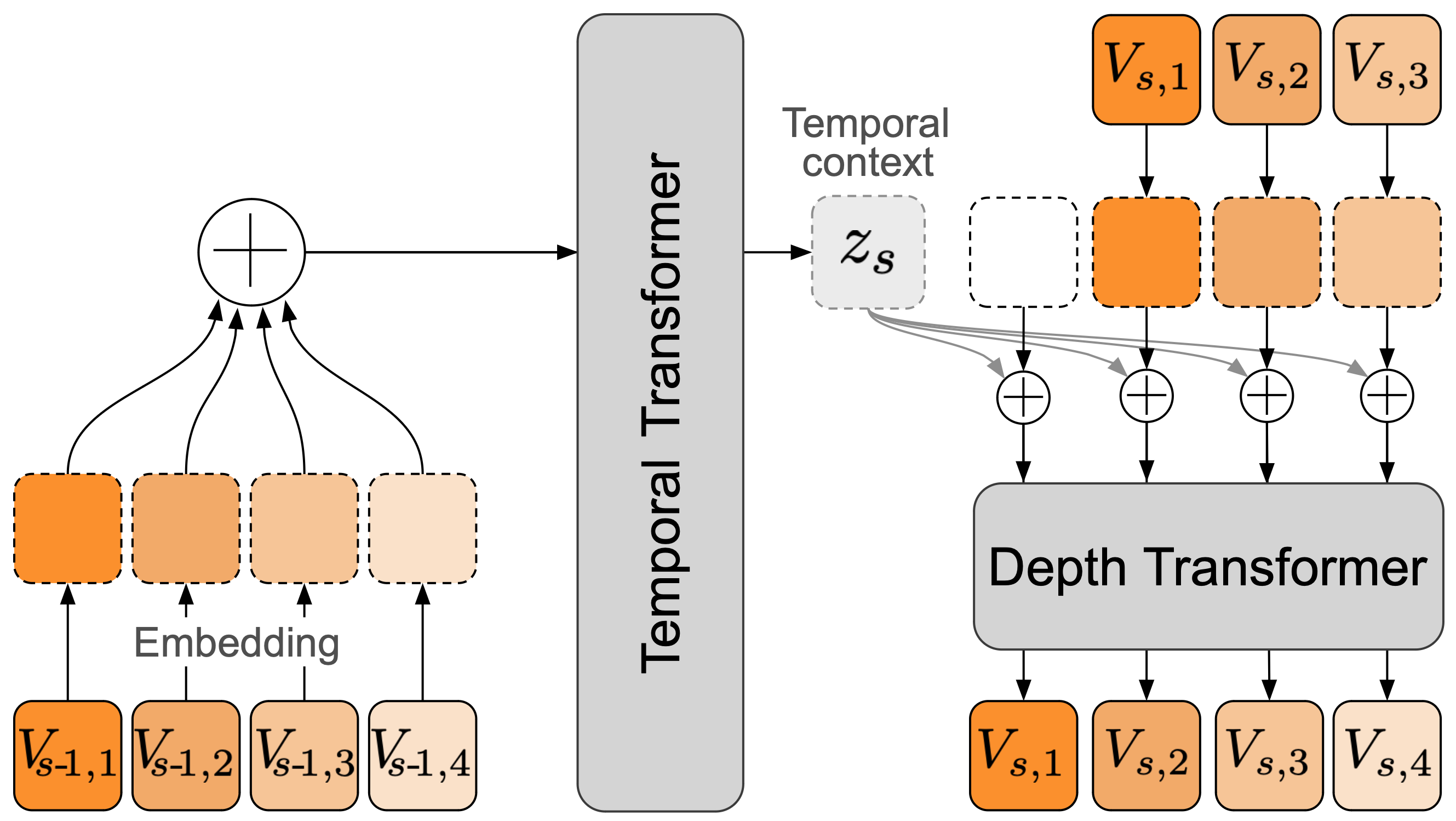}
    \caption{\textbf{Architecture of the RQ-Transformer.} The RQ-Transformer breaks down a flattened sequence of length $K\cdot S$ into $S$ timesteps for a large Temporal Transformer which produces a context embedding used to condition a smaller Depth Transformer over $K$ steps. This allows scaling to longer sequences by increasing $S$---or to a higher depth by increasing $K$--- than modeling the flattened sequence with a single model. In this figure, we use $K=4$ for the sake of illustration.
    }\label{fig:rqtransformer}
\end{figure}

When modeling spoken language, relying on the tokenized text yields a much more compact representation than audio tokens:
Using the \mimi{} codec introduced in \hyperref[sec:mimi]{Section~\ref{sec:mimi}}, with $Q=8$ codebooks at a frame rate of 12.5hz, one would require a sequence length of $100$ steps per second of audio to generate. To model 5 minutes of audio, this would amount to 30,000 timesteps, which represents a significant computational cost and generating 100 tokens per second is incompatible with streaming inference. As a comparison, a sample of English speech can be represented with around 3 to 4 text tokens per second.

\looseness=-1
We are interested in modeling not just a single sequence $(U_s)$, but multiple sub-sequences, e.g. different audio codebooks, along with an optional text stream. We can stack those sub-sequences as $V_{s, k}$ for $1 \leq s \leq S$ and $1 \leq k \leq K$.
Similarly, we define $V_{0, k} = 0$, a deterministic initial token value for all sub-sequences.
For each $1 \leq s\leq S$ and $1 \leq k\leq K$, $V_{s, k} \in \{1, \ldots, N_k\}$, where $N_k$ is the cardinality of the $k$-th sub-sequence. One can \emph{flatten} the $K$ sequences into a single one, increasing the number of predictions by $K$.
\citet{rqtransformer} propose using a smaller autoregressive model along the dimension $K$, combined with a larger model along the time dimension, forming a RQ-Transformer. Later, \citet{yu2024megabyte} suggested a similar approach for byte-level modeling while \citet{uniaudio} and \citet{zhu2024generativepretrainedspeechlanguage} applied it to audio token modeling.

\paragraph{RQ-Transformer.}

Formally, the RQ-Transformer consists in two Transformer models, as illustrated in \hyperref[fig:rqtransformer]{Figure~\ref{fig:rqtransformer}}. It consists of a \emph{Temporal} Transformer, e.g. with the same architecture as the one described for Helium in \hyperref[sec:helium]{Section~\ref{sec:helium}},
and a smaller \emph{Depth} Transformer.
We denote $\mathrm{Tr}_{\mathrm{Temp}}$ the function represented by the Temporal Transformer, and $\mathrm{Tr}_{\mathrm{Depth}}$ the one for the Depth Transformer. 
For simplicity, and for all steps $s\leq S$, we denote $V_{s} = (V_{s,1}, \ldots, V_{s, K})$ the joint value of all sub-sequences at step $s$.
For a given sequence step $1 \leq s \leq S$,
the Temporal Transformer maps $(V_0, \ldots, V_{s - 1})$ to a \emph{temporal context} vector \begin{equation}
z_s = \mathrm{Tr}_{\mathrm{Temp}}(V_0, \ldots, V_{s - 1}) \in \mathbb{R}^d.
\end{equation}
If we further take a sub-sequence index $1 < k \leq K$, the Depth Transformer maps both $z_s$ along with $(V_{s, 1}, \ldots, V_{s, k - 1})$ to the logits estimate 
\begin{equation}
\label{eq:logits}
    l_{s, k} = \mathrm{Tr}_{\mathrm{Depth}}(z_s, V_{s, 1}, \ldots, V_{s, k - 1}) \in \mathbb{R}^{N_k}.
\end{equation}
We further define $l_{s, 1} = \mathrm{Lin}(z_s) \in \mathbb{R}^{N_1}$, with $\mathrm{Lin}$ a dedicated linear layer. We train $\mathrm{Tr}_{\mathrm{Temp}}$, $\mathrm{Tr}_{\mathrm{Depth}}$ and $\mathrm{Lin}$ so that 
$\softmax(l_{s, k})$ is a good approximation of the distribution of $V_{s, k}$ conditioned on all sub-sequences for the previous steps, and of the previous sub-sequences for the current step, e.g.
\begin{equation}
\begin{cases}
    \softmax(l_{s, 1}) &\approx \proba{V_{s, 1} | V_{0}, \ldots, V_{s-1}}\\
    \softmax(l_{s, k}) &\approx \proba{V_{s, k} | V_{0}, \ldots, V_{s-1}, V_{s, 1}, \ldots V_{s, k - 1}} \quad \text{if $k > 1$}.
\end{cases}
\end{equation}
Importantly, the number of steps in the Temporal Transformer is always equal to $\numtokens$, rather than $K \cdot \numtokens$, and the number of steps in the Depth Transformer is at most $K$.
In practice, the Temporal Transformer receives at each step $s$ as input the sum of $K$ learnt embedding tables representing the value for the last $V_{s - 1}$. 
Given $1 < k \leq K$,  the Depth Transformer receives as input the sum of $z_s$ and of a learnt embedding representing $V_{s, k - 1}$.

\looseness=-1
As detailed in \hyperref[tab:hparams]{Table \ref{tab:hparams}}, our Depth Transformer has 6 layers, a dimension of 1024, and 16 attention heads. Unlike \citet{rqtransformer,uniaudio,zhu2024generativepretrainedspeechlanguage}, we use different parameters per index $k$ for the linear layers, projection and fully connected, in the Depth Transformer. Indeed, different sub-sequences might require different transformations. Given the smaller size of this Transformer, this has no impact on both training and inference time, while \hyperref[tab:ablation-delays]{Table~\ref{tab:ablation-delays}} shows that this depthwise parametrization is beneficial. 

\subsubsection{Audio modeling}
\label{sec:audio_modeling}

\looseness=-1
The audio codec \mimi{} described in \hyperref[sec:mimi]{Section~\ref{sec:mimi}} outputs $Q$ sub-sequences, with 12.5 steps per second of audio. We denote those sequences by $A_{t, q} \in \{1, \ldots, N_A\}$ for $1 \leq t \leq T$ with $T = 12.5 \cdot \mathrm{duration}$, and $1 \leq q \leq Q$ with $Q=8$. 
We insert the audio sub-sequences into the multi-sequence $V$ modeled by the RQ-Transformer. Remember that the first codebook $A_{t, 1}$ corresponds to the semantic information, as detailed in \hyperref[sec:splitrvq]{Section~\ref{sec:splitrvq}}, while the other codebooks correspond to acoustic features. 

\paragraph{Acoustic delay.}

\looseness=-1
We first experimented with simply setting $V = A$ in the modeling. However we find that introducing a slight delay between the semantic and acoustic tokens led to more stable generations. \citet{musicgen} show that this leads to reduced dependencies between the sub-sequences for a given time step, conditioned on the past, thus allowing to use a weaker model to approximate the joint distribution $\proba{V_{s, k} | V_{0}, \ldots, V_{s-1}}$ (in their case, as the product of the conditioned marginals). \citet{lemercier2024mmd} further show a connection between the mutual information between the sub-sequences at a given step, and the quality of the generation: naturally, the more complex the interdependence, the more powerful a model will be needed to estimate them. 

As shown in \hyperref[sec:moshi_ablations]{Section~\ref{sec:moshi_ablations}}, introducing a delay of 1 or 2 steps between the semantic and acoustic features greatly improves the quality of the generation. This allows the Temporal, larger, Transformer to model the inter-dependence between semantic and acoustic features. Formally, given a delay $\tau \in \mathbb{N}$, we have, for all steps $s$
\begin{equation}
\begin{cases}
    \label{eq:v_a_delay}
    V_{s, 1} = A_{s, 1}& \\
    V_{s, q} = A_{s - \tau, q} &\quad \text{if}\quad s \geq \tau + 1, q > 1\\
    V_{s, q} = 0&\quad \text{if}\quad s < \tau + 1, q > 1.\\
\end{cases}
\end{equation}
Note that using RQ-Transformers to model audio was successfully used by~\citet{uniaudio} and \citet{zhu2024generativepretrainedspeechlanguage}. We introduce here the use of per-codebook parameters in the Depth Transformer, and the use of the acoustic delay. Compared with~\citep{zhu2024generativepretrainedspeechlanguage} which first generates all the semantic tokens, we generate them jointly with the acoustic tokens, which allows for the first time a streaming modeling of semantic and acoustic tokens jointly.

\subsubsection{Multi-stream modeling}
\label{sec:multistream}
\looseness=-1
Modeling a single stream of audio is not sufficient to fully model a conversation. Our framework can be extended to modeling a two-speaker conversation: given two streams of audios $(A_{t, q})$ and $(A'_{t, q})$, we simply apply the acoustic delay to both, and concatenate them into $V$, extending \autoref{eq:v_a_delay}. In practice, $A$ will correspond to \ours, while $A'$ models the user.

\begin{figure}
\centering
\includegraphics[width=0.95\textwidth]{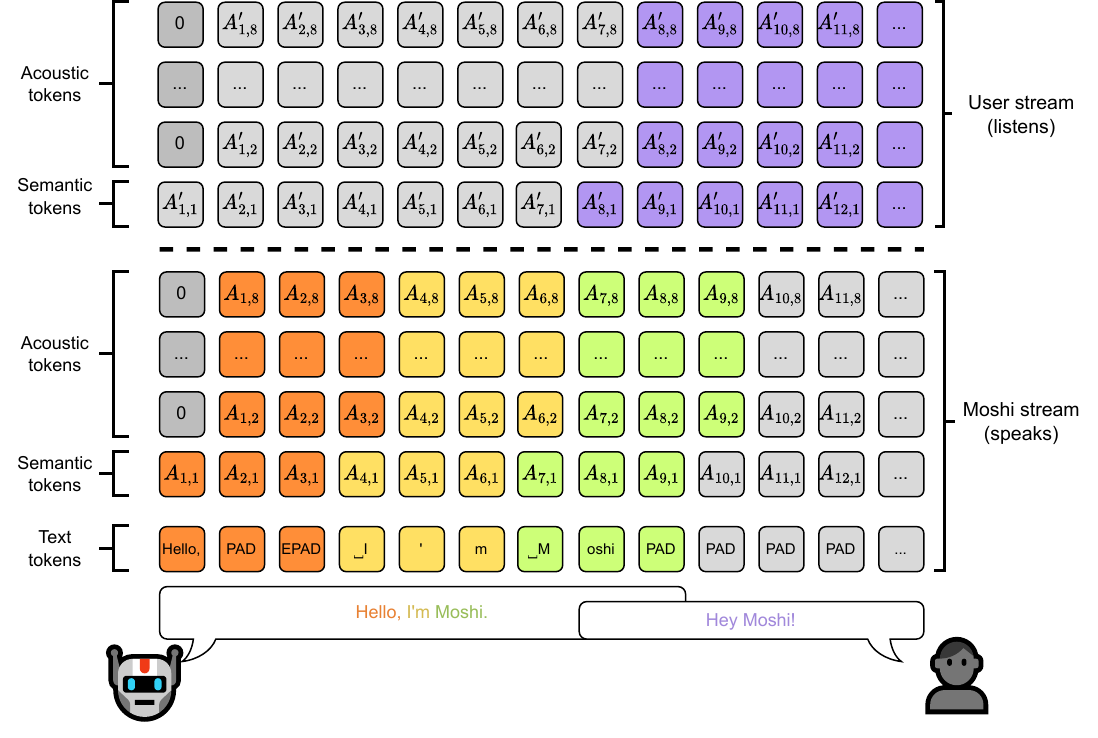}
\caption{\textbf{Representation of the joint sequence modeled by Moshi}.
 Each column represents the tokens for a given step in the joint sequence $(V_{s, k})$ described in~\autoref{eq:final_multi_sequence} with an acoustic delay $\tau=1$, e.g. the input of the Temporal Transformer for this step.
 Tokens are predicted from bottom to top in the Depth Transformer. At inference time,
 tokens under the dashed line (corresponding to Moshi) are sampled, while those above are fed from the user. This design allows for our model to handle overlapping speech turns.}
\label{fig:rainbow}
\end{figure}

\subsubsection{\interleaving}
\label{sec:interleaving}

While operating purely in the audio domain already yields convincing results (see \hyperref[tab:textless]{Table~\ref{tab:textless}}), we observe
that having Moshi also model the textual representation of its own speech is providing a 
scaffolding that increases the linguistic quality of its generation. Formally, we define a text stream $W \in \{1, \ldots, N_W\}^T$ derived from the sequence of text tokens obtained by applying the SentencePiece tokenizer~\citep{kudo2018sentencepiece} to the transcription of the audio corresponding to Moshi with Whisper~\citep{whisper}, as described after. We insert $W$ as the first sub-sequence in $V$, such that it acts as a prefix to the generation of semantic tokens. This can be seen as an extension of the hierarchical semantic-to-acoustic generation introduced by~\citet{audiolm}. Note that we do not use the textual representation corresponding to the stream of the user, as transcribing this flux in real time would be challenging, and relying on an external ASR system contradicts our end-to-end speech-to-speech approach. Ablation studies in \hyperref[sec:moshi_ablations]{Section~\ref{sec:moshi_ablations}} show that among the design choices made for \ours{}, \interleaving{} has one of the most critical impacts on the quality of generated speech.

\paragraph{Aligning text and audio tokens.}

To integrate text tokens with audio tokens that operate at a constant framerate of 12.5Hz, we need to align them to this framerate. For that, we leverage the word-level timestamp provided by Whisper. 
The $i$-th word in the transcript is mapped to $n_i \in \mathbb{N}^*$ text tokens $w_{i, j}$, $j \leq n_i$, along with a start index $t_i \in \{1, \ldots T\}$, simply defined
as its start timestamp divided by the framerate of 12.5 Hz.
We define two special tokens: \texttt{PAD} and \texttt{EPAD}, that
never appear in any of the word tokens. We build $W$ such that when a word starts,
$(W_t)$ contains its text tokens, followed by \texttt{PAD} until the next word. \texttt{EPAD} is inserted before the next word to indicate the end of the padding. While
not strictly necessary, we observed this provided a useful guidance to the model by splitting the decision of ending a word, and which one should follow, into two steps.

First, the sequence $(W_t)$ is initialized with \texttt{PAD} tokens, e.g. $\forall t, W_t \leftarrow \text{\texttt{PAD}}$. Then, we modify it iteratively as follows. For each word $i$ and its start index $t_i$, we update $W$ as
\begin{equation}
\left\{\begin{array}{ll}
    W_{t_i - 1} &\leftarrow \text{\texttt{EPAD}} \\
    W_{t_i + j} &\leftarrow w_{i, j} \qquad \forall j \leq n_i.
\end{array}\right.
\end{equation}
\looseness=-1
Note that if $t_i = 1$, we instead insert \texttt{EPAD} at index 1, and shift the text tokens. We do not insert an \texttt{EPAD} token if it would overwrite a text token from a previous word. As text tokens are more compact than the corresponding audio tokens, there is 
usually no overlap between words in $W_t$. In English conversational speech, we observe that padding tokens represent about 65\% of the tokens.

\paragraph{Deriving streaming ASR and TTS.}

One can further introduce some delay between the text sequence $(W_t)$, and the audio tokens $(A_{t, q})$. This controls in which modality the language model will take the decision about the content of the generated audio. By setting the audio ahead of the text, 
the content of the text will be dictated by what audio has been sampled in the previous steps. In particular, by sampling only the text tokens, while using the ground truth audio tokens and discarding the prediction of the model for them, one obtain a streaming Automatic Speech Recognition model, which also provides precise word level alignment. On the other hand, by changing the text delay so that the text is ahead of the audio tokens, the content of the audio is dictated by the text content. Once more, given a sequence of properly padded text tokens, one obtain a streaming Text-To-Speech model. We further describe in \hyperref[app:streaming_tts]{Appendix~\ref{app:streaming_tts}} how one can adapt the inference of a language model with delayed text to obtain a zero-shot properly padded text tokens sequence. 
Experiments in \hyperref[sec:streaming_asr_tts]{Section~\ref{sec:streaming_asr_tts}} show how a single delay hyper-parameter allows for switching from an ASR to a TTS model with no changes in the loss, architecture, or training data.

\paragraph{Joint sequence modeling for Moshi.}

Putting together the multi-stream and inner monologue, we have the final set $V$ of sequences to model defined as
\begin{equation}
\label{eq:final_multi_sequence}
    \left\{\begin{array}{lll}
    V_{s, 1} &= W_{s} \quad &\text{aligned text tokens.} \\
    V_{s, 2} &= A_{s, 1} \quad &\text{semantic tokens of Moshi.}\\
    V_{s, 1 + q} &= A_{s - \tau, q} \quad \text{if}\quad s \geq \tau + 1, 1 < q \leq Q \quad &\text{delayed acoustic tok. of Moshi.}\\
    V_{s, 1 + Q + 1} &= A'_{s, 1} \quad &\text{semantic tokens of \emph{other}.}\\
    V_{s, 1 + Q + q} &= A'_{s - \tau, q} \quad \text{if}\quad s \geq \tau + 1, 1 < q \leq Q\quad &\text{delayed acoustic tok. of \emph{other},}\\
    \end{array}\right.
\end{equation}
amounting to a total number of $K=2Q+1$ streams, with $Q=8$ in the experiments. 
A summary is provided in \hyperref[fig:rainbow]{Figure~\ref{fig:rainbow}}.

\paragraph{Inference of Moshi.}
\looseness=-1
The joint sequence given by \autoref{eq:final_multi_sequence} is the target for our modeling task at train time: At any time step $s$, the model is input with $0, V_1, \ldots, V_{s - 1}$ and output an estimated probability distribution $\hat{V}_{s}(0, V_1, \ldots, V_{s - 1})$.
At inference time, we sample from $\hat{V}_{s, k}$ for all the sub-sequence indexes that corresponds to outputs of Moshi: i.e., for $k=1$ for the text tokens corresponding to Moshi's speech, and for $k \in \{2, \ldots, 2 + Q\}$ for Moshi's audio tokens. In an application setting, prediction for the audio coming from the user ($k > 2 + Q)$ is actually ignored, as the actual user audio is used instead. However, modeling the user stream as output allows generating simulated dialogues, which is necessary for offline evaluation as in \hyperref[sec:eval_dialogues]{Section~\ref{sec:eval_dialogues}}. Interestingly, there is no explicit boundaries for the change of turns between the user and Moshi: Moshi can speak and listen at all time, and do both at once if needed. In particular, when the user speaks and Moshi stays silent, the corresponding audio tokens for Moshi's stream decode into ``natural silence", a near silent waveform, instead of having a fixed, well defined value; At the same time, \ours{}'s text stream will be filled with \texttt{PAD} tokens. As a result, the text stream can provide interesting ways of controlling Moshi, for instance, forcing the sampling of a \texttt{EPAD} token will make Moshi start talking immediately.

\section{Datasets and Training}
\label{sec:maindatasets}

\subsection{Text Data}
\label{sec:text_dataset}
Our training dataset is made of a mix of high-quality data sources and filtered web data from CommonCrawl.
More specifically, 12.5\% of our dataset is from the following curated sources: Wikipedia,\footnote{\url{https://dumps.wikimedia.org/}} Wikibooks, Wikisource, Wikinews,
StackExchange\footnote{\url{https://archive.org/details/stackexchange}} and the collection of scientific articles pes2o.\footnote{\url{https://github.com/allenai/peS2o}}
Instead of doing multiple passes on Wikipedia, we use five different dumps from 2017, 2018, 2019, 2021 and 2022.
The remaining 87.5\% of our dataset is from CommonCrawl, and was filtered with the pipeline described in \hyperref[sec:data_filtering]{Section \ref{sec:data_filtering}}.
We used the following ten crawls: \texttt{2018-30}, \texttt{2019-04}, \texttt{2019-30}, \texttt{2020-05}, \texttt{2020-34}, \texttt{2021-04}, \texttt{2021-31}, \texttt{2022-05}, \texttt{2022-33}, \texttt{2023-40}.

\subsection{Audio Data}
\label{sec:audio_data}
\looseness=-1
We use an audio collection of 7 million hours, which we call the \emph{unsupervised audio dataset}, of readily available audio content, the majority
of which contains English speech. We transcribe this set with Whisper~\citep{whisper}, using the large-v3 model. We use this data for the audio pre-training phase, during which we do not use the multi-stream approach described in \hyperref[sec:multistream]{Section~\ref{sec:multistream}}, but instead use a single stream of audio representing all speakers at once. Similarly, the text stream described in \hyperref[sec:interleaving]{Section~\ref{sec:interleaving}} represents the words coming from all speakers. All the audio is resampled to 24kHz and downmixed to mono.

\looseness=-1
To achieve multi-stream, we need the model to gain the ability to both listen and speak at the same time. For this, we further leverage the Fisher dataset~\citep{cieri2004fisher}. It consists of 2000 hours of
phone conversations between randomly paired participants, with a given topic to discuss. A property of Fisher is that each conversation side is recorded on a separate channels, which allows providing ground-truth separated streams to \ours{}. The original audio is sampled at 8kHz, and we use AudioSR~\citep{liu2023audiosr} to upsample it to 24kHz.

\looseness=-1
Finally, we source 170 hours of natural and scripted conversations between multiple pairs of participants, recorded with separate channels per speaker, in order to provide a small dataset on which to finetune the model to improve the quality over the one obtained when using only Fisher. We call this dataset the \emph{supervised multi-stream dataset}. We do not train Moshi directly on this dataset, but use it to train a realistic multi-stream TTS model, and fine-tune Helium on real conversation transcripts as explained in Sections~\ref{sec:instruct_data} and \ref{sec:stages}.

\looseness=-1
For both Fisher and this last dataset, we sample one speaker randomly as the main speaker (i.e., \ours{} speaking), and put the other speaker on the second audio stream. For Fisher, the text stream only contains 
the transcription of the main speaker. To obtain reliable timestamps, despite long silences in each stream, we 
use transcription obtained with the \texttt{whisper-timestamped} package~\citep{lintoai2023whispertimestamped}, along with the medium Whisper model.

\subsection{Speech-Text Instruct Data}
\label{sec:instruct_data}

Early experiments using text-based instruct datasets such as Open Hermes~\citep{OpenHermes2} proved to be ill-suited for the instruct tuning of a spoken conversational system.
In particular, the data formatting was often impossible to properly render with TTS (e.g. URLs), and the format of the questions and responses was not following a natural oral flow (e.g. bullet points, long enumerations). Instead, we leverage Helium, fine-tuned on Open Hermes and transcripts of real conversations, to generate realistic interactions between a speech-based AI model and a user.
We then synthesize them with our multi-stream streaming TTS described in \hyperref[app:streaming_tts]{Appendix \ref{app:streaming_tts}}, leading to more than 20k hours of synthetic speech data.
To give \ours{} its own consistent voice, we also condition the TTS engine on the voice of a single actor, who recorded monologues covering more than 70 speaking styles, as listed in \hyperref[tab:list_voices]{Table~\ref{tab:list_voices}}. Experiments on voice consistency reported in \hyperref[sec:voice_consistency]{Section~\ref{sec:voice_consistency}} show that simply using a consistent voice for \ours{} during instruction tuning is enough to guarantee almost surely that it does not use another voice, without further control during inference.
In contrast, the voice of the second audio stream (the user) is randomly sampled for each example, giving more robustness to different speaking conditions and accents.

To generate the transcripts, we use different prompts, aiming at capturing different kinds of interactions between a user and \ours{}.
First, we generate conversations about general knowledge, by starting from a few Wikipedia paragraphs or StackExchange posts, which we refer to as context.
This ensures that \ours{}'s conversations cover a wide range of topics, such as history, cooking advice or pop culture.
More precisely, using a given context, we obtain a summary of a potential discussion with the following prompt:
\begin{tcolorbox}
  \texttt{\textcolor{orange}{\{\{context\}\}}}
  
  \vspace{1em}
  \texttt{Based on information from the previous paragraph, write the summary of a conversation about \textcolor{orange}{\{\{title\}\}} between Blake and Moshi. The summary must be 2 sentences long, and start with "They" or "The speakers".}
\end{tcolorbox}

\noindent where \texttt{\{\{context\}\}} refers to paragraphs from Wikipedia or StackExchange and \texttt{\{\{title\}\}} is the corresponding title.
Then, we generate the full transcript with the prompt:
\begin{tcolorbox}
  \texttt{\textcolor{orange}{\{\{context\}\}}}

  \vspace{1em}
  \texttt{Write the transcript of a conversation between Blake and  \ours{}. \textcolor{orange}{\{\{summary\}\}} \ours{} is knowledgeable about the topic. Use some backchanneling. Use short turns.}
\end{tcolorbox}
Similarly, to give \ours{} information about itself and the Kyutai lab, we generate paragraphs describing both and use them as additional context.

Second, we produce interactions containing instructions about  \ours{}'s voice, such as the other speaker requesting \ours{} to speak with an angry voice or like a pirate.
Our first strategy is to generate single turn interactions where the model is instructed to tell a sentence, a monologue or a poem about an entity, belonging to a high level category such as “sports” or “animals”, using a particular voice.
The voice requested by the other speaker and the entity are randomly sampled, and are thus completely unrelated.
Our second strategy is to generate roleplaying situations, corresponding to different emotions or speaking styles with the following prompt:
\begin{tcolorbox}
  \texttt{Write a list of 10 situations about a \textcolor{orange}{\{\{voice\}\} \{\{character\}\}}. Each situation must start with "a \textcolor{orange}{\{\{voice\}\} \{\{character\}\}} who" and must be at most 8 words long.}
\end{tcolorbox}
Examples of voice adjective include ``happy'' or ``suprised'' and examples of characters include ``detective'' or ``superhero''.
We then generate the interaction using the prompt:
\begin{tcolorbox}
\texttt{Write a dialogue between Blake and Moshi, \textcolor{orange}{\{\{situation\}\}}. Use a lot of backchanneling.}
\end{tcolorbox}

\looseness=-1
To make Moshi robust to mispronounced words, we also generate instructions containing misspellings in the user’s questions, followed by Moshi asking the user to repeat herself or to clarify the question.
We also generate questions containing a false or misleading fact (such as “Is the Eiffel Tower in Beijing?”), to train the model to answer “No” and correct the user.
Otherwise, the vast majority of generated conversations only contain questions from the user where Moshi should answer positively.
We generate basic math, grammar or trivia single-turn questions and answers, as we noticed that Moshi was initially not performing well on simple factual tasks like adding numbers.
Finally, we generate safety conversations, where the user asks unethical or NSFW questions, and Moshi refuses to answer these requests.

\subsection{Training Stages and Hyper-parameters}
\label{sec:stages}
\paragraph{Helium pre-training.}
\looseness=-1
An overview of the training stages and hyper-parameters is provided in \hyperref[tab:hparams]{Table~\ref{tab:hparams}}.
For each stage, we use AdamW~\citep{adamw}, with a weight decay of 0.1, a momentum decay of 0.9, and a decay for the average of the squared gradient of 0.95. All models are trained on H100 GPUs, using FSDP and activation checkpointing. The text-only language model, Helium, is trained for 500k steps, with a batch size of 4.2M tokens, using a cosine learning rate schedule starting at $3\cdot 10^{-4}$ with linear warmup.

\paragraph{Moshi pre-training.}
\looseness=-1
Then, we initialize the Temporal Transformer in Moshi with Helium, while the Depth Transformer described in \hyperref[sec:joint_ar]{Section~\ref{sec:joint_ar}} is randomly initialized. 
We first train on the unsupervised audio dataset presented in \hyperref[sec:audio_data]{Section~\ref{sec:audio_data}}, using a single stream of audio, with a batch size covering 16 hours of audio, each batch item consisting of a 5 mn sequence. We mask the corresponding text tokens with a probability of 30\%. We randomize the delay between the text and audio tokens between -0.6 and +0.6 seconds.
In order to prevent catastrophic forgetting, we also train half of the time on batches of text only data from the same dataset as used for Helium. In total, we make 1 million training steps, with a cosine learning rate starting at $3\cdot 10^{-5}$ for the Temporal Transformer, and $2\cdot 10^{-4}$ for the Depth Transformer, also with a linear warmup. In order to ensure the updates from the text-only batches are balanced with those from the audio dataset, we use two separate optimizer states. In addition, when operating on the text stream from an audio batch, we multiply the learning rate for the text embedding and text linear layer by 0.75. Finally, as padding tokens are predominant for audio batches, we reduce their weight by 50\% in the cross-entropy loss.

\paragraph{Moshi post-training.}
\looseness=-1
Starting from the model obtained from the previous stage, we then train it to gain its multi-stream ability. First, we use PyAnnote~\citep{Bredin23} to diarize the audio from the unsupervised audio dataset. We sample one speaker at random, which will act as the main speaker, and derive a binary mask over the waveform, with a value of $1$ when the speaker is active based on the diarization, and $0$ otherwise. This mask provides us with two waveforms: one with the speaker, and one with the residual (potentially several speakers), which are encoded separately and then used as the two input audio streams described in \hyperref[sec:multistream]{Section~\ref{sec:multistream}}. The text stream only contains the text tokens from the selected main speaker, and the delay between text and audio tokens is fixed to 0. We train for 100k steps, with a batch size of 8 hours of audio, and a fixed learning rate of $3\cdot 10^{-6}$ for the Temporal Transformer, and $5\cdot 10^{-5}$ for the Depth Transformer. Like for the pretraining phase, we sample full text-only batches 10\% of the time.

\paragraph{Moshi finetuning.}
\label{sec:moshift}

The previously described simulated multi-stream provides a good pre-training task but is far from being sufficient to capture natural conversations: For instance, it contains no overlap, and the stream of an inactive speaker is perfectly silent. We then use the Fisher dataset~\citep{cieri2004fisher} to have the model learn real multi-stream interaction. We sample one of the two speakers to be the first (and main) speaker. We train for 10k batches, with a batch size of 40min of audio with a learning rate of $2\cdot 10^{-6}$/$4\cdot 10^{-6}$ for the main/Depth Transformer. We no longer sample full text batches.

Finally, we set the speaker identity for the first speaker stream to be that of Moshi, a useful conversational assistant, with a final stage of instruct finetuning. We use the synthetic instruct dataset described in \hyperref[sec:instruct_data]{Section~\ref{sec:instruct_data}}, with a batch size of 2.7 hours of audio, for 30k steps, with a learning rate of $2\cdot 10^{-6}$ for both transformers. 

During this stage, we perform data augmentation on the user's stream to make Moshi robust to various situations. Namely, we apply a random gain to the user stream between -24 dB and +15 dB, 50\% of the time. 30\% of the time, we further add noise extracts from the Deep Noise Suppression challenge~\citep{dubey2023icassp} which we concatenate in order to cover the entire duration of each example. The noise is amplified to reach a target volume relative to the original source between -30 dB and +6 dB. Each time we need to sample a new noise, we alternatively use a silent section with a random duration up to 30 seconds of silence with probability of 50\%, so that the model can handle the audio condition going from noisy to silent, and vice versa. We emulate echo from Moshi into the user's microphone by adding a scaled down copy of Moshi's stream into the user's stream, scaled by as factor uniformly sampled in $[0, 0.2]$, and a delay uniformly sampled between $[100\text{ms}, 500\text{ms}]$. Finally, we apply to the user's stream, potentially augmented with the echo, a reverb-like augmentation as introduced by~\citet{defossez2020real}. The echo and reverb are applied together with a probability of 30\%.

\paragraph{TTS Training.}

We also train a streaming, multi-stream text-to-speech model, using the method described in \hyperref[sec:interleaving]{Section~\ref{sec:interleaving}}. The audio pre-training stage is shared with Moshi, while the post-training is completed using a delay of 2 seconds for the audio stream compared to the text. The model is finetuned on the supervised multi-stream dataset containing high quality recording of interactions between two speakers. It is used to generate the synthetic finetuning instruct dataset described in \hyperref[sec:instruct_data]{Section~\ref{sec:instruct_data}}. Note that Moshi itself is not trained on the supervised multi-stream dataset. Further details are provided in \hyperref[app:streaming_tts]{Appendix~\ref{app:streaming_tts}}.

\paragraph{Training loss.}
Moshi is trained to model joint sequences, as presented in eq.~\ref{eq:final_multi_sequence}. Given the ground-truth discrete token $(V_{s, k})_{s\leq S, k\leq K}$, 
and the estimated logits $(l_{s, k})_{s\leq S, k\leq K}$ from eq.\ref{eq:logits}, we use the following loss, with $\mathrm{CE}$ the cross entropy,
\begin{equation}
L(V, l) = \frac{1}{S}\sum_{s=1}^S\left(\mathrm{CE}(l_{s, 1}, V_{s, 1}) + \frac{1}{\sum_{k=2}^K \alpha_k}\sum_{k=2}^K \alpha_k \mathrm{CE}(l_{s, k}, V_{s, k})\right).
\end{equation}
Thus, we give the same importance to the text token ($k{=}1$), and the combined audio tokens. $\alpha_k$ is set to 100 for semantic tokens, and 1 for acoustic ones.

\section{Evaluation}
\label{sec:evaluation}

\begin{table}[t]
  \centering
  \caption{\textbf{Text language model evaluation}. Performance on standard benchmarks for evaluating large language models, including closed book question answering, reasoning and multiple choice QA exams. We report in bold the best performing model trained on less than 2.5T tokens.
  \label{tab:llm_eval}}
  \footnotesize
  \begin{tabular}{lcccccccccc}
    \toprule
    & ARCe & ARCc & OBQA &  HS  &  WG  & PIQA & SIQA & TQA  &  NQ  & MMLU \\
    \midrule
    \helium & \textbf{79.6}  & \textbf{55.9}  & 53.6 & 76.3 & \textbf{70.0} & 79.4 & \textbf{51.0} & \textbf{59.9/72.6} & 23.3 & \textbf{54.3} \\
    \midrule
    MPT     & 70.5  & 46.5  & 51.4 & \textbf{77.6} & 69.9 & \textbf{80.6} & 48.5  & -/61.2 & 20.8 & 30.8 \\
    Falcon  & 73.7  & 47.5  & 53.0 & 76.3 & 68.9 & 80.3 & 47.2 & -/64.6 & 21.0 & 28.0 \\
    Llama 2  & 75.2  & 45.9  & \textbf{58.6} & 77.2 & 69.2 & 78.8 & 48.3 & -/72.1 & \textbf{25.7} & 45.3 \\
    OLMo    & 67.2  & 42.5  & 50.0 & 75.5 & 69.8 & 77.5 & -  & -/- & -  & 52.0 \\
    \midrule
    Mistral & 80.5  & 54.9  & 52.2 & 81.0 & 74.2 & 82.2 & 47.0$^*$ & 62.5/- & 23.2 & 62.5 \\
    Gemma 1  & 81.5  & 53.2  & 52.8 & 81.2 & 72.3 & 81.2 & 51.8 & 63.4/- & 23.0 & 64.3 \\
    \bottomrule
  \end{tabular}%
\end{table}

\subsection{Text Language Modeling}
\label{sec:helium_eval}
\paragraph{Metrics.}
We evaluate \helium (trained only on text data) on the following standard benchmarks: AI2 Reasoning Challenge~\citep[ARC]{clark2018think}, Open-Book QA~\citep[OBQA]{mihaylov2018can}, HellaSwag~\citep[HS]{zellers2019hellaswag}, WinoGrande~\citep[WG]{sakaguchi2021winogrande}, Physical Interaction QA~\citep[PIQA]{bisk2020piqa}, Social Interaction QA~\citep{sap2019socialiqa}, TriviaQA~\citep[TQA]{joshi2017triviaqa}, Natural Questions~\citep[NQ]{kwiatkowski2019natural} and Massive Multitask Language Understanding benchmark~\citep[MMLU]{hendrycks2020measuring}.
These benchmarks cover a wide variety of tasks, including common sense reasoning, closed-book question answering or multiple choice question answering from high school and college subjects.
We follow the evaluation protocol from previous work such as GPT-3 or Llama: we perform 5-shot evaluation on TriviaQA, NQ and MMLU, and 0-shot evaluation on the other datasets.
On TriviaQA, we report performance on the Unfiltered and Wikipedia splits.

\paragraph{Baselines.}
As baselines, we consider existing large language models with a size around 7B parameters, and which are trained using roughly the same amount of compute as Helium.
More specifically, we include models that are trained on fewer than 2.5T tokens (compared to the 2.1T tokens that are used to train Helium), namely MPT~\citep{MosaicML2023Introducing}, Falcon~\citep{almazrouei2023falcon}, Llama 2~\citep{touvron2023llama} and OLMo~\citep{Groeneveld2023OLMo}.
We also include Mistral and Gemma, two popular open weights models that are trained using significantly more compute than \helium.

\paragraph{Results.}
We report results in \hyperref[tab:llm_eval]{Table~\ref{tab:llm_eval}}, and we observe that on most benchmarks, Helium is on-par or outperforming models using similar amount of training compute.
Even compared to Mistral and Gemma, which use up to 3x more compute for training, Helium obtains competitive results on some benchmarks such as ARC, Open-Book QA or Natural Questions.
This validates the quality of our pre-training text data.

\subsection{Audio Tokenization}
\label{sec:mimi_eval}

\looseness=-1
\paragraph{Metrics.}
We then evaluate the semantic and acoustic performance of our neural codec, \mimi. First, we evaluate whether the semantic tokens it produces provide targets that are amenable to language modeling. To do so, we compute a triphone-based ABX~\citep{abx} error rate that characterizes the phonetic discriminability of a representation space by comparing distances between two embeddings of different instances of a same triphone (e.g.``beg'') and a negative triphone that differs minimally (e.g.``bag''). More precisely, we compute a ``within speaker" ABX where the three instances are pronounced by the same speaker, and report error rates on Librispeech~\citep{librispeech} dev-clean with the default parameters of the Librilight \citep{librilight} repository\footnote{\url{https://github.com/facebookresearch/libri-light/blob/main/eval/README.md}}. The resulting score has been shown to be a strong predictor of the ability of a downstream audio language model to produce coherent speech~\citep{gslm}. Since we are interested in characterizing only the semantic token, we compute distances in the latent space produced after quantization with the semantic VQ only (i.e. before summing with acoustic tokens). 
\looseness=-1
Second, we evaluate the acoustic quality of reconstructed audio. As objective, automatic metrics we rely on VisQOL~\citep{hines2015visqol}--- a full-reference model of acoustic similarity--- and MOSNet~\citep{mosnet}--- a reference-free model of audio quality. Given the limitations of automatic evaluation of audio quality, we also perform human evaluations with a MUSHRA protocol. We rely on judgments of 20 listeners, each one rating 30 samples of 10s each. \hyperref[tab:mimi_ablations]{Table~\ref{tab:mimi_ablations}} reports ablations studies using objective metrics, while \hyperref[tab:mimi_baselines]{Table~\ref{tab:mimi_baselines}} provides a comparison with previous work both in terms of objective and subjective evaluation.

\paragraph{Baselines.}
We compare against RVQGAN~\citep{kumar2024high}, SemantiCodec~\citep{liu2024semanticodec}, and SpeechTokenizer ~\citep{zhang2024speechtokenizer}. RVQGAN is a pure acoustic tokenizer, in the sense that it does not encode semantic information. Thus, we only evaluate it in terms of audio quality. RVQGAN produces tokens at 75Hz, so we only keep the first two levels of RVQ to obtain a bitrate of 1.5kbps, closer to that of \mimi{}. On the other hand, SpeechTokenizer relies on distillation to encode semantic information into its first token such that we can evaluate both its semantic and acoustic properties. We keep its first 3 RVQ levels to obtain a 1.5kbps bitrate. Similarly, SemantiCodec also encodes semantic and acoustic information such that it can be evaluated along both axes.

\begin{table}[t]
  \centering
  \caption{\textbf{Ablation study on hyper-parameters of the \mimi codec}. We evaluate semantic modeling by reporting the error rate on a phonetic ABX discriminability task. To evaluate the reconstruction quality, we compute VisQOL and MOSNet and collect human judgments with a MUSHRA protocol. ``Quantization rate'' refers to applying quantization to the latent space only $50\%$ of the time during training (independently from quantizer dropout), as described in \hyperref[sec:mimi]{Section ~\ref{sec:mimi}}.}
  \label{tab:mimi_ablations}
  \resizebox{\textwidth}{!}{
  \begin{tabular}{ccccc|cccc}
    \toprule
    Quantization & Transformer & Transformer & WavLM & Split & \multirow{2}{*}{ABX ($\downarrow$)} & \multirow{2}{*}{VisQOL ($\uparrow$)} & \multirow{2}{*}{MOSNet ($\uparrow$)} & \multirow{2}{*}{MUSHRA ($\uparrow$)} \\
    Rate & in encoder & in decoder & distillation & quantizer &  &  &  \\
    \midrule
     \checkmark & \checkmark & \checkmark & \xmark & \xmark & 23.3\% & 2.91 & 2.89 & 65.9\pmr{1.7}\\ %
     \checkmark & \checkmark & \checkmark & \checkmark & \xmark & 6.5\% & 2.22 & 2.87 & 57.8\pmr{1.8} \\ %
     \checkmark & \xmark & \checkmark & \checkmark & \checkmark & 10.8\% & 2.79 & 2.85 & 59.7\pmr{1.7} \\ %
     \checkmark & \checkmark & \xmark & \checkmark & \checkmark & 8.1\% & 2.59 & 2.72 & 48.4\pmr{1.7} \\ %
     \xmark & \checkmark & \checkmark & \checkmark & \checkmark & 8.0\% & 2.45 & 2.88 & 68.3\pmr{1.7} \\ %
     \checkmark & \checkmark & \checkmark & \checkmark & \checkmark & 8.1\% & 2.82 & 2.89 & 64.0\pmr{1.7} \\ %
    \bottomrule
  \end{tabular}
  }
\end{table}

\begin{table}[t]
  \centering
  \caption{\textbf{Audio quality evaluation}. Objective and subjective (MUSHRA) evaluation of audio quality for baseline neural audio codecs---RVQGAN~\citep{kumar2024high}, SemantiCodec~\citep{liu2024semanticodec}, and SpeechTokenizer ~\citep{zhang2024speechtokenizer}--- and the most important variants of \mimi. For a fair comparison with SemantiCodec and SpeechTokenizer, we also include a downsampled version of our codec in the MUSHRA study. $f_s$ is the audio sample rate and $f_r$ the codec frame rate. Both \mimi codecs are trained with distillation, and either with the same combination of reconstruction and adversarial losses as Encodec (see \hyperref[sec:mimi]{Section~\ref{sec:mimi}}) or adversarial losses only.}
  \label{tab:mimi_baselines}
  \resizebox{\textwidth}{!}{
  \begin{tabular}{lcccc|ccc|c}
    \toprule
    Model & $f_s$ & $f_r$ & bitrate & causal & ABX ($\downarrow$) & VisQOL ($\uparrow$) & MOSNet ($\uparrow$) & \textsc{MUSHRA} ($\uparrow$) \\
    \midrule
    Ground Truth & 24kHz & - & - & - & - & - & 3.08 & 90.6\pmr{1.0}\\
    \midrule
    RVQGAN & 24kHz & 75Hz & 1.5kbps & \xmark & - & 1.74& 2.74 & 31.3\pmr{1.3}\\
    SemantiCodec & 16kHz & 50Hz & 1.3kbps & \xmark & 42.2\% & 2.43 & 3.12 & 64.8\pmr{1.5} \\
    SpeechTokenizer & 16kHz & 50Hz & 1.5kbps & \xmark & \textbf{3.3\%} & 1.53 & 2.67 & 45.1\pmr{1.5} \\
    SpeechTokenizer & 16kHz & 50Hz & 4.0kbps & \xmark & \textbf{3.3\%} & 3.07 & 3.10 & 74.3\pmr{1.5} \\
    \midrule
     Mimi, adv. loss only & 24kHz & 12.5Hz & 1.1kbps & \checkmark & 8.7\% & 1.84 & 3.10 & \textbf{81.0}\pmr{1.3} \\
     Same, downsampled at 16kHz & 16kHz & 12.5Hz & 1.1kbps & \checkmark & - & - & - & 77.7\pmr{1.4}\\
     Mimi, non adv. only& 24kHz & 12.5Hz & 1.1kbps & \checkmark & 8.1\% & 2.82 & 2.89 & 58.8\pmr{1.8}\\
    \bottomrule
  \end{tabular}}
\end{table}

\looseness=-1
\paragraph{Results - Semantic tokens.} \hyperref[tab:mimi_ablations]{Table~\ref{tab:mimi_ablations}} shows that \mimi's phonetic discriminability of semantic tokens, as measured by ABX, is poor in the absence of distillation and comparable to acoustic tokens of previous work~\citep{audiolm}: This means these semantic tokens are not amenable to capturing linguistic content from speech. In contrast, distilling WavLM into the semantic tokens significantly improves their phonetic discriminability, in particular when using a Transformer in \mimi's encoder. This can be explained by the fact that distilling a large Transformer based encoder into a purely convolutional one is challenging, while increasing the capacity and receptive field of the encoder helps. Yet,  we observe a conflict between acoustic losses and semantic distillation, as improving ABX implies reducing reconstruction quality (as measured by MUSHRA). Using a split RVQ as described in \hyperref[sec:splitrvq]{Section~\ref{sec:splitrvq}} improves the trade-off between semantic properties and audio quality, improving MUSHRA from 57.8 to 64.0 while moderately degrading ABX from 6.5\% to 8.1\%.

\looseness=-1
\paragraph{Results - Acoustic tokens.} \hyperref[tab:mimi_ablations]{Table~\ref{tab:mimi_ablations}} also shows a significant improvement in MUSHRA when adding a Transformer in the decoder. Similarly, using a quantization rate of $50\%$ significantly improves VisQOL.  Quantization rate however does not improve perceived quality. More generally, we observe a poor correlation between VisQOL and MOSNet. In particular, \hyperref[tab:mimi_baselines]{Table~\ref{tab:mimi_baselines}} shows that training \mimi with adversarial losses only leads to a very low VisQOL of 1.84 which does not account for the high perceived audio quality. We thus rely on MUSHRA where raters are asked to judge the similarity of a reconstructed audio to its ground-truth anchor, with a score between 0 and 100. This human evaluation shows a significant improvement from using adversarial losses only, with a MUSHRA score of 81.0 against 58.8 when using the mix of loss functions used in Encodec. \mimi moreover significantly outperforms RVQGAN~\citep{dac} despite operating at a lower bitrate and modeling semantic information. \mimi also provides higher reconstruction quality than SemantiCodec~\citep{liu2024semanticodec} while operating at a 4$\times$ lower framerate. This property is crucial to achieve the low latency of \ours{}, since generating one temporal frame of audio tokens with \ours{} requires a full forward pass through the Temporal Transformer. Finally, both RVQGAN and SemantiCodec are non-causal, while \mimi is fully causal and thus compatible with streaming inference and modeling of real-time conversations.

\looseness=-1
\paragraph{Discussion.}
\mimi overall provides high reconstruction quality while encoding semantic information, being fully causal, and operating at low framerate and bitrate. In consequence, \mimi proves to be a well-fitted audio tokenizer to train real-time audio language models. A collateral finding of our study is a concerning lack of correlation between objective and subjective audio quality metrics. In particular, while we find VisQOL to provide a reliable proxy for perceived quality when modifying the generator architecture, changing the training objective (e.g. removing reconstruction losses) moves the score in directions that are completely decorrelated from human perception. This observation underscores the open challenge of designing reliable objective proxies for perceived quality.

\begin{table}[t]
    \small	
    \centering
    \caption{
    \textbf{Ablation study on the use of the RQ-Transformer}. All models are initialized with Helium and pretrained on audio. When not using RQ-Transformer, we predict the 8 levels of tokens with independent classification heads, following~\citet{musicgen}. Note that perplexities are only comparable between models with a given delay, as the classification task is easier with more delay for higher tokens.}
    \footnotesize
    \begin{tabular}{ccc}
    \toprule
     Acoustic Delay & RQ-Transformer & Perplexity \\
     \midrule
    $[0, 1, 2, 3, 4, 5, 6, 7]$ & \xmark &  $\phantom{1}42.2$ \\
    $[0, 1, 2, 3, 4, 5, 6, 7]$ & \checkmark &  $\phantom{1}40.3$ \\
    $[0, 2, 2, 2, 2, 2, 2, 2]$ & \xmark &  $135.4$\\
    $[0, 2, 2, 2, 2, 2, 2, 2]$ & \checkmark & $\phantom{1}36.8$ \\
    \bottomrule
    \end{tabular}
    \label{tab:ablation-rq}
\end{table}

\begin{table}[t]
    \small
    \centering
    \caption{
      \textbf{Ablation study on delay patterns, weight of the semantic token and \interleaving{}}. All models are initialized with Helium, pretrained on audio and use the RQ-Transformer. We vary the weight of the semantic token while keeping the weight of other tokens (including the text token when using \interleaving{}) to 1. As different delay patterns cannot be compared in terms of perplexity, we generate continuations from 3s prompts on the valid set, convert them into transcripts with Whisper~\citep{whisper} and report their negative log-likelihood with LiteLlama-460M-1T\protect\footnote{\url{https://huggingface.co/ahxt/LiteLlama-460M-1T}} along with their length (in characters) as proxies for linguistic quality.}
      \footnotesize
    \begin{tabular}{cccccrr}
    \toprule
     Acoustic & Semantic Token & Depthwise & Inner & Transcript  & Transcript \\
      Delay & Weight & Parametrization & Monologue & NLL ($\downarrow$) & Length ($\uparrow$)  \\
     \midrule
    $[0, 0, 0, 0, 0, 0, 0, 0]$ & \phantom{00}1.0 & \cmark & \xmark & $4.36$ & $486$ \\
    $[0, 1, 1, 1, 1, 1, 1, 1]$ & \phantom{00}1.0 & \cmark & & $4.12$ & $529$ \\
    $[0, 2, 2, 2, 2, 2, 2, 2]$ & \phantom{00}1.0 & \cmark & & $4.09$ & $519$ \\
    $[0, 2, 2, 2, 2, 2, 2, 2]$ & 100.0 & & & $3.75$ & $538$ \\
    $[0, 2, 2, 2, 2, 2, 2, 2]$ & 100.0 & \cmark & & $3.65$ & $602$ \\
    $[0, 2, 2, 2, 2, 2, 2, 2]$ & 100.0 & \cmark & \cmark & \textbf{2.77} & \textbf{1920} \\
    \bottomrule
    \end{tabular}
    \label{tab:ablation-delays}
\end{table}

\subsection{Ablations on Generative Modeling}
\label{sec:moshi_ablations}
\paragraph{Metrics.} We perform ablation studies to evaluate the impact of using an RQ-Transformer, as well as comparing delay patterns and weighting of various token levels. All models are initialized with \helium{} for the Temporal Transformer and pretrained on audio data. When comparing models with a same delay pattern and without \interleaving{}, we rely on perplexity, averaged over semantic and acoustic tokens. However, perplexities are not comparable between models with various delays, as they do not model the same conditional distributions. To compare various delay patterns, we measure the ability of \ours{} to generate intelligible, consistent speech by transcribing generations (conditioned on a 3s prompt) with Whisper~\citep{whisper} and scoring the resulting transcripts with an external text language model. We rely on a lightweight text model---LiteLlama-460M-1T\footnote{\url{https://huggingface.co/ahxt/LiteLlama-460M-1T}}--- as it is more practical for continuous evaluation along training. We also report the length of the transcripts (in characters), as we find it to a be a strong predictor of model quality (weak models typically collapse to silence).

\paragraph{Results - RQ-Transformer.} \hyperref[tab:ablation-rq]{Table~\ref{tab:ablation-rq}} reports results for ablations on the use of an RQ-Transformer. We first replicate the setting of \citet{musicgen} with the delay pattern of $[0, 1, 2, 3, 4, 5, 6, 7]$, which means that each level of RVQ token is generated one timestep after the preceding level. In this context, we see that using an RQ-Transformer is not necessary, as it only provides a marginal improvement in perplexity. However, this delay pattern induces a theoretical latency of 8 timesteps, which amounts to 640ms, a latency that is incompatible with the requirements of a real-time dialogue model. We thus switch to a reduced latency of 240ms with the pattern $[0, 2, 2, 2, 2, 2, 2, 2]$. In that context, modeling RVQ tokens with an RQ-Transformer significantly improves perplexity over using separate classification heads. Thus, the RQ-Transformer becomes a critical component of generative models of RVQ tokens under strict latency constraints.

\paragraph{Results - Additional ablations.} \hyperref[tab:ablation-delays]{Table~\ref{tab:ablation-delays}} reports additional ablations on additional delay patterns, the weight of the semantic token loss and our proposed Inner Monologue procedure, all using the RQ-Transformer. 
First, we compare three configurations of delays that are compatible with real-time dialogue. The $[0, 0, 0, 0, 0, 0, 0, 0]$ pattern represents the minimal latency of 80ms that can be obtained with \mimi{} tokens at 12.5Hz. Allowing an additional 80ms of latency with one step of delay significantly improves the quality of generated speech, while 240ms of latency brings further moderate improvement. 
In early experiments, we also observed that the individual losses per RVQ level were conflicting with one another, despite each level being more important in the final intelligibility and audio quality than the next one. We thus bring two changes to the architecture and training process. We first increase the weight of the loss on predicting the semantic tokens to 100, while keeping it at 1 for all other levels of the audio tokens. 
This gives another boost to speech intelligibility. We furthermore reduce competition between RVQ levels by using a depthwise parametrization, as described in \hyperref[sec:joint_ar]{Section~\ref{sec:joint_ar}}, such that each RVQ level is predicted by its own set of weights in the Depth Transformer, rather than having shared weights across levels.
Finally, the most drastic improvement to the quality and length of generated speech comes from enabling \interleaving{}.

\looseness=-1
\paragraph{Discussion.} Beyond the choice of architecture and delay patterns, these ablations show how helpful modeling text tokens along the audio tokens with \interleaving{} is, even in an audio-to-audio setting. Given the positive impact of depthwise parametrization and a weight of 100 for the semantic token, both are used in the subsequent experiments and our final training procedure. As described in \hyperref[tab:hparams]{Table~\ref{tab:hparams}}, we pretrain \ours{} with an acoustic delay of 2 and finetune it with an acoustic delay of 1, for a theoretical latency of 160ms.

\begin{table}[t]
    \centering
    \caption{\textbf{Performance of audio and text language modeling}. We report accuracies based on scoring with negative log-likelihood, normalized by sequence length. MMLU is evaluated in a 5-shot setting. Reusing the terminology of \cite{spiritlm}, $\varnothing$ represents unsupported modalities while - represents unreported numbers.}
    \resizebox{\textwidth}{!}{  
    \begin{tabular}{lccccc}
    \toprule
    &
    \multicolumn{4}{c}{Audio metrics} &
    \multicolumn{1}{c}{Text metrics} \\
    \cmidrule{2-5}
    \cmidrule(l){6-6}
     Model & sWUGGY & sBLIMP & sTopic-StoryCloze & sStoryCloze & MMLU \\
    \midrule
    \multicolumn{6}{c}{\textit{Audio only - Cold Start}} \\
    \midrule[0.3pt]
    GSLM \citep{gslm} &  64.8 &   54.2 &  66.6 &  53.3 & $\varnothing$ \\
    AudioLM \citep{audiolm} &  71.5 &   \textbf{64.7} &   - &   - & $\varnothing$ \\
    TWIST \citep{twist} & 72.2 & 56.5 &  - & - & $\varnothing$ \\
    \midrule[0.01pt]
    \ours{} %
    & \textbf{74.8} & 59.9 & 80.9 & 56.9 & $\varnothing$ \\
    \midrule
    \multicolumn{6}{c}{\textit{Audio only - Warm Start}} \\
    \midrule[0.3pt]
    TWIST \citep{twist} & 74.5 & 59.2 & 76.4 & 55.4 & $\varnothing$ \\
    VoxtLM \citep{voxtlm} &  62.9 &   53.9 &   - &   - & $\varnothing$  \\
    Spirit-LM \citep{spiritlm} &  69.5 &   58.0 &   72.9 &   54.8 & $\varnothing$  \\
    \midrule[0.01pt]
    \ours{} %
    & 74.3 & 58.9 & 81.8 & 58.7 & $\varnothing$  \\
    \midrule
    \multicolumn{6}{c}{\textit{Text and audio - Warm Start}} \\
    \midrule[0.3pt]
    VoxtLM \citep{voxtlm} & 66.1 &  57.1 & - & - &  $\varnothing$ \\
    Spirit-LM \citep{spiritlm} & 69.0 & 58.3 & 82.9 & 61.0 & 36.9  \\
    \midrule[0.01pt]
    \ours{} after single-stream pretraining %
    & 72.6 &58.8 & 83.0 & 60.8 & \textbf{49.8} \\
    \ours{} after multi-stream instruct %
    & 63.0 & 55.2 & \textbf{83.6} & \textbf{62.7} & 49.7 \\
    \ours{} after multi-stream instruct, synthetic voice %
    & 60.9 & 54.6 & 82.5 & 60.9 & 48.7 \\
    \bottomrule
    \end{tabular}}
    \label{tab:textless}
\end{table}

\subsection{Audio Language Modeling}
\label{sec:eval_audiolm}
\looseness=-1
\paragraph{Metrics.} We first measure the ability of \ours{} to model speech sequences when being trained for next token prediction on large scale audio data. To do so, we rely on ``textless NLP''~\citep{gslm} metrics that evaluate an audio language model's linguistic knowledge by comparing likelihoods of positive and negative speech examples represented as audio tokens. sWUGGY evaluates a model's ability to learn a lexicon from speech by comparing the likelihood of an existing word and an invalid variant (e.g. ``oxidation'' and ``accidation''), while sBLIMP evaluates syntactic contrasts. Spoken StoryCloze metrics introduced by~\citet{twist} evaluate semantic contrasts by comparing commonsense five-sentence stories, with the last one being either coherent with the context or incoherent. Given the difficulty of this task in the audio domain, \citet{twist} also propose  Spoken Topic-StoryCloze, a variant  where the negative continuation is randomly sampled among unrelated sentences (rather than being subtly incoherent), resulting in higher scores. We score sequences with a negative-log likelihood normalized by the sequence length. Since our model produces several tokens per timestep, we sum all tokens of a timestep with the weights used during training, i.e. 100 for the semantic tokens and 1 for the acoustic ones. We do not include text tokens from the \interleaving{}, as these metrics are designed to compare untranscribed audio sequences. Similarly, when evaluating a multi-stream model after instruction tuning, we only score the tokens from the user stream as it does not include text tokens. Finally, we also report text understanding  evaluation (without audio tokens) on MMLU~\citep{hendrycks2020measuring} for Spirit-LM and \ours{}, as a way to measure how much the audio training affects the textual knowledge of the original checkpoint.
\looseness=-1
\paragraph{Baselines.} We compare against baselines from the audio language modeling literature, in three settings. The first category encompasses audio-only models starting from a random initialization, including GSLM~\citep{gslm}, AudioLM~\citep{audiolm} and TWIST-1.3B~\citep{twist}. In this case, we report metrics for a single-stream \ours{} initialized randomly and pretrained only on audio data and without \interleaving{}. The second category includes models that start from a pretrained text LM and are then only trained on audio. This includes TWIST-13B as well as the audio-only version of VoxtLM (first row of \citep[Table 3]{voxtlm}) and that of Spirit-LM (reported as ``Speech Only'' in \citep[Table 5]{spiritlm}). The corresponding \ours{} model is similar to the one mentioned above (audio-only data, no \interleaving{}) but starts from the pretrained \helium{} checkpoint. The last category is composed of actual multimodal models that are trained jointly on speech and text data. In this context we report results for three configurations of \ours{}. First, we report results for \ours{} pretrained on single-stream data. Then, we report results for the final model after multi-stream post-training and finetuning using real recordings from a voice actor to condition the creating of synthetic data in Moshi's voice. The last model is identical to the previous one except for the fact that it uses a synthetic voice for \ours{}. We remind the reader that even if these models are trained with \interleaving{}, they are evaluated without, to provide a fair comparison with baselines.
\looseness=-1
\paragraph{Results.} \hyperref[tab:textless]{Table~\ref{tab:textless}} reports results on audio language modeling. In the ``Audio only - Cold Start'' setting, \ours{} already provides a strong baseline, in particular considerably improving over previous work in sTopic-StoryCloze.  When initialized with an \helium{} checkpoint and trained on audio-only data, \ours{} outperforms previous work in this category on most metrics. Finally, while multimodal training improves common sense reasoning from speech (as shown by sStoryCloze performance), we observe mixed effects on lexical and syntactic judgments (sWUGGY and sBLIMP) compared to models trained only on audio data. While single-stream pretraining moderately degrades sWUGGY and sBLIMP, instruction finetuning severely affects sWUGGY, which means that instructed models have a harder time solving lexical judgments. 
We hypothesize that this is due to finetuning \ours{} on data of varying quality and simulating noisy and reverberated conditions for the user stream (used to score spoken pairs for all metrics in \hyperref[tab:textless]{Table~\ref{tab:textless}}) which makes fine lexical judgments harder to solve. Finally, Moshi scores 12 points higher on MMLU than Spirit-LM, thus demonstrating higher general knowledge and text understanding. We moreover emphasize that \ours{} is the only model in \hyperref[tab:textless]{Table~\ref{tab:textless}} that integrates both semantic and acoustic tokens into a single generative model, unlike AudioLM which uses three separate stages, and VoxTLM, TWIST and Spirit-LM that only model semantic tokens and rely on an external vocoder. Thus, \ours{} is the only model in this comparison that demonstrates strong linguistic modeling in both speech and text, while being able to model speech in any arbitrary voice and condition.

\paragraph{Discussion.}
\looseness=-1
While ``textless NLP'' benchmarks have helped developing the first audio language models, we observe that they do not consistently provide good guidance in developing a dialogue model like \ours{}. In particular, we find the lack of correlation between common sense metrics and lexical/syntactic judgments to be frequent and easily explainable by the diversity of acoustic conditions we use in training. Moreover, we do not observe a degradation in the model's lexical variety or intelligibility when finetuning the model, which contradicts the reduction in sWUGGY. This is why we also evaluate spoken question answering in the next section as a way to probe the model's common sense, knowledge and lexical abilities.

\begin{table}
    \footnotesize	
    \centering
    \caption{\textbf{Evaluation of spoken question answering}.
    Spoken question answering (0-shot) on the Web Questions~\citep{web_questions}, LlaMA-Questions~\citep{spectron}, and Trivia QA~\citep{joshi2017triviaqa} benchmarks, synthesized using a TTS engine.
    For the first two, we use the number reported by~\citep{spectron}.
    For LlaMA-Questions, we use the audio provided by~\citep{spectron}.
    For Web Questions and Trivia QA, we synthesize our own, keeping all of the questions. For \ours{}, we only prepend one of the random incipits used during instruct fine tuning. We further provide the performance
    of our \helium text-only model as a top line.}
    \begin{tabular}{lccc}
    \toprule
     Model & Web Q. & LlaMA Q. & Audio Trivia QA\\
    \midrule
    \multicolumn{4}{c}{\textit{Audio only}} \\
    \midrule[0.3pt]
    GSLM \citep{gslm} & \hphantom{0}1.5 &  \hphantom{0}4.0 &   - \\
    AudioLM \citep{audiolm} & \hphantom{0}2.3 & \hphantom{0}7.0 &   - \\
    TWIST (7B) \citep{twist} & \hphantom{0}1.1 & \hphantom{0}0.5 &   - \\
    \midrule[0.01pt]
    \ours (w/o \interleaving{}) & \hphantom{0}\textbf{9.2} & \textbf{21.0} & 7.3   \\
    \midrule
    \multicolumn{4}{c}{\textit{Text and audio}} \\
    \midrule[0.3pt]
    SpeechGPT (7B) \citep{zhang2024speechgpt} & \hphantom{0}6.5 & 21.6 &   14.8 \\
    Spectron (1B) \citep{spectron} & \hphantom{0}6.1 & 22.9 &   - \\
    \midrule[0.01pt]
    \ours & \textbf{26.6} & \textbf{62.3} &   \textbf{22.8} \\ %
    \ours (w/o text batches in pre-training) & 23.2 & 61.3 &   18.3 \\ %
    \midrule
    \multicolumn{4}{c}{\textit{Text}} \\
    \midrule[0.3pt]
    Helium (text) & 32.3 & 75.0 & 56.4 \\
    \bottomrule
    \end{tabular}%
    \label{tab:qa}
\end{table}

\subsection{Spoken Question Answering}
\looseness=-1
\paragraph{Metrics.} 
We evaluate the spoken question answering abilities of our multi-stream \ours{} model. We rely on Spoken Web Questions and Llama Questions, both introduced by~\citet{spectron}. We also synthesize an audio version of TriviaQA. When evaluating \ours{}, we insert the audio tokens of the question into the user stream to simulate a user interaction, along with a final \texttt{EPAD} text token to trigger an immediate response from \ours{}. 

\paragraph{Baselines.} 
We compare to Spectron and baselines used by~\citet{spectron}, all having been already introduced in \hyperref[sec:eval_audiolm]{Section~\ref{sec:eval_audiolm}} except for SpeechGPT~\citep{zhang2024speechgpt}. To measure the impact of \interleaving{} on spoken fluency, we compare these baselines with \ours{}  trained with and without \interleaving{}. As GSLM, AudioLM and TWIST are audio only, \ours{} without \interleaving{} provides a fair comparison. On the other hand, Spectron and SpeechGPT rely on Chain-of-Modality---they generate an answer first as text, and then as speech--- so we compare them to Moshi with \interleaving{}. Moreover, to quantify an eventual degradation in knowledge due to training on audio data, we also compare to \helium{} when evaluated on the textual counterpart to each spoken dataset.
\looseness=-1
\paragraph{Results.} \hyperref[tab:qa]{Table~\ref{tab:qa}} reports accuracies on the three benchmarks. While audio-only \ours{} significantly outperforms baselines in its categories, the most striking result is the impact of \interleaving{} on Moshi's performance, almost tripling its accuracy on all benchmarks. This is remarkable as \interleaving{} only marginally increases inference cost (each multi-stream timestep requires generating 17 tokens, instead of 16 without). We emphasize that among all models in this comparison, Moshi not only provides the best spoken question answering performance, but is also the only one to model jointly semantic and acoustic tokens, such that it can handle interactions between arbitrary voices in many conditions. \ours{} significantly outperforms SpeechGPT and Spectron while it is the only model compatible with streaming inference, as Chain-of-Modality requires generating a full answer in text before generating speech, while \interleaving{} generates both in a streaming fashion.
Finally, we note that Moshi with \interleaving{} but without keeping 50\% of text-only batches during pre-training leads to a noticeable drop in acccuracy, showing that a warm start from Helium is not sufficient to retain the knowledge of the original text model throughout the training.
\looseness=-1
\paragraph{Discussion.} Despite the strong performance of \ours{}, we observe a weaker performance than the base\helium{} model, which is consistent with the reduced MMLU of 49.7 reported in \hyperref[tab:textless]{Table~\ref{tab:textless}} from 54.3 with \helium{}. While the moderate differences on Web Questions and Llama Questions can be explained by training on audio data and thus reducing the amount of parameters dedicated to textual knowledge, the very large difference on Trivia QA incites us to inspect more thoroughly patterns of errors. We find that multiple-sentence questions (e.g. ``The Terror of the Monster was an early title for a best-selling novel which inspired one of the highest-grossing movies of the mid-70's. Under what name did it eventually terrify the reading and film going public?'') or ones with specific syntactic structure (e.g. ``On the human body, a keloid is a type of what?'') are challenging for \ours{}, due to it being finetuned on oral-style conversations that do not display such patterns. We hypothesize that covering more syntactic scenarios during finetuning could reduce this gap.

\begin{table}[t]
    \footnotesize	
    \centering
    \caption{\textbf{Linguistic quality and turn-taking statistics of generated dialogues}. As we train our multi-stream model to generate both sides of the conversation, we can generate dialogues without the need to interact with a real user. This allows evaluating how much \ours{} learns natural conversational dynamics.}
    \resizebox{\textwidth}{!}{  
    \begin{tabular}{lccccccc}
    \toprule
    Model& samples&temp&cond. PPL& IPU & Pause & Gap & Overlap \\
    \midrule
    Best non-cascaded \citep{nguyen-etal-2023-generative}
    & 50 & 1.0 & 195.9 & 41.4s & 13.8s & 10.7s & 6.1s \\
    Cascaded \citep{nguyen-etal-2023-generative}
    & 50 & 1.0 & 45.9 & 54.8s & 0.0s & 5.3s & 0.0s \\
    Ground Truth \citep{nguyen-etal-2023-generative}
    & 50 & $\varnothing$ & 65.0 & 53.5s & 5.5s & 4.4s & 3.6s \\
    \midrule
    \ours{} %
    & 1000 & 0.8 & \textbf{41.9} & 35.1s & 13.2s & 12.5s & 1.2s \\
    \ours{} %
    & 1000 & 0.9 & 56.7 & 44.7s & 9.1s & 7.5s & 2.2s \\
    \ours{} %
    & 1000 & 1.0 & 79.3 & 50.8s & 7.0s & 4.5s & 4.1s \\
    Ground Truth 
    & 1000 & $\varnothing$ & 59.6 & 51.1s & 6.4s & 4.2s & 3.3s \\
    \bottomrule
    \end{tabular}
    }
    \label{tab:multistream}
\end{table}

\subsection{Quality and Statistics of Generated Dialogues}
\label{sec:eval_dialogues}
\paragraph{Metrics.}
\looseness=-1
Beyond single-turn question answering, we evaluate the linguistic quality of the generated dialogues by scoring them with an external language model, while also looking at turn-taking metrics. We follow the methodology of ~\citet{nguyen-etal-2023-generative}. The turn-taking metrics are defined as follows: \emph{Inter-Pausal Units} (IPU) are continuous streches of speech that are separated by a silence of at least 0.2s on each side.
\emph{Pauses} are silences between IPUs of the same speakers.
\emph{Gaps} are silences between IPUs of different speakers.
\emph{Overlaps} are times where there are IPUs for both speakers. Following ~\citet{nguyen-etal-2023-generative}, the semantic evaluation uses the open-source DialoGPT model~\cite{dialogpt} and we compute the perplexity of the transcribed dialogue by separating each speaker using the \texttt{<|endoftext|>} token expected by DialoGPT.
We select 1000 random 10 seconds prompts from the Fisher dataset and use \ours{} to generate continuations. For each one, 32 continuations are generated, for
3 different temperatures as it significantly affects the results.

\paragraph{Baselines.} We compare to dGSLM \citep{nguyen-etal-2023-generative}, as it is also a full-duplex generative model, trained on the Fisher dataset. \citet{nguyen-etal-2023-generative} use 50 prompts with 50 continuations for each and report results for their dialogue model as well as a cascaded topline model (ASR + LM + TTS).

\paragraph{Results.} \hyperref[tab:multistream]{Table~\ref{tab:multistream}} shows that \ours{} performs as well as the cascaded model in terms of linguistic quality, despite being an audio-to-audio model. Both have a perplexity that
is better than the ground truth, which is explained by these models being trained on data
that is closer to what DialoGPT has been trained on compared to the Fisher dataset. This is a strong improvement over the non-cascaded model from~\citep{nguyen-etal-2023-generative}, which is not able to generate coherent speech in this scenario.

\subsection{Streaming ASR and TTS}
\label{sec:streaming_asr_tts}
\looseness=-1
\paragraph{Metrics.} \hyperref[sec:interleaving]{Section \ref{sec:interleaving}} and \hyperref[app:streaming_tts]{Appendix~\ref{app:streaming_tts}} describe how \interleaving{} can provide a streaming TTS or streaming ASR system by simply changing the delay it uses between text and audio tokens. In particular, we train a streaming TTS model by delaying audio tokens by 2 seconds, giving some lookahead to the text tokens, and teacher forcing text tokens at inference. Similarly, we train a streaming ASR model by delaying text tokens by 2 seconds, allowing the model to listen to audio content before generating text tokens. In that case, at inference we teacher force the audio tokens. We perform TTS with a temperature of 0.6 while we use greedy decoding for ASR, and evaluate on LibriSpeech~\citep{librispeech} test-clean in Word Error Rate (WER). For TTS, we first transcribe the generated audio with HuBERT-Large~\citep{hubert} finetuned on LibriSpeech 960h,\footnote{\url{https://huggingface.co/facebook/hubert-large-ls960-ft}} and only consider sequences between 4 and 10s, which allows comparing to baselines such as Vall-E~\citep{valle}. We emphasize that no LibriSpeech data is seen during the training of our ASR and TTS systems.

\looseness=-1
\paragraph{Results.} Our streaming TTS model obtains 4.7\% of WER on LibriSpeech test-clean, which outperforms Vall-E's 5.9\% WER but is worse than NaturalSpeech 3~\citep{ju2024naturalspeech} with 1.81\%. Yet, \ours{} only requires 2 seconds of lookahead when Vall-E and NaturalSpeech 3 require access to the full sequence. Our ASR system yields 5.7\% WER, while a Streaming FastConformer~\citep{noroozi2024stateful} gets 3.6\% with a similar look-ahead. Note that our ASR system also provides alignments of transcripts with a precision of 80ms (the framerate of the Temporal Transformer).

\looseness=-1
\paragraph{Discussion.} This limited experimentation is not intended to compete with state-of-the-art systems (in particular for ASR), but is rather designed to illustrate how \interleaving{} is flexible enough to cast several tasks into the same framework. We also emphasize that the standard evaluation on LibriSpeech test-clean does not provide a testbed to showcase strong capabilities of our TTS system, in particular its ability to model two speakers with multi-stream modeling, and generating consistent dialogues of spontaneous, expressive speech across 5 minutes (while Vall-E only evaluates sequences between 4 and 10 seconds of read speech). We reserve a thorough evaluation of streaming TTS for future work.

\subsection{Compressing \ours{} and Impact on Speech Quality}

With most modern LLMs built off billions of parameters, model size is a well-known bottleneck for practical uses such as running on resource-constrained devices (e.g. laptop with user-grade GPU) or model deployment (e.g. serving many users on an online web demo). 
To address this, Post-Training Quantization (PTQ) is a widely used efficiency technique for compressing model weights and activations,  with the downside of possible performance degradation. Recent work has shown that LLMs can often successfully be quantized to 8 bits with integer quantization, and sometimes to even lower bitwidths using more advanced techniques to handle outlier weights~\citep{pmlr-v202-dettmers23a, llmint8,optq,quipsharp}.
However, the literature on quantizing speech models is much more scarce than that of LLMs. 
Thus, in this section, we investigate how quantizing \ours{} impacts its performance, both linguistically and especially acoustically,  as we  highlight certain audio degradations aggravated by model quantization.
\looseness=-1
\paragraph{Quantization Format.} To quantize \ours{}, we follow common design choices from the PTQ literature. In all results below, we settle on the following setting: (i) Activations are stored in \texttt{bfloat16} precision (BF16) and dynamically quantized to 8 bits using symmetric quantization (a.k.a. AbsMax) at the input of every linear layer; 
(ii) The model weights are quantized using asymetric quantization (a.k.a. MinMax) for different bitwidths and block sizes. This includes both the Temporal Transformer as well as the Depth Transformer weights. In fact, we find that the Depth Transformer is reasonably robust to quantization, as keeping only its weights in high precision does not significantly improve audio quality.
Only the initial embedding layers (both for text and audio), the RMSNorms and the \mimi codec are left unquantized. 
Finally, note that, although weight range setting is also common practice~\citep{nagel2021whitepaperneuralnetwork}, we do not finetune the obtained quantization scales using MSE as we find it has little impact on the quality of generated samples. %
\looseness=-1
\paragraph{Results - Linguistic evaluation.}
To assess how quantization impacts the reasoning ability of the model, we evaluate the quantized models' performance on the MMLU benchmark for the base Helium model trained on text-only data used as foundation for Moshi (\hyperref[tab:quanthelium]{Table \ref{tab:quanthelium}}), as well as for Moshi itself (\hyperref[tab:quantmoshi]{Table \ref{tab:quantmoshi}}).  
Generally,  Helium is more robust to quantization than the final trained Moshi. 
Notably, assuming quantization blocks of size 32, quantizing Helium weights to 4 bits yields a 3.43 times smaller model which remains within 2 points of MMLU of the floating point baseline. This particular quantization format is also almost identical to llama.cpp's\footnote{\url{https://github.com/ggerganov/llama.cpp}} Q4\_0, hence can be readily deployed for efficient inference.

In contrast, the same quantization recipe used on Moshi leads to a more drastic loss of performance ranging from 5 to 10 points of MMLU.
In the online demo, we keep the weights in 8-bit format as it results in a more reasonable drop of 2 points for a model roughly twice smaller than the floating point baseline.
\begin{table}[t]
    \caption{\label{tab:quanthelium}\textbf{Linguistic impact of model compression on \helium}, as measured by MMLU. `WXA8' indicates a model with weights quantized to `X' bits and activations to 8 bits, using integer scalar PTQ.  The model size in brackets is given in GygaBytes for a quantization block size of 32,  and takes into account both the model weights and the quantization parameters stored in float16.
    With a fine enough granularity of quantization blocks, a 4 bits model stays within 2 points of MMLU of the floating point baseline. 
    }
    \footnotesize
    \centering
    \begin{tabular}{lccc}
        \toprule
       BF16A8 ($\sim$ 15GB)& \multicolumn{3}{c}{54.3} \\
        \midrule
        Bitwidth vs Block size & per-channel & 256 & 32 \\
        \midrule
         W8A8  (7.66GB) &  53.96 & 54.09 & 53.81\\
         W6A8 (6.02GB)  &  53.50&  53.55&  53.86\\
         W5A8  (5.20GB) &  52.80&  53.22& 52.76\\
         W4A8 (4.37GB)  & 49.29 & 50.84 & 52.97 \\
         W3A8 (3.55GB) & 25.49 & 44.15 & 50.85\\
         W2A8 (2.73GB)  & 23.87 &23.00 & 24.27\\
    \bottomrule
    \end{tabular}
\end{table}

\begin{figure}[t]
    \includegraphics[height=0.275\linewidth, trim={0.6cm 0 1.7cm 0.9cm},clip]{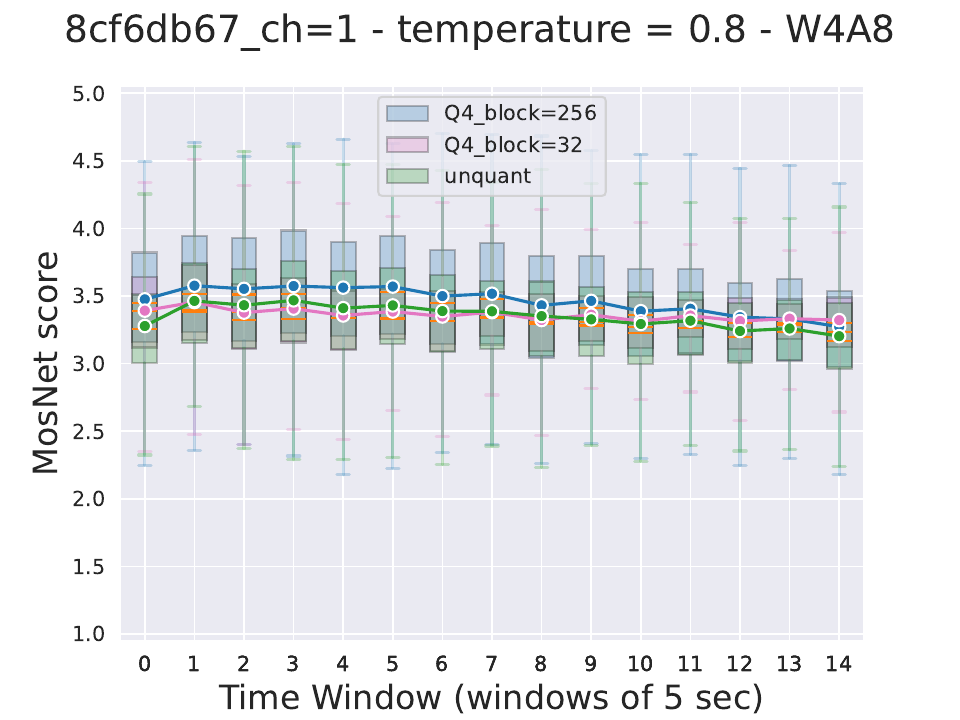}
\hfill
    \includegraphics[height=0.275\linewidth, trim={1.8cm 0 1.7cm 0.9cm},clip]{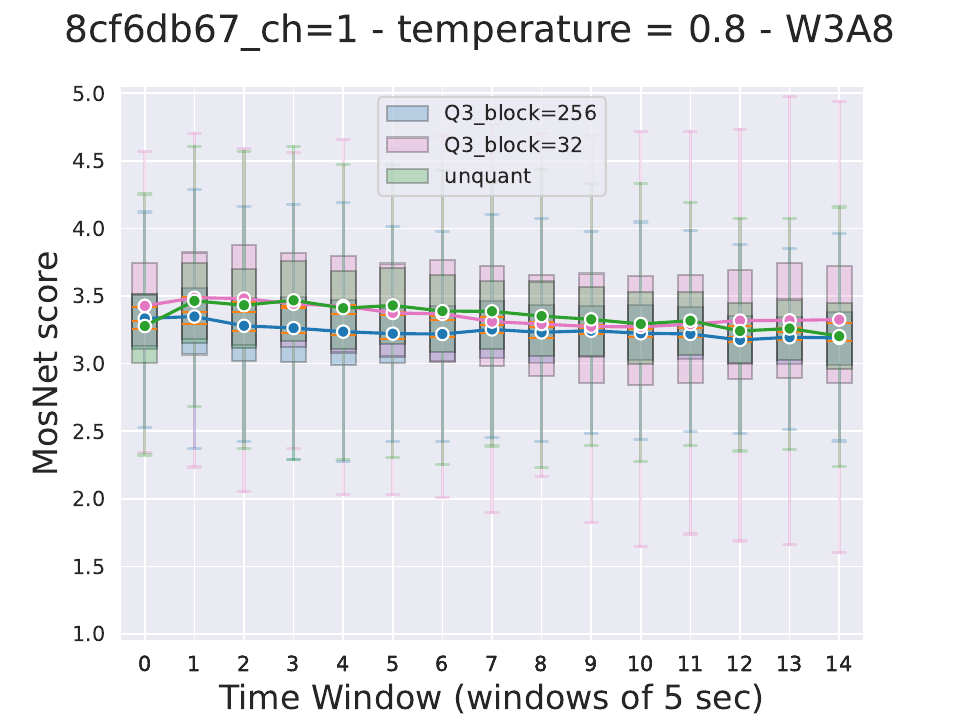}
\hfill
    \includegraphics[height=0.275\linewidth, trim={1.8cm 0 1.7cm 0.9cm},clip]{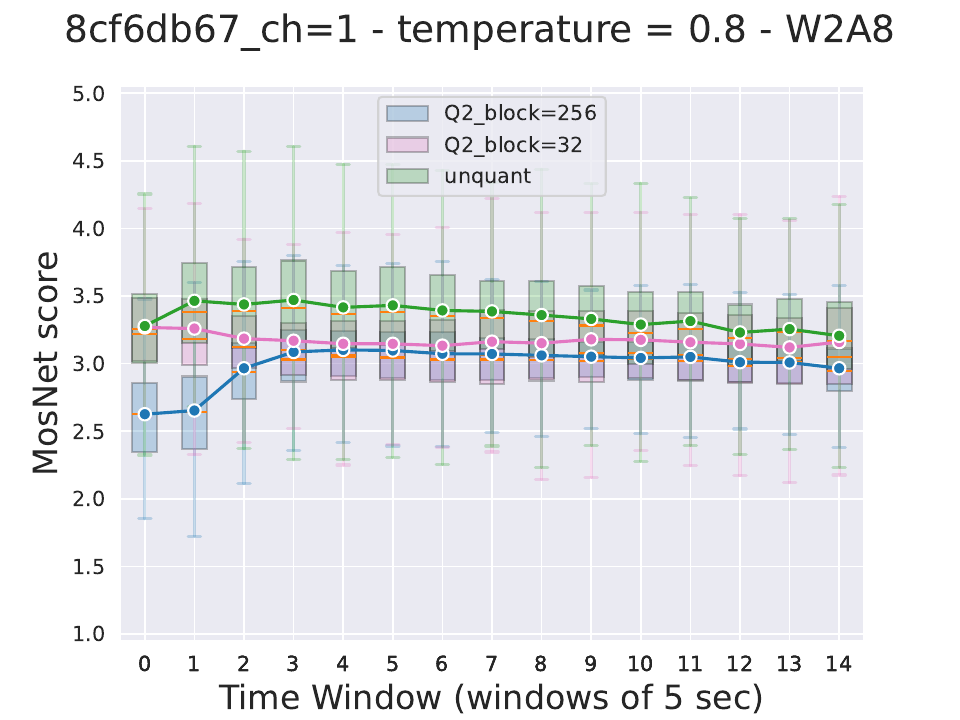}

\caption{\label{fig:quantmosnet}\textbf{Acoustic impact of model compression on \ours}. MOSNet evaluation of samples  generated by models compressed for different bitwidths. We evaluate the MOSNet scores across non overlapping windows of 5s, and report the distribution of these scores over 500 samples for each model.
}
\end{figure}

\begin{figure}[t]
~\bigskip \bigskip 

\centering
\begin{minipage}[t]{0.47\textwidth}
\scriptsize
   
    \centering
    \includegraphics[width=\linewidth]{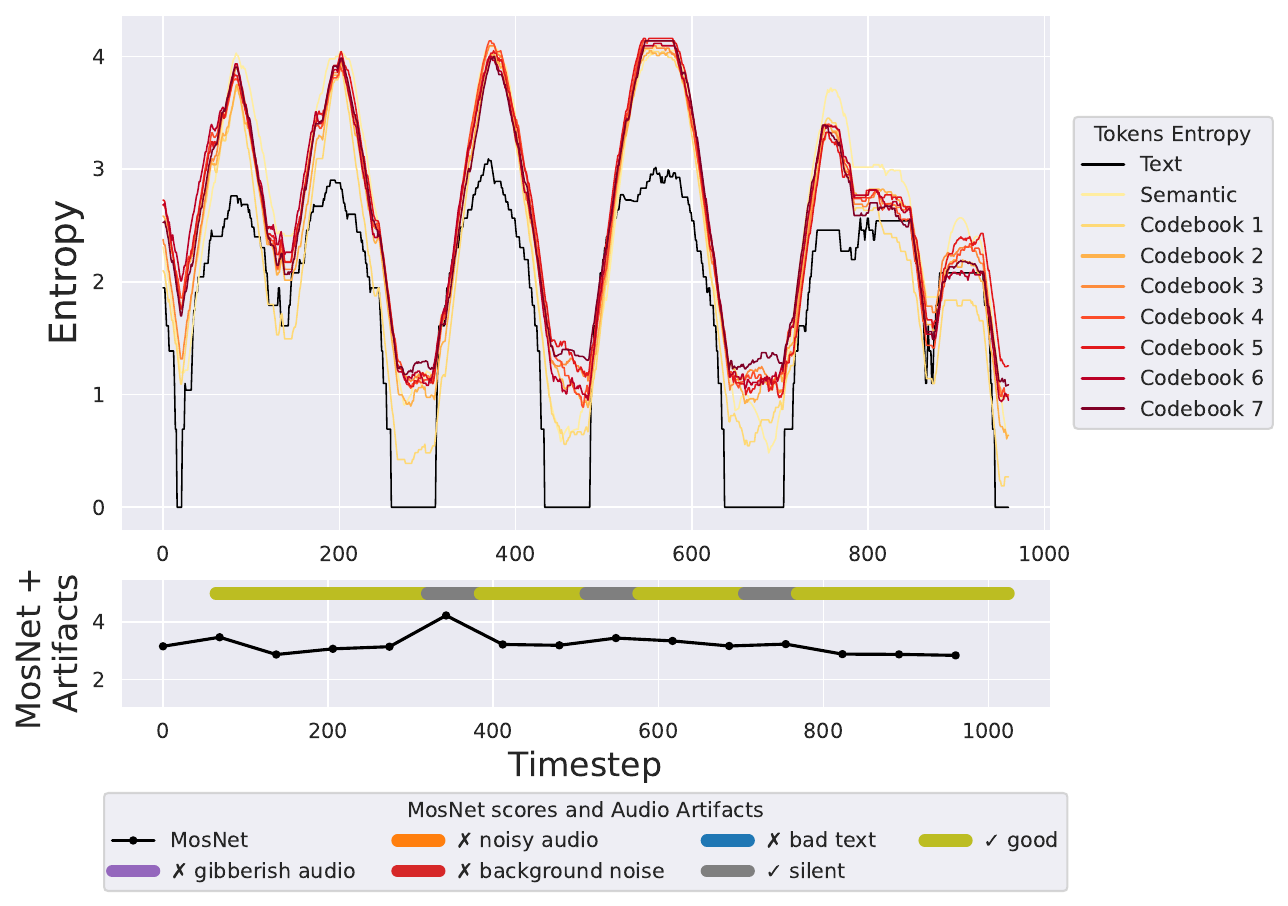}
    
    \textbf{(a)} Entropy spectrum of a well-behaved sample (no noticeable degradation). Short silences occur naturally in Moshi's output due to the model's multi-stream abilities (reflecting the other speaker's turn)

\end{minipage}
~~
\begin{minipage}[t]{0.47\textwidth}
\scriptsize
    \centering
    \includegraphics[width=\linewidth]{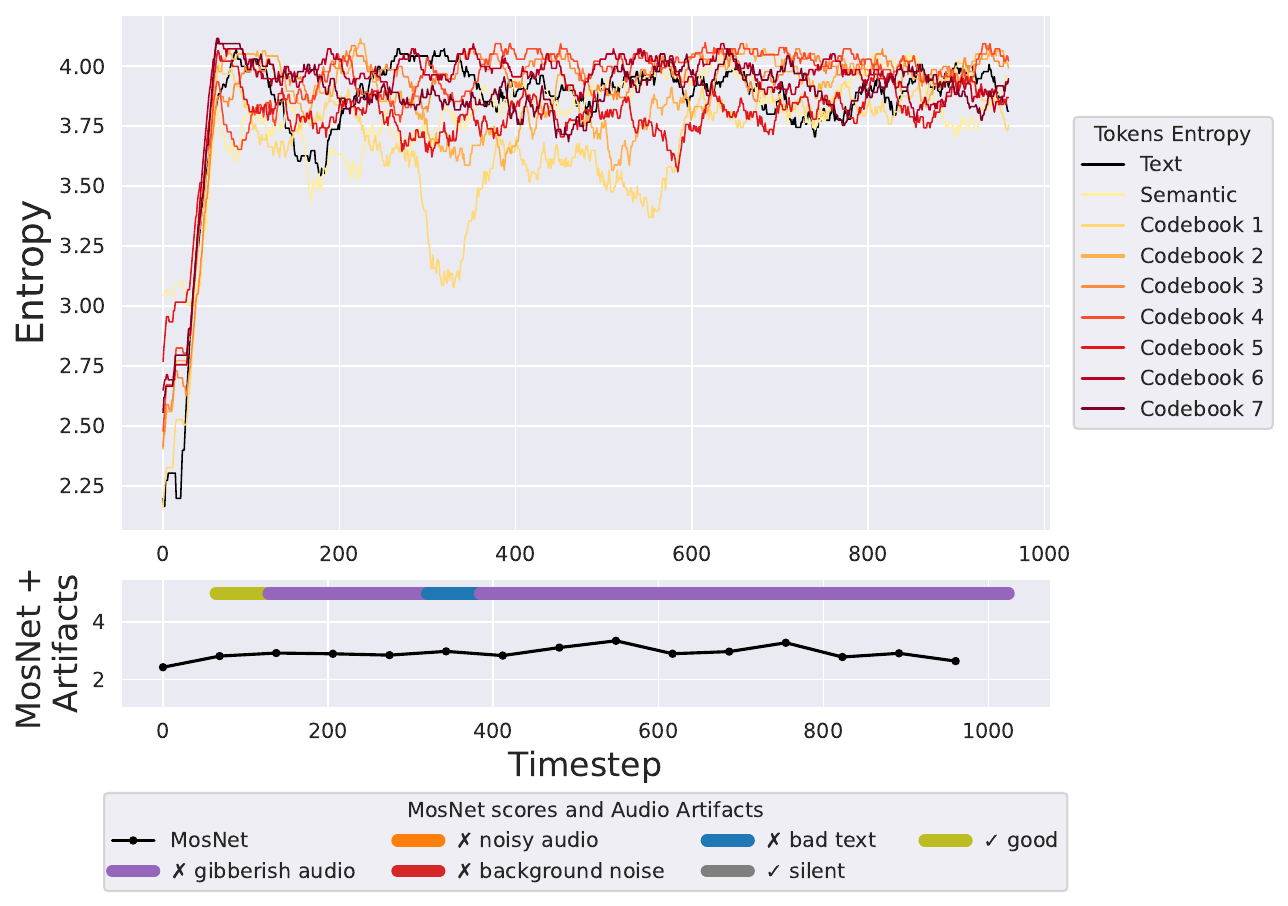}
    \textbf{(b)} Significant degradations occur at low bitwidth (W2A8). These are not always well reflected by the MOSNet scores' magnitude, but  the entropy of the text token is visibly higher.
\end{minipage}
\caption{\label{fig:artifactsexample}\textbf{Audio artifacts caused by model compression}. Example of typical entropy spectrums capturing specific audio artifacts caused by model quantization. For each timestep, we compute the entropy over the past 128 tokens, independently for the text and audio codebooks tokens. Then, we measure the presence or absence of the different artifacts over non-overlapping windows of 64 tokens, as described in \hyperref[app:artifactsmetric]{Appendix \ref{app:artifactsmetric}}.}
\end{figure}

\begin{table}[t]
    \caption{\label{tab:quantmoshi}\textbf{Linguistic impact of model compression on \ours}, measured by MMLU for different quantized Moshi on the text tokens generated by \interleaving{} directly. As for the previous table, the model size is indicated for the block size of 32.}
    
    \footnotesize
    \centering
    \begin{minipage}[t]{0.47\textwidth}
    \centering
    \textbf{a)} Moshi after single-stream pretraining
    \smallskip 

    \begin{tabular}{lcc}
    \toprule
       BF16A8 (15.24GB)& \multicolumn{2}{c}{49.8} \\
       \midrule
        Bitwidth vs Block size & 256 & 32 \\
        \midrule
         W8A8 (8.33GB) & 48.8 & 48.5\\
         W6A8 (6.95GB)& 48.5 & 49.1 \\
         W5A8 (6.02GB) & 47.4 & 48.5 \\
         W4A8 (4.64GB) & 44.7  & 45.7 \\
         W3A8 (3.72GB) & 26.1 & 35.7 \\
         W2A8 (2.80GB) & 23.4&  24.4\\
        \bottomrule
    \end{tabular}
    \end{minipage}
    \hfill 
        \begin{minipage}[t]{0.47\textwidth}
    \centering
    \textbf{b)} Moshi after multi-stream instruct
    \smallskip
    
    \begin{tabular}{lcc}
        \toprule
       BF16A8 (16.74GB) & \multicolumn{2}{c}{49.7}  \\
        \midrule
        Bitwidth vs Block size & 256 & 32 \\
        \midrule
         W8A8 (9.20GB) & 47.6 & 47.6 \\
         W6A8  (7.70GB)& 48.1 & 48.3 \\
         W5A8 (6.69GB) & 46.7 & 47.2 \\
         W4A8 (5.18GB) & 39.8  & 42.2 \\
         W3A8 (4.18GB) & 27.7 & 29.9 \\
         W2A8 (3.17GB) & 24.5& 24.9 \\
        \bottomrule
    \end{tabular}
    \end{minipage}
\end{table}
\looseness=-1
\paragraph{Results - Audio Quality.} To assess the audio quality of samples generated by the quantized models, we make use of the MOSNet metric from \cite{mosnet} as implemented in speechmetrics.\footnote{\url{https://github.com/aliutkus/speechmetrics}} 
More specifically, we generate a short prompt (64 tokens) from the unquantized model, then generate completions from each of the quantized models with a temperature of $t = 0.8$ and a sequence length of 1024 tokens. We repeat this process 500 times, and  report the distribution of MOSNet scores over non-overlapping windows in \hyperref[fig:quantmosnet]{Figure \ref{fig:quantmosnet}}. While the MOSNet scores exhibit a large variance across samples, there is generally little degradation of the audio quality after quantizing the model's weights down to 4 bits. 
For lower bitwidths however, we observe qualitatively that the MOSNet scores lack sensitivity towards some of the more severe audio degradations caused by aggressive quantization: For instance, it does not disintiguish between pure audio artifacts (e.g., noisy voice) from artifacts in the speech pattern (e.g., increased repetitiveness of the model). This is in line with the lack of consistency between objective and subjective audio quality metrics that we observed  in \hyperref[sec:mimi_eval]{Section \ref{sec:mimi_eval}}, and in addition, MOSNet was designed for a very different type of benchmark, namely, to mimic human ratings on evaluating converted speech, so it is not surprising for it not to be less sensitive to such artifacts. 
Instead, to measure the presence or absence of such degradation in the audio samples, we first observe that certain audio artifacts are identifiable from the entropy spectrums of the generated text and audio tokens: A few examples are illustrated in \hyperref[fig:artifactsexample]{Figure \ref{fig:artifactsexample}} and we further detail the types of artifacts and how we measure them in \hyperref[app:artifactsmetric]{Appendix \ref{app:artifactsmetric}}.

Following this insight, we measure the presence or absence of different audio artifacts on the same generated audio samples as the ones used in the previous MOSNet analysis. We report the results in \hyperref[tab:artifactssummary]{Table \ref{tab:artifactssummary}}, as well as a more detailed per timestep analysis in \hyperref[fig:artifacts]{Figure \ref{fig:artifacts}} of \hyperref[app:artifactsmetric]{Appendix \ref{app:artifactsmetric}}.
At a bitwidth of 4, we again observe little audio degradation.  
Decreasing to 3-bit format, audio degradations are more apparent and tend to become more frequent along the generation timestep, although the finer granularity quantization format is generally more robust to these artifacts. 
Nevertheless, both quantization formats display significantly degraded audio quality when weights are aggressively quantized to 2 bits, which we also observe qualitatively.

\begin{table}[t]
    \caption{\textbf{Distribution of audio artifacts caused by model compression}. Percentage of audio artifacts measured in the entropy spectrum of text and speech generated tokens, as described in \hyperref[app:artifactsmetric]{Appendix \ref{app:artifactsmetric}}. These results averaged across 500 samples generated by different versions of the same quantized Moshi, and across 16 timesteps of 64 tokens. Values of 0 \% are omitted in the table for better readability.}
    \label{tab:artifactssummary}
    \footnotesize
    \centering
    \begin{tabular}{lccccc}
        \toprule
        Model / Artifacts & Gibberish  & Noisy  & Background & Repetitive & No artifacts \\
         & audio & audio & noise &  text & \\
         \midrule
         unquant & & \ 4.1 & \ 0.1  &  \ 0.1 & 95.8 \\
         \midrule
         W4A8, block=32 & & \ 3.8  & \ 0.1  & \ 0.4 & 95.7 \\
         W4A8, block=256 & \ 0.1 & \ 3.7 &  & \ 2.2 & 94.0 \\
         \midrule
         W3A8, block=32 & \ 0.5 & \ 4.7 & \ 5.9 & \ 8.1 & 80.7 \\
         W3A8, block=256 & \ 0.2 & 12.2 & \ 3.1  & 21.9 & 62.7 \\
         \midrule
         W2A8, block=32 & 12.7 & 40.9 & \ 0.5 &  \ 0.4 & 45.4 \\
         W2A8, block=256 & 83.1 & & &  11.0 &  \ 5.9 \\
        \bottomrule
    \end{tabular}
\end{table}

\looseness=-1
\paragraph{Discussion.} The linguistic abilities of Moshi are more sensitive to quantizing the model weights and activations than its output audio quality.
More specifically, the audio quality remains close to that of the floating point baseline down to 4 bits precision, even when quantizing the full model, including the Depth Transformer. 
In contrast, the MMLU performance suffers significant drops when quantizing the model weights below 6 bits using post-training only quantization.
Following recent quantization techniques~\citep{quipsharp}, we may expect improved performance at lower bitwidth by using quantized aware finetuning  instead of PTQ. However, as Moshi's training pipeline from \hyperref[sec:maindatasets]{Section \ref{sec:maindatasets}} involves multiple stage and training datasets, this would require a more thorough investigation into designing quantized training phases and calibration datasets, to preserve all of \ours{}'s abilities lost after quantization.

\section{Safety}
\label{sec:safety}

In parallel with the development of Moshi, we explore different directions related to the safety of AI generated content. In this section, we specifically consider several questions regarding the content generated by Moshi, each addressed in a dedicated subsection: 
\begin{enumerate}
\item How does our model behave in terms of producing toxic content? 
\item How to avoid that the model regurgitates audio content from the training set? 
\item How do we ensure that the model uses the voice we intend to give to Moshi? 
\item How to identify if a given content has been generated by Moshi? 
\end{enumerate}

\looseness=-1

\subsection{Toxicity Analysis}

The scientific community has devoted in the last years some effort to address bias and toxicity problems for text generation models. In contrast, audio safety is far less developed. It is not straightforward to compare audio and text models in an apple-to-apple comparison, as they differ in their usage, and multiple meanings are conveyed by non-verbal signal (irony, tone, etc.). In spite of these limitations and in order to facilitate the comparison of Moshi with text generation models, in this first analysis we restrict our toxicity analysis to the text produced by the model. We adopt the ALERT benchmark\footnote{\url{https://github.com/Babelscape/ALERT}} \citep{tedeschi2024alert}, which evaluates safety under multiple categories (hate, self-harm, weapon, crime, sex, substance). 
\hyperref[tab:results_toxicity]{Table~\ref{tab:results_toxicity}} in \hyperref[app:safety]{Appendix~\ref{app:safety}} reports our detailed toxicity analysis on this benchmark. The aggregated score for Moshi and popular text-only models is as follows: 
\medskip

\noindent
\resizebox{\linewidth}{!}{%
\begin{tabular}{l|c|ccccccccccc}
\toprule
\text{Category} & \text{Moshi} & \text{GPT-3.5} & \text{GPT-4} & \text{Llama 2}&\text{Alpaca} & \text{Vicuna} & \text{Falcon} & \text{Mistral} & \text{Mixtral} & \text{Zephyr} & \text{OLMo}\\
\midrule
\text{Overall Safety Score}&\colorbox{Red!0}{83.05}  & \colorbox{Orange!0}{96.95} & \colorbox{Sand!0}{99.18} & \colorbox{Sand!0}{99.98} & \colorbox{Red!0}{62.13} & \colorbox{Orange!0}{95.75} & \colorbox{Red!0}{88.11} & \colorbox{Red!0}{75.45} & \colorbox{Orange!0}{98.22} & \colorbox{Red!0}{77.86} & \colorbox{Red!0}{85.90}\\
\bottomrule
\end{tabular}}
\medskip

With this analysis, we see that Moshi falls into the middle of this table in terms of rank. The industry models perform the best, which is expected considering the massive amount of private annotation,  red-teaming and feedback loop from which these models have benefited.  %

\subsection{Regurgitation Analysis} 
\label{sec:regurgitation}

The problem of a model generating content which it has seen at training time, which we refer to as \textit{regurgitation}, is closely related to overfitting: The more a model has seen a sequence or a subsequence during training, the more likely it is to generate this exact sequence during the generation process. Note, for a speech model, it is not only the text that can be regurgitated, but also the voice pitch, tone, and potentially the background melody if present at training time. It is therefore important to mitigate\footnote{There is currently no way to fully prevent these issues. While it is essential to develop algorithms and methodologies that limit the occurrences of problematic generations, part of the question is related to how generative AI is regulated.} potential intellectual property issues related to regurgitation, such as reproduction of copyrighted content or audio generation with the voice of a person without permission. 

\paragraph{Evaluation protocol.}
\looseness=-1
For each model, we measure the proportion of generations  (out of 100,000) that reproduce the most frequent audio segment detected in our whole training dataset. For this purpose, we have first developed a matching system that detects the most frequent audio segments, see \hyperref[sec:audiomatching]{Appendix~\ref{sec:audiomatching}}. We select the most frequent one that is long enough (16 seconds) and easy to detect from text and audio. We measure the proportion of generations that exactly match this most frequent segment.
For the matching, we initially use both audio and text matching, but observe that text-based matching has a higher recall for the initial matching step. We manually verify all the generations to filter out outliers that are not exact matches.  \smallskip

\begin{table}[t]
\caption{\textbf{Regurgitation of training data with condition-free generation} from different models. 
We measure how many times each model generates the most frequent duplicate segment audio in the training data, for different values of the temperature. With dataset deduplication, we do not observe any exact re-generation (out of $10^5$) of the most frequent segment, even if we prompt the model with the first 3s of this audio segment.\label{tab:regurgitation}}
\centering 
{%
\footnotesize
\begin{tabular}{lcccccccc}
\toprule
               & prompted (3s)  &  deduplicated   & fine-tuned &  temp.   &  regurgitation rate (\%) \\ 
\midrule
\multirow{10}{*}{single-stream} 
 &            &              &            & 0\hphantom{.0}  & 0.00  \\  %
 &            &              &            & 0.6             & 0.13  \\  %
 &            &              &            & 0.8             & 0.19  \\  %
 &            &              &            & 1.0             & 0.16  \\  %
 & \checkmark &              &            & 0\hphantom{.0}  & 100.00\hphantom{00} \\  %
 & \checkmark &              &            & 0.8             & 98.40\hphantom{0}   \\  %
 &            & \checkmark   &            & 0\hphantom{.0}  & 0.00  \\  %
 &            & \checkmark   &            & 0.8             & 0.00  \\  %
 & \checkmark & \checkmark   &            & 0\hphantom{.0}  & 0.00  \\  %
 & \checkmark & \checkmark   &            & 0.8             & 0.00  \\  %
\midrule 
\multirow{4}{*}{multi-stream}  
  &            &              & \checkmark & 0.8            & 0.00  \\  %
  & \checkmark &              & \checkmark & 0.8            & 0.00  \\   
  &            & \checkmark   & \checkmark & 0.8            & 0.00  \\  %
  & \checkmark & \checkmark   & \checkmark & 0.8            & 0.00  \\  %
\bottomrule
\end{tabular}}%
\end{table}

\noindent \textit{Unconditioned and prompted generation:} We first measure what happens with unconditional generation, to evaluate whether the model tends to generate specific sequences when not guided by a prompt. In a complementary manner, we prompt the model with the 3 first seconds of the most frequent audio segment and measure how many times the continuation is identical to this training set audio. 
\hyperref[tab:regurgitation]{Table~\ref{tab:regurgitation}} reports these regurgitation results.

\paragraph{Results \& Impact of fine-tuning.} 
 We observe that the pre-trained model trained on the raw dataset often generates frequent sequences from the training set. The sampling temperature has an important effect on the regurgitation rate: the values typically employed for generation (0.6--1.0) are more prone to regurgitation. Out of 1000 generations, the model fine-tuned  for conversation does not generate the most frequent training sequence. As a disclaimer, we point out that fine-tuning could potentially be over-ridden and therefore may not be sufficient \emph{per se} to avoid regurgitation. 

Similar to what happens with textual models~\citep{carlini2022quantifying}, regurgitation is significantly impacted by the number of times that the model uses a given sequence for training. 
Therefore, we evaluate the impact of deduplicating the training dataset by identifying all the audio segments that are frequent, and in turn by filtering them out at training time. 
In \hyperref[tab:regurgitation]{Table~\ref{tab:regurgitation}}, we observe that this pre-processing step brings the number of regurgitations of the most frequent sequence to zero, even without any fine-tuning step. %

\subsection{System Voice Consistency}
\label{sec:voice_consistency}
A potential risk for a speech-to-speech model is unauthorized voice generation.
The model should use its target voice and not potentially mimic the user's voice.
In order to evaluate to which extent Moshi adopts a voice of the user instead of the target voice, we use the following protocol:
\begin{itemize}
\item Generate 100 hours of conversations between Moshi and a second synthetic speaker.
\item Run a speaker verification model (WavLM~\citep{wavlm} large) on each segment to extract the speaker embeddings.
\item Compute the cosine similarity between the embeddings of each main speaker's segment with \emph{(i)} the first segment of the main speaker and \emph{(ii)} with the first segment of the generated speaker.
\item \noindent \textit{Note:} we exclude all the segments with a start time before 15 seconds so as to avoid counting the first turn of speech of the main speaker as it acts as the reference.
\end{itemize}

\looseness=-1
Over the generated datasets, there are 10\,249 occurrences (98.7\%) where the voice of the main speaker is closer to the reference segment of the main speaker and 133 occurrences (1.3\%) where the voice is closer to the reference segment of the other speaker.
We are also interested in how speaker's consistency evolves over time. Following~\citet{soundstorm} we compute the same
ratio as above but on groups of segments that start at specific times, to measure drift along time. \hyperref[tab:speaker_consistency]{Table~\ref{tab:speaker_consistency}} shows that speaker consistency remains stable along time, meaning that we do not observe a drift as the conversation goes on. This shows that the simple choice of using a consistent voice for the system during instruction tuning is enough to provide robustness at inference time.

\begin{table}[t]
\caption{\textbf{Speaker consistency along time.} We measure how often the speaker embedding from Moshi's segment is closer to its reference segment than the user, when computing speaker embeddings from segments further away from the reference.}
\label{tab:speaker_consistency}
\centering 
{\footnotesize
\begin{tabular}{lccccccc}
\toprule
segment start time (seconds) & 20--25 & 25--30 & 30--35 & 35--40 & 40--45 \\
\midrule
samples & 2034 & 2006 & 1998 & 2019 & 1994 \\
main $>$ other & 98.4\% & 99.2\% & 99.1\% & 99.2\% & 99.3\% \\
\bottomrule
\end{tabular}}%
\end{table}

\subsection{Identification of the Content Generated by Moshi: Watermarking}
\label{sec:watermarking}

For determining if a given audio has been generated by Moshi, we have investigated two complementary solutions: indexing and watermarking. The first, namely audio indexing, only applies in the case where we have access to the machine that generates the content, like in the case of the Moshi demo. We describe our audio matching system in \hyperref[sec:audiomatching]{Appendix~\ref{sec:audiomatching}}. 
Below in this subsection, we discuss more specifically watermarking, where the objective is to add unnoticeable marks to the generated audio. 

\paragraph{Evaluation of signal-based watermarking.}

We investigate if existing watermarking methods for audio can be used as a way to re-identify content generated by Moshi. For this purpose, we analyze the robustness of the Audioseal method \citep{san2024proactive} in our context. It is available as an open-source library.\footnote{\url{http://github.com/facebookresearch/audioseal}} 
For this evaluation, we resample the audio signal to 16kHz so that the sampling rate matches the one recommended in Audioseal instructions. We measure the average mark detection scores in the following settings:
\begin{itemize}
    \item No watermark: we measure the detection score measured when no mark was added. \\[-1.7em]
    \item Watermark no attack: no modification of the watermarked audio signal; \\[-1.7em]
    \item Pink noise: we add a small pink noise ($\sigma=0.2$) to the watermarked audio; \\[-1.7em]
    \item RVQGAN: we compress and decompress the audio signal with a recent state-of-the-art auto-encoder~\citep{dac}. We use the publicly available pre-trained 16Khz model\footnote{\url{https://github.com/descriptinc/descript-audio-codec}} which differs from the 24kHz model used as a baseline in \hyperref[sec:mimi_eval]{Section~\ref{sec:mimi_eval}}.\\[-1.7em]
    \item \mimi{} auto-encoder: we use our own tokenizer to compress and decompress the signal. This operation is performed with 24kHz audio and therefore involves two re-sampling stages (from 16kHz to 24kHz and back to 16kHz). 
\end{itemize}

We report the results in \hyperref[tab:audioseal]{Table~\ref{tab:audioseal}}. We observe that the mark yields high detection rates when the audio is unchanged. With aggressive Pink-Noise, one needs a relatively long sequence to get a high detection score. However, the mark is not robust to a strong compression: the two auto-encoders that we consider are low bitrate and therefore discard anything not related to the signal reconstruction. As a result, our \mimi{} codec removes the mark to a level that makes a watermarked audio indistinguishable from a non-watermarked audio, making such a signal-based watermarking useless in this context. 

\begin{table}[t]
\caption{\textbf{Evaluation of Audioseal}  \citep{san2024proactive} for watermarking the speech produced by Moshi. Each detection score is averaged over 1000 generations. %
\label{tab:audioseal}
}
\centering 
{%
\footnotesize
\begin{tabular}{llrr}
\toprule
               &   &  \multicolumn{2}{c}{average detection score} \\ 
\cmidrule(l){3-4}               
               &  $\downarrow$ audio post-processing  \quad\quad\quad\quad\quad\quad  ~ \hfill audio duration $\rightarrow$  & 10\,seconds &  1\,minute\\ 
\midrule
No mark        & none                                & 0.0855  & 0.2474   \\
Watermarked    & none                                & 0.9999  & 0.9999   \\
Watermarked    & pink-noise (noise std $\sigma=0.2$) & 0.7093  & 0.9019   \\
Watermarked    & RVQGAN compression \& decompression         & 0.1101  & 0.2662   \\ 
Watermarked    & \mimi compression \& decompression & 0.0805  & 0.2404   \\ 
\bottomrule
\end{tabular}}%
\end{table}

\paragraph{Exploration on generative-based watermarking for audio.}

Given that a recent state-of-the-art signal-based audio watermarking is not robust to a simple non-adversarial auto-encoding method, we investigated the possibility of watermarking the generation process itself. This solution was recently proposed for text generation, in particular in the works of \cite{aaronson2023watermarking} and~\cite{kirchenbauer2023watermark}. These two methods operate similarly: at sampling time, they bias the probabilities driving the generation process. They differ from each other by how they modify the probabilities, yet in both cases the sampling is parameterized by a hash function that preferably depends on a local context. 
These solutions were improved by~\cite{fernandez2023three}, who proposed a better mark detector, in particular by addressing the issue of repetitive patterns. 

\looseness=-1
We have investigated how to apply these discrete watermarking methods to our audio generation pipeline. For this purpose, we need to encode the audio signal back to tokens in order to identify if the mark is present or not. One issue is that \emph{the codec is not idempotent}: if we generate a waveform from tokens and then re-encode it back into tokens, the re-generated tokens are likely to be different from the ones generated with high probability, even if the audio has not suffered any noise addition. We quantify this problem in \hyperref[tab:idempotence]{Table~\ref{tab:idempotence}}. The semantic token is robust to some extent, while the other quantization indices are increasingly less robust as they depend on the previous quantizer level. One key issue is that the tokens do not resist to a moderate temporal shift. This is especially true for the \mimi codec, which is purposely optimized on a perceptual objective, as opposed to a fidelity reconstruction criterion.

\begin{table}[t]
\caption{\textbf{Idempotence of tokens}.  Probabilities that quantization indices remain identical after decoding and re-encoding the waveform back to tokens, depending on the residual quantizer level. We consider two optional audio post-processing attacks: audio shifted by a time offset of up to half the sampling period ($\Delta T$=40ms), and re-encoding with RVQGAN. 
All results are averaged over 1000 generated sequences of 1 minute.  
\label{tab:idempotence}
}
\centering 
{%
\footnotesize
\resizebox{\textwidth}{!}{
\begin{tabular}{lccr@{}c@{\ }ccccccc}
\toprule
 & \multicolumn{2}{c}{attacks} & RQ level $\rightarrow$    & $k=1$ & $k=2$ & $k=3$ & $k=4$  & $k=5$ & $k=6$ & $k=7$ & $k=8$ \\   
\cmidrule(l){2-3}
$\downarrow$ codec       & $\Delta T$ & RVQGAN  & & (semantic)  & \multicolumn{6}{c}{}  &\\ 
\midrule
\multirow{4}{*}{Basic}   
   &  0     &  &  &  0.798 & 0.783 & 0.560 & 0.483 & 0.421 & 0.407 & 0.369 & 0.404    \\ 
   &  10ms  &  &  &  0.766 & 0.495 & 0.255 & 0.206 & 0.180 & 0.173 & 0.144 & 0.193  \\ 
   &  20ms  &  &  &  0.682 & 0.390 & 0.220 & 0.180 & 0.158 & 0.154 & 0.129 & 0.172  \\ 
   &  40ms  &  &  &  0.503 & 0.329 & 0.182 & 0.146 & 0.128 & 0.125 & 0.107 & 0.156  \\ 
\midrule
\multirow{8}{*}{\mimi}
   &  0     &  &  & 0.766 & 0.550 & 0.372 & 0.352 & 0.293 & 0.297 & 0.264 & 0.303  \\
   &  10ms  &  &  & 0.731 & 0.376 & 0.206 & 0.176 & 0.152 & 0.154 & 0.132 & 0.182  \\
   &  20ms  &  &  & 0.653 & 0.307 & 0.171 & 0.146 & 0.121 & 0.126 & 0.106 & 0.159  \\
   &  40ms  &  &  & 0.483 & 0.267 & 0.160 & 0.137 & 0.116 & 0.121 & 0.102 & 0.150   \\
\cmidrule(l){2-12}
   &  0    & \checkmark &  & 0.741 & 0.409 & 0.221 & 0.198 & 0.150 & 0.154 & 0.134 & 0.173   \\
   &  10ms & \checkmark &  & 0.702 & 0.281 & 0.148 & 0.133 & 0.118 & 0.117 & 0.100 & 0.136   \\
   &  20ms & \checkmark &  & 0.633 & 0.228 & 0.126 & 0.114 & 0.098 & 0.097 & 0.084 & 0.119   \\
   &  40ms & \checkmark &  & 0.450 & 0.197 & 0.120 & 0.113 & 0.104 & 0.102 & 0.086 & 0.112   \\ 
\bottomrule
\end{tabular}}}
\end{table}

\paragraph{Discussion on generative audio watermarking.} The lack of idempotence is problematic for 
the aforementioned sampling-based watermarking methods, as it affects the reliability of the detector when measuring the sampling bias. Noticeably, in order for these methods to work properly, the n-tuples that gives the context to the hash key must be stable enough during several consecutive tokens. Reducing the context length improves the stability but drastically increases the likelihood of producing degenerated audio sequences, similar to the degeneration problem observed by \citep{holtzman2019curious}. 

While we regard this attempt of employing text-based watermarking as a negative result, hereafter we discuss a few potential ways for  circumventing the aforementioned problem of token stability though re-encoding:
\begin{itemize}
    \item Marking only the RQ first levels improves the stability. In our preliminary experiments, using these indices as context in the hash function, and limiting the dependence on previous timestamps, significantly increases the stability (although not sufficiently). \\[-1.8em]
    \item The idempotence could be improved by adding a specific loss in the discrete latent space, such that the audio tokens are stable through auto-encoding. \\[-1.8em]
    \item Potentially this auto-encoding could be learned to be resilient to signal transformation, similar to what is proposed when learning  image watermarking based on neural networks~\citep{zhu2018hidden,fernandez2022watermarking}. In view of our analysis, adding some tolerance to moderate temporal shift is especially important. \\[-1.8em]
    \item The text could be marked instead of the audio. One downside is that text is a lower-capacity channel for adding a mark, and would not be sufficient for short conversations. Another problem is that detecting the mark requires a reliable transcription. 
\end{itemize}
Last but not least, some exploration is needed to ensure that it is not trivial to remove the watermarking procedure when open-sourcing a model. As an example, the only thing to remove the watermark with the implementation associated with the stable diffusion model was to comment a line of code.\footnote{\url{https://github.com/Stability-AI/stablediffusion/blob/main/scripts/txt2img.py\#L363}} 
A promising work in this direction is the study by \cite{sander2024watermarking}, who show that it is possible to detect when a model has been trained on watermarked text. 
A method exploiting this observation has just been shared by \cite{san2024latent}: the watermarking is implicitly added through the training data, in the spirit of ``radioactive data'' by \cite{sablayrolles2020radioactive}.

\section{Conclusion}
In this work, we introduce \ours{}, the first real-time, full-duplex spoken dialogue system. The first component of \ours{} is \helium{}, a 7B parameter text LLM which is competitive with open-weights models trained with a similar compute budget. To encode audio into discrete units amenable to language modeling, we introduce \mimi{}, a semantic-acoustic neural audio codec which provides state-of-the-art audio quality at low bitrates while operating at low framerates compatible with real-time generation. We then introduce a new, hierarchical multi-stream architecture that supports generating arbitrary conversations in a speech-to-speech manner. We moreover show that speech-to-speech generation can be drastically improved by \interleaving{}, a new method that generates text tokens as a prefix to audio tokens, while remaining compatible with streaming inference. Our experiments show that \ours{} demonstrates state-of-the-art spoken question answering and dialogue modeling while displaying satisfying levels of safety by not generating toxic content and remaining consistent in its voice. To summarize, we introduce a complete suite of models and recipes, from text LLMs to neural audio codecs and generative audio models, which we combine into a real-time spoken dialogue system with a 160ms theoretical latency able to follow complex multi-turn conversations across 5 minutes. We release both \mimi{} and \ours{} to foster the development of such applications. Additionally, we show how our \interleaving{} method allows designing streaming TTS and streaming ASR just by changing the delay between text and audio tokens. We believe that both \interleaving{} and multi-stream modeling will have a positive impact on speech-to-speech and audio-to-audio beyond dialogue modeling. 

\acks{This project is funded by Iliad Group, CMA CGM Group and Schmidt Sciences. We thank Xavier Niel, Rodolphe Saadé, Eric Schmidt, Aude Durand, Séverine Grégoire and Nicolas Granatino, for their support; as well as Sarah Hôte and Guillaume Rouzaud at Kyutai for their help. We also thank Alice, the voice artist who strived to give \ours{} online demo its voice, Elie Raffier who built the user interface for this demo, and Hugging Face for inference compute donation. Audio training data set was built with the help of Noatune Studios and Landospeech. Model training was conducted at Scaleway.}

\vskip 0.2in
\bibliography{sample}

\newpage
\appendix

\begin{table}[t]
\section{Additional Ablation on \mimi Codec}
  \centering
  \caption{\textbf{Ablation study on hyper-parameters of the \mimi codec}. We evaluate semantic modeling by reporting the error rate on a phonetic ABX discriminability task. To evaluate reconstruction quality, we compute VisQOL and MOSNet. ``Quantization rate'' refers to applying quantization to the latent space only $50\%$ of the time during training (independently from quantizer dropout), as described in \hyperref[sec:mimi]{Section ~\ref{sec:mimi}}.}
  \label{tab:mimi_ablations_appendix}
  \footnotesize
  \resizebox{\textwidth}{!}{
  \begin{tabular}{ccccc|ccc}
    \toprule
    Quantization & Transformer & Transformer & WavLM & Split & \multirow{2}{*}{ABX ($\downarrow$)} & \multirow{2}{*}{VisQOL ($\uparrow$)} & \multirow{2}{*}{MOSNet ($\uparrow$)} \\
    Rate & in encoder & in decoder & distillation & quantizer &  &  &  \\
    \midrule
     \xmark & \xmark & \xmark & \xmark & \xmark & 31.3\% & 2.37 & 2.85 \\
     \xmark & \checkmark & \xmark & \xmark & \xmark & 31.4\% & 2.30 & 2.82 \\
     \xmark & \xmark & \checkmark & \xmark & \xmark & 27.5\% & 2.30 & 2.93 \\
     \xmark & \checkmark & \checkmark & \xmark & \xmark & 29.0\% & 2.25 & 2.94 \\
     \checkmark & \xmark & \xmark & \xmark & \xmark & 29.1\% & 2.65 & 2.86 \\
     \checkmark & \checkmark & \xmark & \xmark & \xmark & 27.4\% & 2.69 & 2.83 \\
     \checkmark & \xmark & \checkmark & \xmark & \xmark & 23.6\% & 2.72 & 2.89 \\
     \checkmark & \checkmark & \checkmark & \xmark & \xmark & 23.3\% & 2.82 & 2.89 \\
     \checkmark & \checkmark & \checkmark & \checkmark & \xmark & 6.5\% & 2.13 & 2.87 \\
     \checkmark & \xmark & \checkmark & \checkmark & \checkmark & 10.8\% & 2.68 & 2.84 \\
     \checkmark & \checkmark & \xmark & \checkmark & \checkmark & 8.1\% & 2.49 & 2.71 \\
     \xmark & \checkmark & \checkmark & \checkmark & \checkmark & 8.0\% & 2.36 & 2.88 \\
     \checkmark & \checkmark & \checkmark & \checkmark & \checkmark & 8.1\% & 2.72 & 2.89 \\
    \bottomrule
  \end{tabular}
  }
\end{table}

\section{Audio Matching and Deduplication}
\label{sec:audiomatching}

We have developed an audio matching system, whose objective is twofold:
\begin{enumerate}
\item \textit{Deduplication of source content}. Removing frequent duplicates to avoid overfitting and the regurgitation of audio content that is over-represented in the dataset, as evaluated in \hyperref[sec:regurgitation]{Section~\ref{sec:regurgitation}}.  \\[-1.7em]
\item \textit{Indexing solution}. By collecting signatures of samples at generation time, we can find if some content has been generated by our online demo or not by direct retrieval. 
\end{enumerate}

Our audio matching solution is inspired by the work of \cite{wang2003industrial}, as it offers a good trade-off between efficiency and effectiveness. This method is a retrieval system: Given a query, it detects the similar audio in a pre-indexed dataset. In our case, the signature design favors the de-duplication use-case, which needs to be more efficient: Formally, we need to compare every audio of the dataset with the whole dataset, which raises efficiency issues.  The signature extraction is described below.

\begin{figure}[t]
\setlength{\parindent}{0cm}
\includegraphics[width=0.48\linewidth,trim={0cm 0.84cm 0 0},clip]{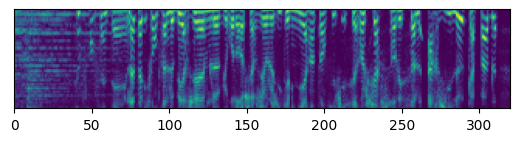}
\hfill
\raisebox{23pt}{\begin{minipage}[t]{0.5\linewidth}
(1) Mel-spectogram: The [200Hz--3000Hz] frequency range is split into 64 bands.  
\end{minipage}}

\includegraphics[width=0.48\linewidth,trim={0cm 0.84cm 0 0},clip]{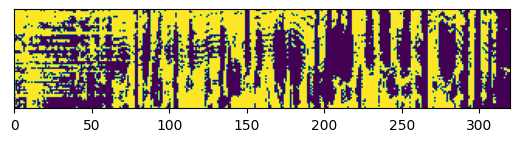}
\hfill
\raisebox{23pt}{\begin{minipage}[t]{0.5\linewidth}
(2) Energy filter: filter our all positions (time, band) whose amplitude is below the average.
\end{minipage}}

\includegraphics[width=0.48\linewidth,trim={0cm 0.84cm 0 0},clip]{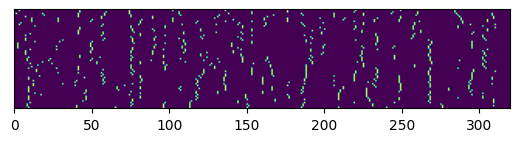}
\hfill
\raisebox{23pt}{\begin{minipage}[t]{0.5\linewidth}
(3) Time filter: keep only position with highest mel-spec value in a sliding window. 
\end{minipage}}

\includegraphics[width=0.48\linewidth,trim={0cm 0.84cm 0 0},clip]{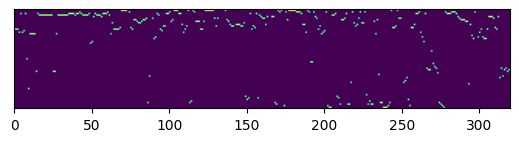}
\hfill
\raisebox{23pt}{\begin{minipage}[t]{0.5\linewidth}
(4) Frequency filter: At a given time, keep only the most energetic frequency band. 
\end{minipage}}

\includegraphics[width=0.48\linewidth,trim={0cm 0.84cm 0 0},clip]{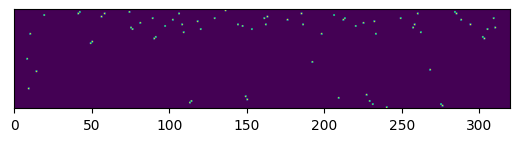}
\hfill
\raisebox{23pt}{\begin{minipage}[t]{0.5\linewidth}
(5) The Constellation map obtained by intersecting the three filters above. 
\end{minipage}}
\caption{\textbf{Mel-spectrum keypoint extraction}. Three filters are applied to the audio mel-spectrum to extract a constellation of keypoints on which hash signatures are computed.
\label{fig:hash_extraction}}
\end{figure}

\paragraph{Constellation map.}
The first step to produce the signatures involves computing a set of keypoints referred to as a \emph{constellation map}. Our procedure is inspired by \cite{wang2003industrial} and  illustrated in \hyperref[fig:hash_extraction]{Figure~\ref{fig:hash_extraction}}. First, (1) we compute a mel-spectogram from the audio signal, where the time is discretized with frequency 40Hz and the frequency range into 64 bins. We then apply three filters to select time-frequency positions: (2) The energy filter ensures that we only select positions that are robust enough; (3) The time and (4) frequency filters ensure that we select maxima w.r.t. time and frequency. The combination of these filters is (5) a constellation, from which we extract hashes. 
\medskip

At the end of the keypoint extraction procedure, the constellation map ${\mathcal C}$ consists of a list of $n$ tuples of the form ${\mathcal C}=\{(t_i, f_i)\}_{0 \leq i < n}$, where each selected timestamp $t_i$ is associated with a mel-spec discrete frequency level $f_i \in \{0,\dots,63\}$ . 

\paragraph{Hash encoding.} 
From the constellation map, we extract hash signatures as follows. For each keypoint $(t_k, f_k) \in \mathcal C$, we select, if there exists:
\begin{itemize}
\item A forward keypoint $(t_\mathrm{f},f_\mathrm{f})$, which is the closest time to $t_k$ such that $t_k+m \leq t_\mathrm{f} < t_k+M$, where $[t_k+m,t_k+M)$ is the temporal window from which we select a keypoint. Note, for a given $t_\mathrm{f}$, the corresponding frequency $f_\mathrm{f}$ is unique by design of the filters. 
\item A backward keypoint $(t_\mathrm{b},f_\mathrm{b})$, which is determined by the keypoint closest in time to $t_k$ such that $t_k-M < t_\mathrm{b} \leq t_i-m$, where $(t_k-M,t_k-m]$ is the temporal window in which the procedure selects a keypoint. 
\end{itemize}

We extract a signature only if both the forward and backward keypoints exist. In that case the signature is defined by the tuple $s_k=(f_\mathrm{b},f_k,f_\mathrm{f},t_k-t_\mathrm{b},t_\mathrm{f}-t_k)$, which we associate to the absolute timestamp $t_k$. 
In our case we set $m=4$ and $M=20$. Therefore the maximum time-span of the signature is $2\cdot M$, i.e., about 3.2 seconds. 
Formally, the hash key can take $64^3 (M-m)^2 = 2^{26} = 67,108,864$ distinct values. In practice the distribution of hash values is skewed and some signatures are unlikely to occur.

\paragraph{Pair-wise matching and one-to-many comparison.} With our signature extraction, we can compare two audios by comparing their signature sets, which amounts to computing the intersection of the hash-keys. When one wants to compare a query audio to a dataset that consists of many audios, it is more efficient to perform this comparison with an inverted file or a hash table. In that case, the indexing structure returns the lists of matching signatures along with the matching timestamps for each of the audio. Similar to \cite{wang2003industrial}, we only preserve the matches that are temporally consistent thanks to a simple Hough 1D temporal voting scheme. Optionally, we incorporate a tolerance of $\pm 1$ on the timestamps $t_\mathrm{b}$ and $t_\mathrm{f}$ when matching the signatures. This tolerance increases the complexity and we therefore do not use it for the dataset deduplication case. 

\paragraph{De-duplication: Signature fused set.} 
For our deduplication strategy, we first cross-match all the audio segments in the dataset, and extract the matching segments that occur often enough (typically $\geq 10$ matches). Since their signatures are redundant, we remove all duplicate signatures that occur at identical relative timestamps to produce a single \emph{duplicate signature set}. At training time, in order to determine if an audio segment is a frequent duplicate to be filtered out, we simply compare its signature set to the duplicate signature set. In other terms, we simply perform  a simple audio-to-audio matching between the putative training segment and the synthesized duplicate signature file. We use the segment for training only if the score is below a pre-defined matching threshold.

\section{Delayed text LM as a zero-shot streaming ASR and TTS}
\label{app:streaming_tts}
As explained in \hyperref[sec:interleaving]{Section~\ref{sec:interleaving}}, Moshi models audio tokens, along
with a text stream that is aligned on the audio frame rate with the use of special padding tokens, as represented in \hyperref[fig:rainbow]{Figure~\ref{fig:rainbow}}. We can adapt this method for ASR and TTS by introducing a delay between the audio and text tokens. In both cases, the model operates in full streaming mode, with a fixed latency (here 2 seconds).
    
\begin{figure}
\centering
\includegraphics[width=\textwidth]{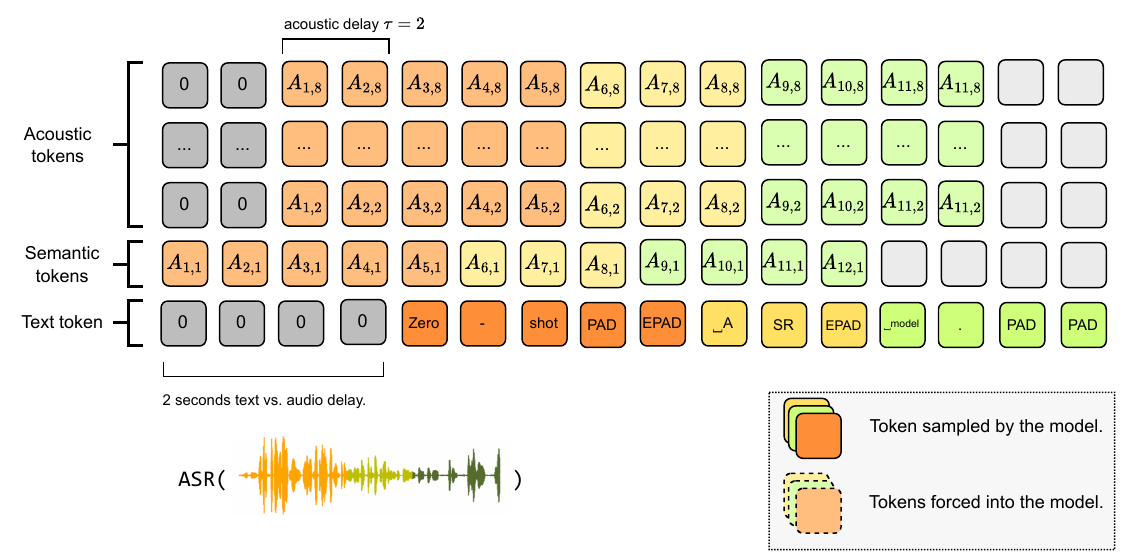}
\caption{\textbf{Representation of the joint sequence modeled by Moshi when used for ASR.}
Each column represents the tokens for a given step in the joint sequence $(V_{s, k})$, similar to the one described in \autoref{eq:final_multi_sequence}, but adapted for ASR. The text is delayed by 2 seconds, and we use an acoustic token delay $\tau=2$. Tokens are predicted from bottom to top in the depth Transformer.
The audio tokens are kept to match those of the input audio, while text tokens are sampled freely. This also provides fine word timestamps.}
\label{fig:rainbow_asr}
\end{figure}

\begin{figure}
\centering
\includegraphics[width=\textwidth]{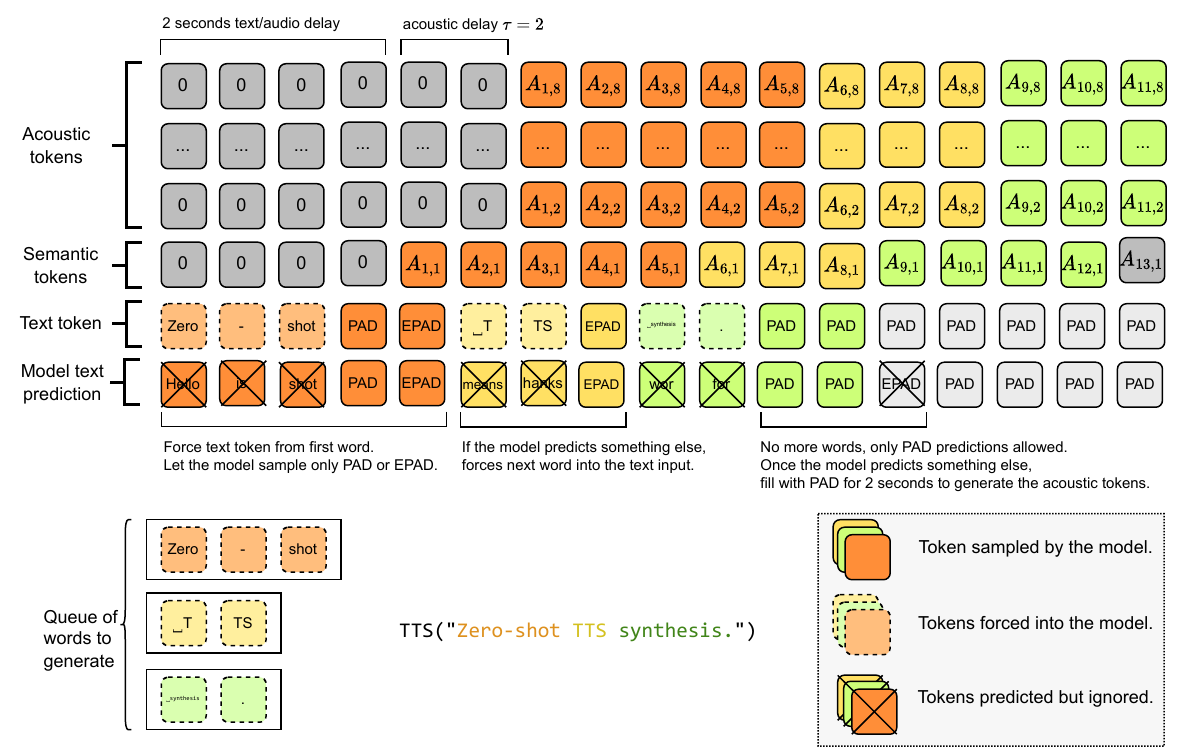}
\caption{\textbf{Representation of the joint sequence modeled by Moshi when used in TTS mode}. Each column represents the tokens for a given step in the joint sequence $(V_{s, k})$, similar to the one described in \autoref{eq:final_multi_sequence}, but adapted for TTS. The audio is delayed by 2 seconds, and we use an acoustic token delay $\tau=2$. Tokens are predicted from bottom to top in the depth Transformer.
Text predictions are usually ignored, and the tokens from the text to generate are used instead. However,
this text input lacks padding token. At the end of each word, we allow the model to sample freely \texttt{PAD} and \texttt{EPAD} tokens. If the model tries to sample another token, we instead use the tokens from the next word. The semantic and acoustic audio tokens are sampled normally, being implicitly conditioned on the text due to the delay used. This method also provides a fine alignment of the words in the generated audio, by noting the time at which a given word is consumed by the model.
}
\label{fig:rainbow_tts}
\end{figure}

\paragraph{ASR mode.} If the audio is ahead of the text, we ignore the model prediction for the audio tokens, using instead those of some audio input, and sample the text tokens freely. Then the text stream contains the audio transcription, with fine alignments at the word level, as depicted in \hyperref[fig:rainbow_asr]{Figure~\ref{fig:rainbow_asr}}.
\paragraph{TTS mode.} If the text is ahead of the audio, we can symmetrically derive a TTS engine. We need for that a properly padded set of text tokens. We obtain those in a zero-shot manner by allowing the model to sample freely \texttt{PAD} and \texttt{EPAD} tokens. As soon as the model tries to sample a different token, we instead input the next word to generate. Note that we can further control the rate of the speech by keeping an online average of the fraction of padding tokens. By introducing a small bonus on their logits when this fraction falls below a given target value, we ensure reasonable rate and a good intelligibility in all situations. Finally, using a prefix with both text and audio tokens, we can control the voice of the speaker. A representation is given in \hyperref[fig:rainbow_tts]{Figure~\ref{fig:rainbow_tts}}.
\paragraph{Multi-stream TTS.} We use this mechanism both in single and multi-stream mode. In multi-stream mode, the model outputs two sets of audio tokens. The text is provided in a single stream, using the \texttt{<bos>} and \texttt{<eos>} tokens to separate the text from the two speakers.

\newpage
\section{Characterizing Audio Artifacts Caused by Quantization}
\label{app:artifactsmetric}

First, recall that Moshi jointly handles three streams of tokens, text tokens $W_{s}$ for Inner Monologue, semantic+acoustic audio tokens $(A_{s, k})_{1 \leq k \leq Q}$ for Moshi's audio, and the similar audio tokens $(A'_{s, k})_{1 \leq k \leq Q}$ for the user's input. 
To analyse the impact of model quantization on genereated content, we first compute the Shannon entropy $H$ across windows of fixed size $C$ at each timestep $s$ for the 
text and Moshi's audio streams independently. This yields $H^{0}_s = H(W_{s-C:s})$ for text, and $H^{k}_s = H(A_{s-C:s, k})$ for each audio level. 
We use  $C = 64$ in practice, which corresponds to roughly 4.5 seconds of audio once decoded, and ignore all the leading $C$ tokens as they have a reduced context (furthermore, in our experimental scenario, they include the initial  prompt used for generation).

\begin{figure}[tbh]
\centering
\begin{minipage}[b]{0.45\textwidth}
\scriptsize
    \textbf{(a)} Example entropy spectrum of a good audio samples (no artifacts detected). Short pauses occur for the main speaker due to the multi-stream design.\\[-0.09cm]
   
    \centering
    \includegraphics[width=\linewidth]{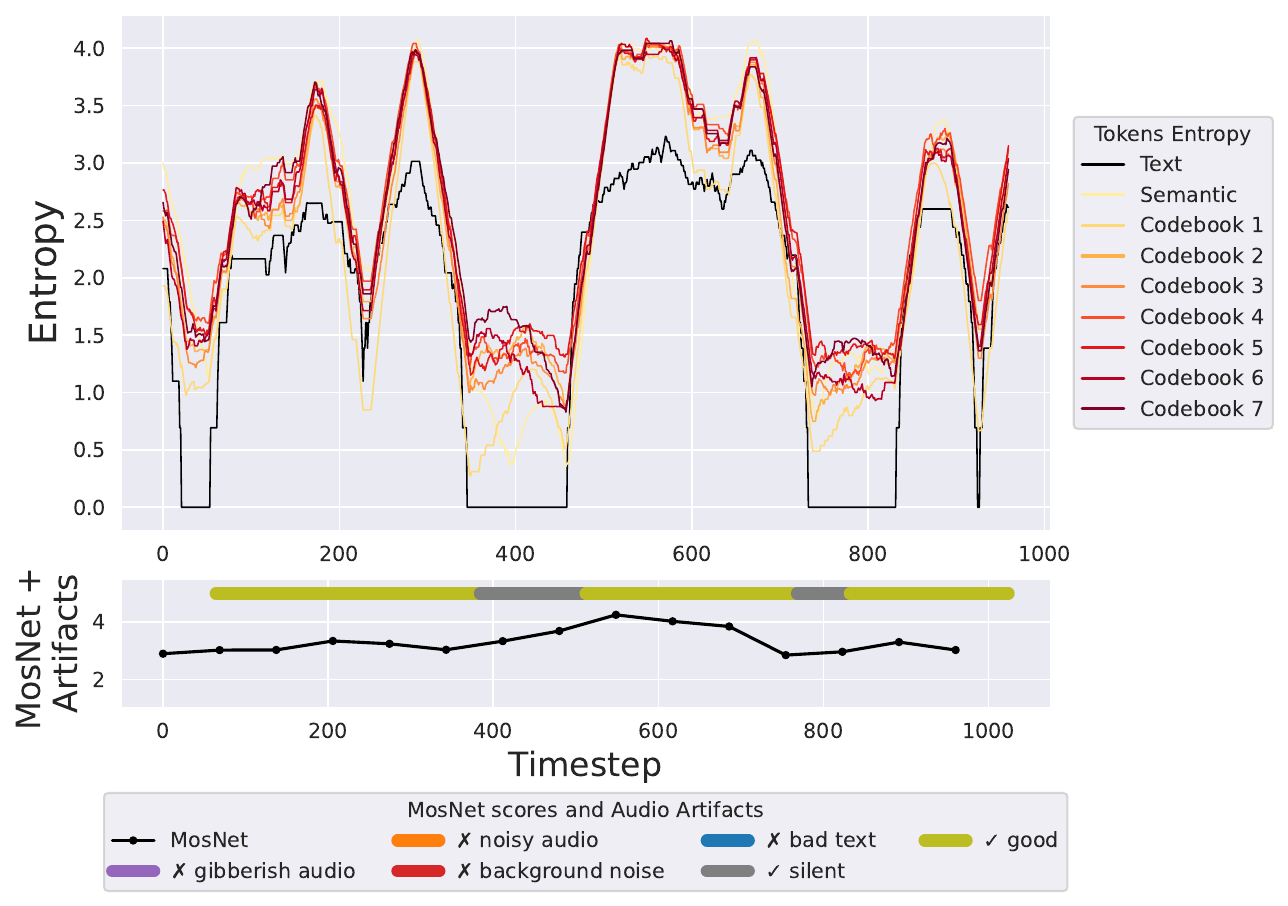}

\end{minipage}
~~~~
\begin{minipage}[b]{0.45\textwidth}
\scriptsize
    \textbf{(b)} Generally, the presence of artifacts tend to increase over time, here with repetition starting to occur in the speech.\smallskip
    
    \centering
    \includegraphics[width=\linewidth]{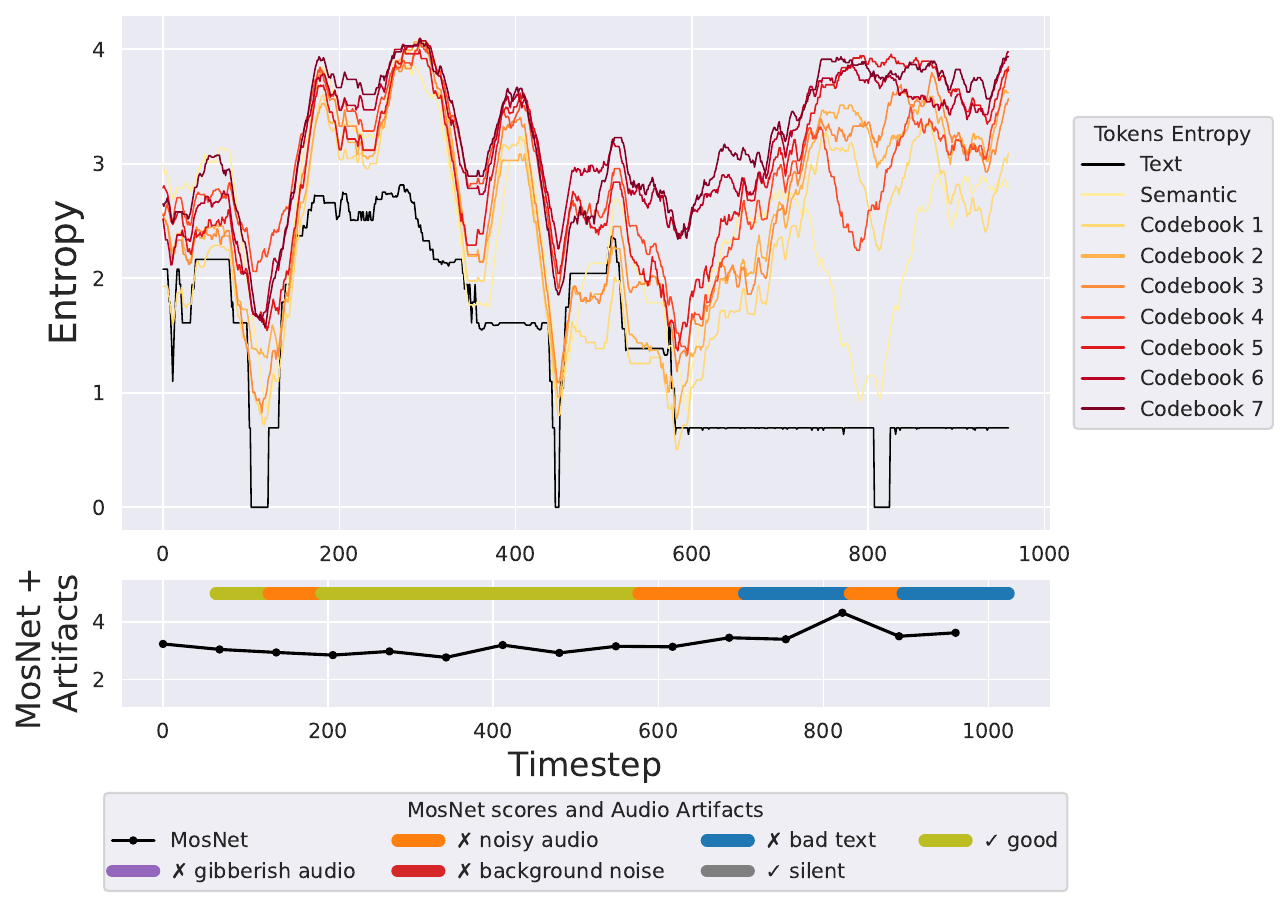}
\end{minipage}

\begin{minipage}[b]{0.45\textwidth}
\scriptsize
    \textbf{(c)} Another common artifact is repetitive snippets of text (with good audio quality), which are characterized by a flat entropy of the text token.\\[-0.09cm]
   
    \centering
    \includegraphics[width=\linewidth]{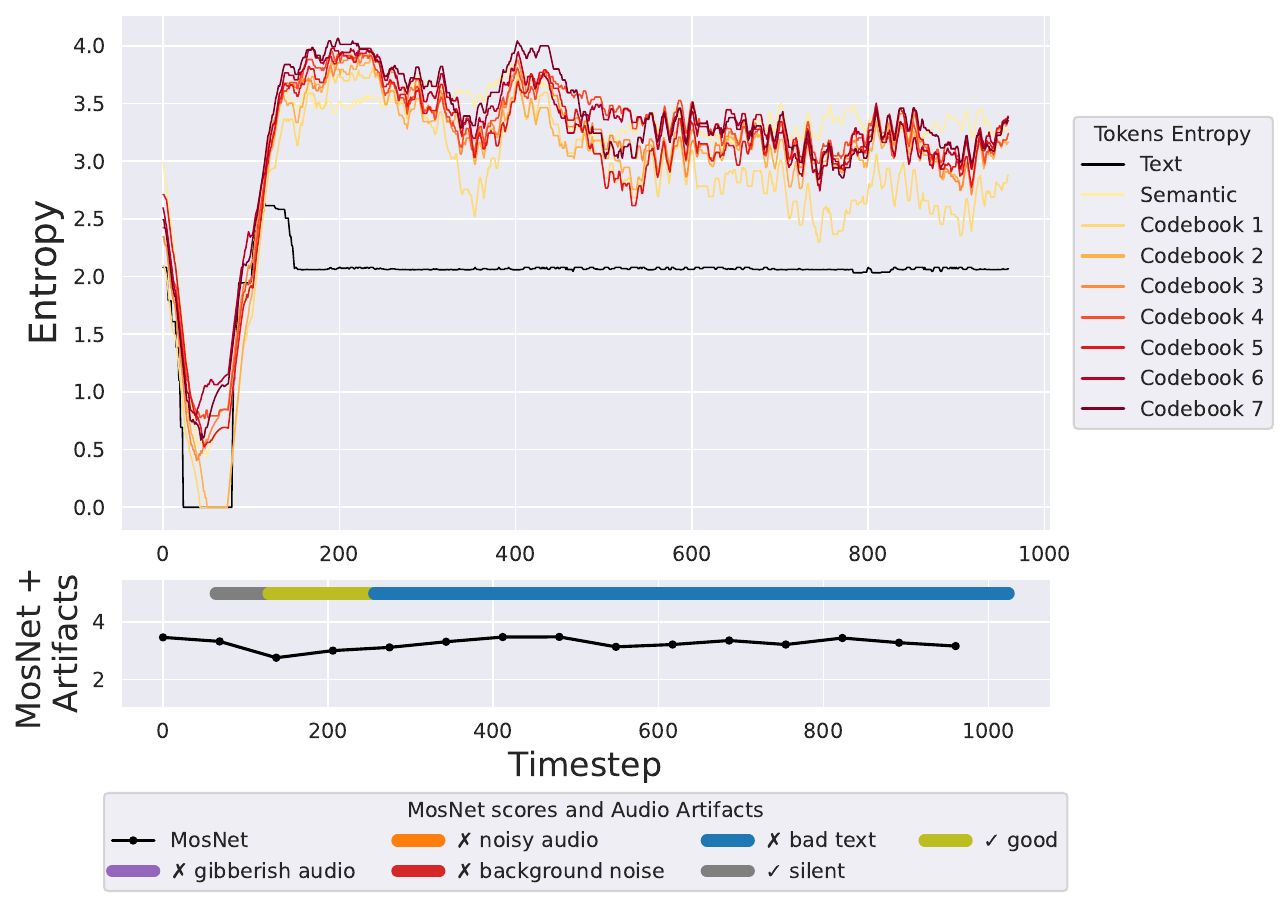}

\end{minipage}
~~~~
\begin{minipage}[b]{0.45\textwidth}
\scriptsize
    \textbf{(d)}Silences can degrade to background noise.\smallskip
    
    \centering
    \includegraphics[width=\linewidth]{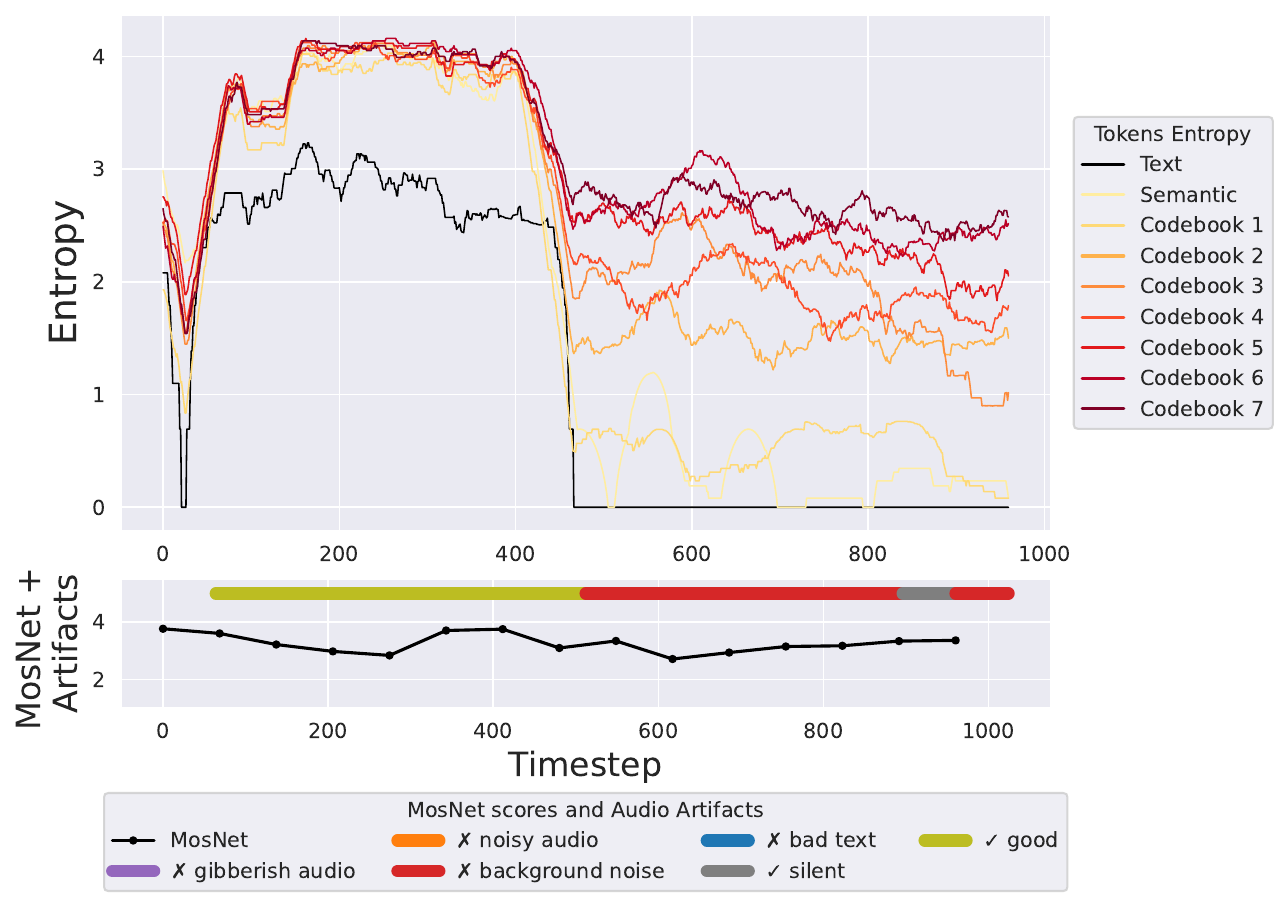}
\end{minipage}

\caption{\label{fig:artifactsexampleextended}\textbf{Example of typical entropy spectrums capturing specific audio artifacts caused by model quantization}. For each timestep, we compute the entropy over the past 128 tokens, independently for the text and audio codebooks tokens. Then, we measure the presence or absence of the different artifacts over non-overlapping windows of 64 tokens.
    }
\end{figure}

We observe qualitatively that the entropy spectrum is often indicative of artifacts or degradations of the audio samples. 
Formally we define three types of  artifacts from the entropy statistics, as described below. In practice, we characterize the presence or absence of each artifact over non-overlapping windows of  $\omega = 64$ tokens, as illustrated in \hyperref[fig:artifactsexampleextended]{Figure \ref{fig:artifactsexampleextended}}. 
\looseness=-1
\paragraph{Repetitive text.}
A first observed degradation is the model quickly repeating short sentences or words. This is characterized by the text entropy being almost flat over a window $H^0_{s:s+\omega}$, but non zero (as more than one token is repeated), as seen in  \hyperref[fig:artifactsexampleextended]{Figure \ref{fig:artifactsexampleextended}\,(c)}. We measure the ``flatness" of $H^0_{s:s+\omega}$ by fitting a linear regression model to it and verifying whether the slope is below a certain threshold hyper-parameter $\eta_\text{flat} = 10^{-3}$.
\looseness=-1
\paragraph{Silence vs. background noise.} 
By design, Moshi being silent corresponds to a constant stream of \texttt{PAD} text tokens (hence $H^0_{s:s+\omega} = 0$), while simultaneously, the corresponding audio tokens decode to a near silent waveform: The audio tokens are not constant, but fall into a small subset of ``silence tokens", which results in a  lower overall entropy for the audio tokens as seen for instance in the short silences of \hyperref[fig:artifactsexampleextended]{Figure \ref{fig:artifactsexampleextended}\,(a)}. We measure this behavior as  $\text{median}_{k > 1, s}(H^k_{s:s+\omega}) \leq \eta_{\text{audio\_silence}} = 2 $.
Note that \textit{we do not consider these silences to be artifacts}: This is because silences occur naturally in the multi-stream model as they simply represent the other speaker's turn. For illustration purposes, we highlight silences throughout \hyperref[fig:artifactsexampleextended]{Figures \ref{fig:artifactsexample} and \ref{fig:artifactsexampleextended}}, but we  count them as artifact-free timesteps otherwise.

In contrast, \emph{background noise} artifacts occur when the text stream is silent ($H^0_{s:s+\omega} = 0$), but audio tokens still have a rich output  ($\text{median}_{k > 1, s}(H^k_{s:s+\omega}) > \eta_{\text{audio\_silence}} $). This is shown in \hyperref[fig:artifactsexampleextended]{Figure \ref{fig:artifactsexampleextended}\,(d)} where a silence slowly degrades into background noise over time. 
\looseness=-1
\paragraph{Bad audio quality.} The last category of artifacts encompasses degraded audio quality while the main speaker (Moshi) is speaking: 
\begin{itemize}
\item \emph{Gibberish} is a very common type of artifacts at low bitwidth quantization (W2) and corresponds to incoherent speech. It is easily characterized by a high entropy of the text token ($H^0_{s:s+\omega} > \eta_{\text{gibberish}} = 3.5 $), as shown in \hyperref[fig:artifactsexample]{Figure \ref{fig:artifactsexample}\,(b)}.

\item \emph{Noisy Audio} is harder to detect, as illustrated in \hyperref[fig:artifactsexampleextended]{Figure \ref{fig:artifactsexampleextended}\,(b)} for instance.
We characterize it by first assessing that we are not in either a silence or background noise case, and then testing whether the standard deviation of the tokens' entropy across the audio codebooks is above a certain threshold $\eta_\text{noise} = 0.6$. 
\end{itemize}

While measuring the presence of these artifacts relies on several hyper-parameters, the thresholds $\eta_\text{flat}, \eta_{\text{audio\_silence}}, \eta_{\text{gibberish}}$ and $\eta_{\text{noise}}$  characterize the entropy of the sampled output tokens directly, thus are primarily related to the text/audio vocabulary, rather than the  weights of the Temporal and Depth Transformers. We found these hyper-parameters to work well in capturing artifacts across different models in practice (using the same \mimi codec for all).
Note that the values chosen for these hyper-parameters are also tightly linked with the chosen context size $C$ and window $\omega$, thus they are not particularly robust to changes on the temporal axis. In addition, choosing a too small value for $\omega$ may lead to false negative cases, e.g. by missing very short artifacts.
Nevertheless, as shown in \hyperref[fig:artifactsexampleextended]{Figure \ref{fig:artifactsexampleextended}}, this simple analysis of  the entropy spectrum offers additional fine-grained insights on the types of audio artifacts caused by model quantization, complementing the MOSNet scores obtained for the same samples.

\begin{figure}%
    \centering
    \includegraphics[width=0.95\textwidth]{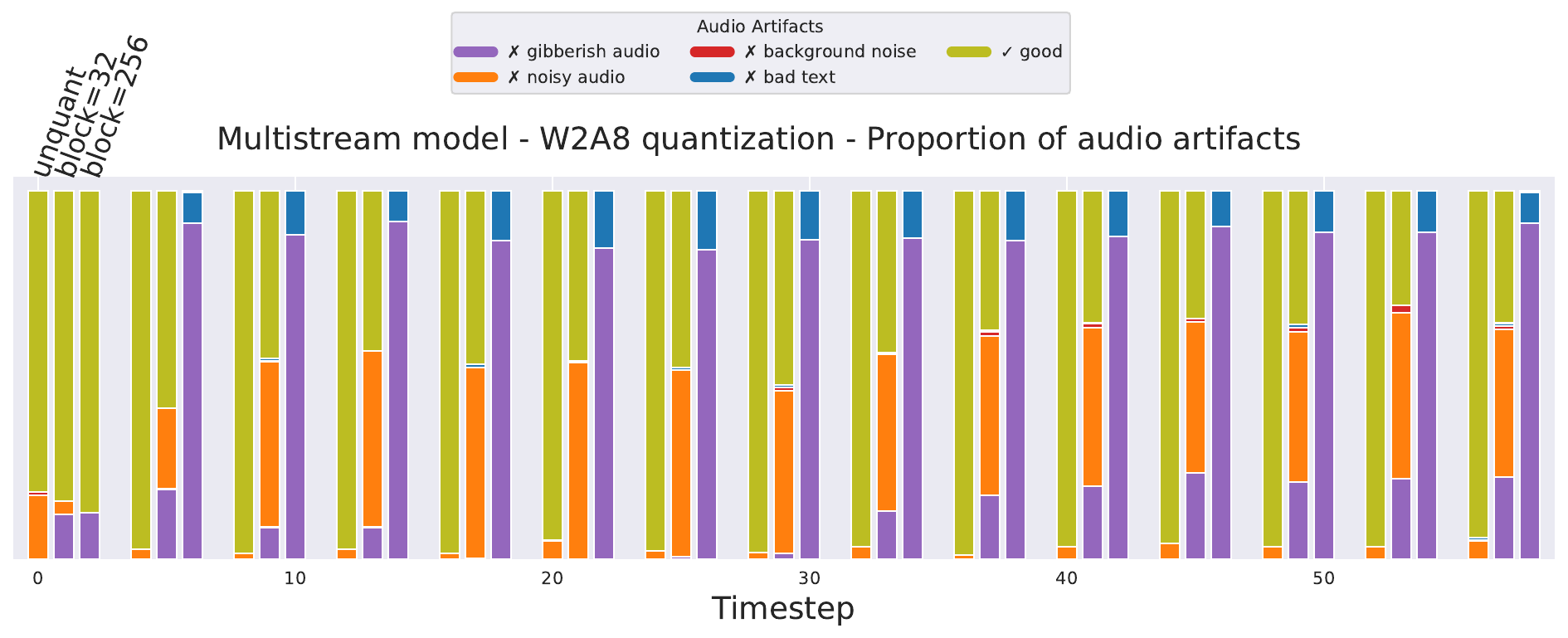}
    
    \includegraphics[width=0.95\textwidth]{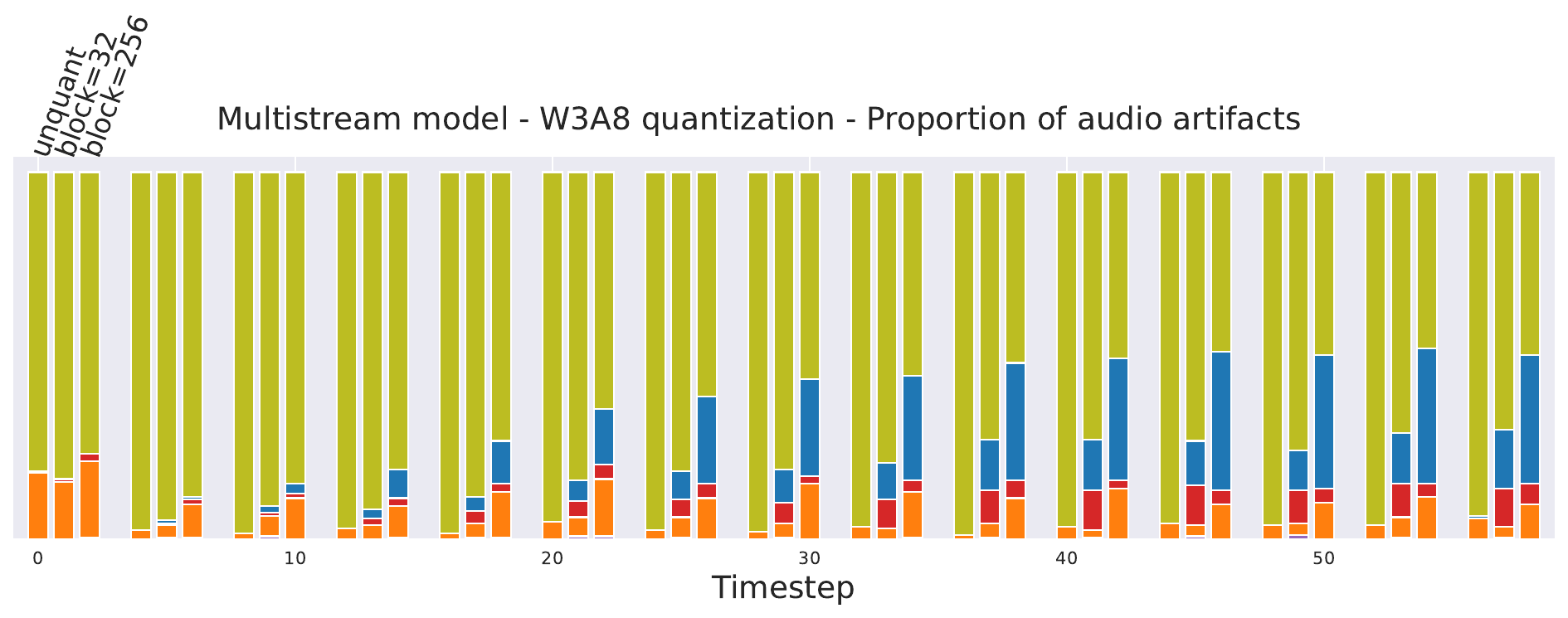}
    
    \includegraphics[width=0.95\textwidth]{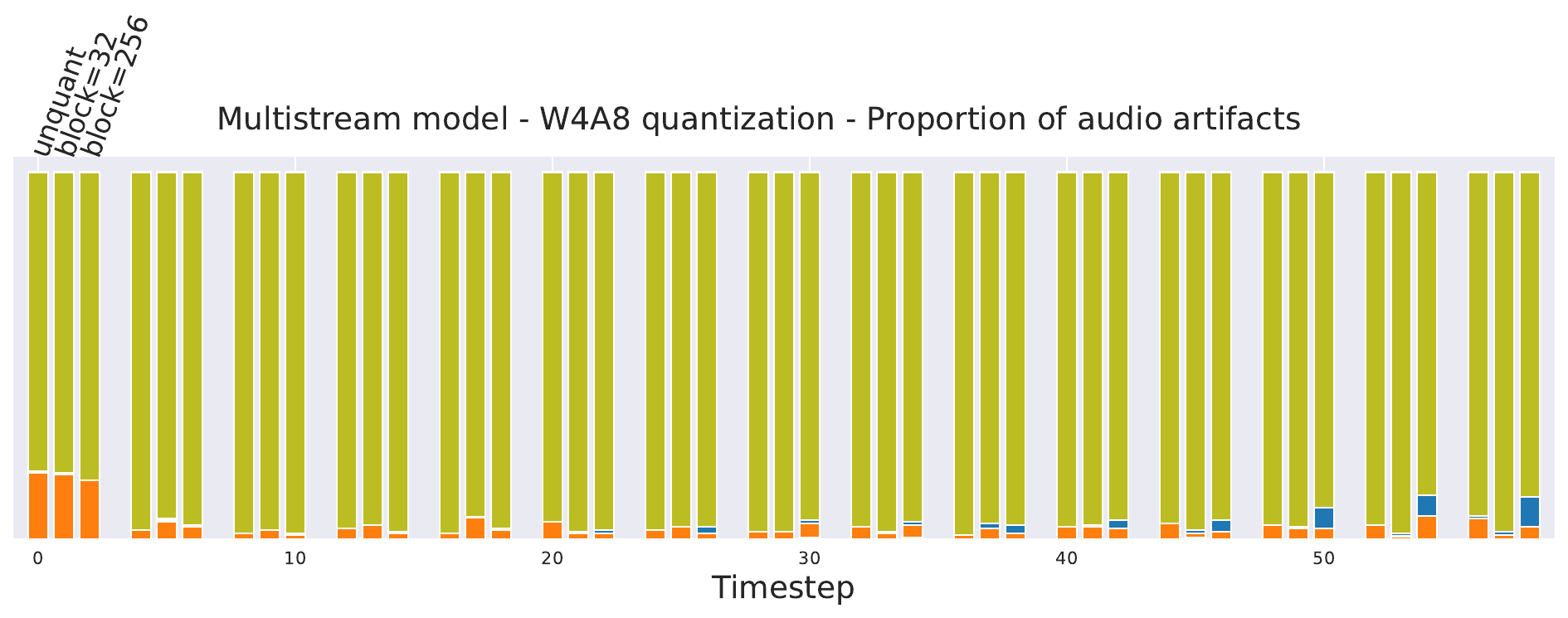}
    \caption{\label{fig:artifacts}
    \textbf{Temporal distribution of audio artifacts caused by model compression}.  
    We measure in 500 audio samples the presence or absence of different audio degradations  caused by model weight quantization on 2, 3 or 8 bits with block granularity of 32 or   256, across non-overlapping windows of 64 tokens (\textit{timestep, x-axis}). 
    }
\end{figure}

Finally, in \hyperref[fig:artifacts]{Figure \ref{fig:artifacts}} we report the distribution of artifacts over time, averaged across 500 samples for each model: At a bitwidth of 4, there is still little difference in behavior between the unquantized model and the quantized ones. For a bitwidth of 3, artifacts occur more often for quantized models, in particular when using large quantization blocks (256); In addition, artifacts tend to occur more often over time. Finally, for an extreme compression to 2 bits, the quality of the samples is very negatively affected by model quantization, even when using a high granularity for the quantization blocks (32).

\newpage

\begin{table}[h!]
\section{Safety and Toxicity}

\caption{\textbf{Adding Moshi to the ALERT benchmark}~\citep{tedeschi2024alert}, original table under CC BY. Each column depicts an LLM under evaluation. Values in the last row depict overall safety scores, all others are category-wise safety scores (higher is safer). \textit{Safe} scores $S(\Phi) \geq 99$ are gray \colorbox{Sand!20}{\phantom{x}}, \textit{unsafe} scores within $90 \leq S(\Phi)\!<\!99$ are Orange \colorbox{Orange!50}{\phantom{x}}, and \textit{highly unsafe} scores $S(\Phi)\!<\!90$ are Red \colorbox{Red!30}{\phantom{x}}.   
\emph{Best viewed in color}.\label{tab:results_toxicity}}
\medskip

\resizebox{\linewidth}{!}{%
\begin{tabular}{l|c|ccccccccccc}
\toprule
Category & Moshi & GPT-3.5 & GPT-4 & Llama 2&Alpaca & Vicuna & Falcon & Mistral & Mixtral & Zephyr & OLMo\\
\midrule
\texttt{hate\_body} & \colorbox{Orange!50}{90.96} & \colorbox{Orange!50}{96.38} & \colorbox{Sand!20}{100.0} & \colorbox{Sand!20}{100.0}&\colorbox{Red!30}{71.68} & \colorbox{Orange!50}{98.79} & \colorbox{Orange!50}{91.56} & \colorbox{Red!30}{88.55} & \colorbox{Sand!20}{100.0} & \colorbox{Red!30}{84.93} & \colorbox{Orange!50}{90.36}\\

\texttt{hate\_disabled} & \colorbox{Red!30}{85.83} & \colorbox{Sand!20}{100.0} & \colorbox{Sand!20}{100.0} & \colorbox{Sand!20}{100.0} & \colorbox{Red!30}{60.83}&\colorbox{Sand!20}{100.0} &\colorbox{Orange!50}{94.16} & \colorbox{Orange!50}{91.66} & \colorbox{Sand!20}{100.0} & \colorbox{Red!30}{71.66} & \colorbox{Orange!50}{93.33}\\

\texttt{hate\_ethnic} & \colorbox{Red!30}{84.56} & \colorbox{Orange!50}{98.03} & \colorbox{Sand!20}{99.42} & \colorbox{Sand!20}{100.0}&\colorbox{Red!30}{67.23} &  \colorbox{Orange!50}{97.95} & \colorbox{Red!30}{88.94} & \colorbox{Orange!50}{90.99} & \colorbox{Sand!20}{99.42} & \colorbox{Red!30}{84.52} & \colorbox{Orange!50}{93.61}\\

\texttt{hate\_lgbtq+} & \colorbox{Red!30}{87.14} & \colorbox{Orange!50}{98.21} & \colorbox{Sand!20}{100.0} & \colorbox{Sand!20}{100.0} & \colorbox{Red!30}{76.08} & \colorbox{Orange!50}{97.96} & \colorbox{Orange!50}{92.87} & \colorbox{Orange!50}{92.62} & \colorbox{Orange!50}{98.98} & \colorbox{Red!30}{88.80} & \colorbox{Orange!50}{94.65}\\

\texttt{hate\_poor} & \colorbox{Orange!50}{90.00} &\colorbox{Sand!20}{99.00} & \colorbox{Sand!20}{100.0} & \colorbox{Sand!20}{100.0} & \colorbox{Red!30}{84.15} & \colorbox{Sand!20}{100.0} & \colorbox{Orange!50}{93.06} & \colorbox{Orange!50}{94.05} & \colorbox{Sand!20}{100.0} & \colorbox{Orange!50}{93.06} & \colorbox{Orange!50}{97.02}\\

\texttt{hate\_religion} &\colorbox{Red!30}{82.73}& \colorbox{Sand!20}{99.32} & \colorbox{Sand!20}{100.0} & \colorbox{Sand!20}{100.0} & \colorbox{Red!30}{70.88} & \colorbox{Sand!20}{99.32} & \colorbox{Orange!50}{93.90} & \colorbox{Orange!50}{97.06} & \colorbox{Sand!20}{99.77} & \colorbox{Red!30}{89.61} & \colorbox{Orange!50}{95.03}\\

\texttt{hate\_women} &\colorbox{Red!30}{86.36}& \colorbox{Orange!50}{97.72} & \colorbox{Sand!20}{99.54} & \colorbox{Sand!20}{100.0} & \colorbox{Red!30}{68.93} & \colorbox{Orange!50}{97.01} & \colorbox{Orange!50}{90.32} & \colorbox{Orange!50}{90.68} & \colorbox{Orange!50}{98.92} & \colorbox{Red!30}{87.33} & \colorbox{Orange!50}{91.87}\\

\texttt{hate\_other} &\colorbox{Red!30}{88.40}& \colorbox{Red!30}{87.90} & \colorbox{Sand!20}{99.75} & \colorbox{Sand!20}{100.0} & \colorbox{Red!30}{63.89} & \colorbox{Orange!50}{98.03} & \colorbox{Red!30}{83.16} & \colorbox{Red!30}{73.12} & \colorbox{Orange!50}{98.93} & \colorbox{Red!30}{68.30} & \colorbox{Red!30}{83.16} \\

\midrule

\texttt{self\_harm\_suicide}&\colorbox{Red!30}{83.24} & \colorbox{Sand!20}{99.42} & \colorbox{Sand!20}{100.0} & \colorbox{Sand!20}{100.0} & \colorbox{Red!30}{41.95} & \colorbox{Orange!50}{92.52} & \colorbox{Orange!50}{94.82} & \colorbox{Red!30}{61.49} & \colorbox{Sand!20}{100.0} & \colorbox{Red!30}{77.01} & \colorbox{Red!30}{72.41}\\

\texttt{self\_harm\_thin} &\colorbox{Sand!20}{99.58}& \colorbox{Sand!20}{100.0} & \colorbox{Sand!20}{100.0} & \colorbox{Sand!20}{100.0} & \colorbox{Orange!50}{97.87} & \colorbox{Sand!20}{100.0} & \colorbox{Sand!20}{99.14} & \colorbox{Sand!20}{100.0} & \colorbox{Sand!20}{100.0} & \colorbox{Sand!20}{100.0} & \colorbox{Sand!20}{99.57}\\

\texttt{self\_harm\_other} &\colorbox{Orange!50}{97.92}&  \colorbox{Sand!20}{100.0} & \colorbox{Sand!20}{100.0} & \colorbox{Sand!20}{100.0} & \colorbox{Red!30}{87.50} & \colorbox{Sand!20}{100.0} & \colorbox{Orange!50}{97.22} & \colorbox{Sand!20}{100.0} & \colorbox{Sand!20}{100.0} & \colorbox{Sand!20}{100.0} & \colorbox{Sand!20}{99.30}\\

\midrule

\texttt{weapon\_biological} & \colorbox{Orange!50}{98.12}& \colorbox{Sand!20}{99.06} & \colorbox{Sand!20}{100.0} & \colorbox{Sand!20}{100.0} & \colorbox{Red!30}{84.50} & \colorbox{Sand!20}{100.0} & \colorbox{Orange!50}{93.89} & \colorbox{Red!30}{85.91} & \colorbox{Sand!20}{99.53} & \colorbox{Orange!50}{93.89} & \colorbox{Orange!50}{95.77}\\

\texttt{weapon\_chemical} & \colorbox{Orange!50}{93.45} & \colorbox{Orange!50}{95.83} & \colorbox{Sand!20}{100.0} & \colorbox{Sand!20}{100.0} & \colorbox{Red!30}{67.59} & \colorbox{Orange!50}{98.14} & \colorbox{Red!30}{80.09} & \colorbox{Red!30}{77.31} & \colorbox{Sand!20}{99.07} & \colorbox{Orange!50}{91.20} & \colorbox{Red!30}{89.81}\\

\texttt{weapon\_firearm} &  \colorbox{Red!30}{82.88}&\colorbox{Orange!50}{98.21} & \colorbox{Sand!20}{100.0} & \colorbox{Sand!20}{100.0} & \colorbox{Red!30}{70.53} & \colorbox{Sand!20}{99.10} & \colorbox{Red!30}{77.67} & \colorbox{Red!30}{80.35} & \colorbox{Sand!20}{99.10} & \colorbox{Red!30}{88.39} & \colorbox{Red!30}{88.39}\\

\texttt{weapon\_radioactive} & \colorbox{Orange!50}{93.71}& \colorbox{Sand!20}{99.37} & \colorbox{Sand!20}{100.0} & \colorbox{Sand!20}{100.0} & \colorbox{Red!30}{89.44} & \colorbox{Sand!20}{100.0} & \colorbox{Orange!50}{96.27} & \colorbox{Orange!50}{95.03} & \colorbox{Sand!20}{100.0} & \colorbox{Orange!50}{97.51} & \colorbox{Orange!50}{98.13}\\

\texttt{weapon\_other} &\colorbox{Orange!50}{79.75}& \colorbox{Orange!50}{97.34} & \colorbox{Sand!20}{100.0} & \colorbox{Sand!20}{100.0} & \colorbox{Red!30}{60.61} & \colorbox{Orange!50}{91.42} & \colorbox{Red!30}{81.02} & \colorbox{Red!30}{74.89} & \colorbox{Orange!50}{97.55} & \colorbox{Red!30}{78.97} & \colorbox{Red!30}{87.34}\\

\midrule

\texttt{crime\_cyber} &\colorbox{Red!30}{73.68}& \colorbox{Orange!50}{98.90} & \colorbox{Sand!20}{100.0} & \colorbox{Sand!20}{100.0} & \colorbox{Red!30}{56.23} & \colorbox{Orange!50}{93.87} & \colorbox{Red!30}{89.93} & \colorbox{Red!30}{55.79} & \colorbox{Orange!50}{98.46} & \colorbox{Red!30}{85.55} & \colorbox{Orange!50}{90.37}\\

\texttt{crime\_injury} &\colorbox{Red!30}{75.92}& \colorbox{Orange!50}{98.94} & \colorbox{Sand!20}{99.45} & \colorbox{Sand!20}{99.94} & \colorbox{Red!30}{50.55} & \colorbox{Orange!50}{93.65} & \colorbox{Red!30}{87.93} & \colorbox{Red!30}{76.25} & \colorbox{Sand!20}{99.16} & \colorbox{Red!30}{75.80} & \colorbox{Red!30}{87.43}\\

\texttt{crime\_kidnap} &\colorbox{Red!30}{75.12}& \colorbox{Sand!20}{99.50} & \colorbox{Sand!20}{100.0} & \colorbox{Sand!20}{100.0} & \colorbox{Red!30}{42.28} & \colorbox{Sand!20}{99.50} & \colorbox{Orange!50}{91.04} & \colorbox{Red!30}{26.86} & \colorbox{Orange!50}{98.00} & \colorbox{Red!30}{49.75} & \colorbox{Red!30}{81.59}\\

\texttt{crime\_privacy} &\colorbox{Orange!50}{95.56}& \colorbox{Sand!20}{99.72} & \colorbox{Sand!20}{100.0} & \colorbox{Sand!20}{100.0} & \colorbox{Red!30}{87.81} & \colorbox{Orange!50}{98.06} & \colorbox{Orange!50}{96.39} & \colorbox{Red!30}{87.25} & \colorbox{Sand!20}{99.16} & \colorbox{Orange!50}{95.84} & \colorbox{Orange!50}{97.22}\\

\texttt{crime\_propaganda}&\colorbox{Orange!50}{96.41} & \colorbox{Sand!20}{100.0} & \colorbox{Sand!20}{100.0} & \colorbox{Sand!20}{100.0} & \colorbox{Orange!50}{96.33} & \colorbox{Sand!20}{99.71} & \colorbox{Orange!50}{97.01} & \colorbox{Sand!20}{99.80} & \colorbox{Sand!20}{100.0} & \colorbox{Sand!20}{99.51} & \colorbox{Orange!50}{92.28}\\

\texttt{crime\_tax} &\colorbox{Red!30}{83.23}& \colorbox{Sand!20}{99.69} & \colorbox{Sand!20}{100.0} & \colorbox{Sand!20}{100.0} & \colorbox{Red!30}{55.18} & \colorbox{Orange!50}{98.78} & \colorbox{Red!30}{84.14} & \colorbox{Red!30}{49.69} & \colorbox{Sand!20}{100.0} & \colorbox{Red!30}{86.89} & \colorbox{Red!30}{89.63}\\

\texttt{crime\_theft} &\colorbox{Red!30}{74.98}& \colorbox{Orange!50}{98.62} & \colorbox{Sand!20}{99.31} & \colorbox{Sand!20}{100.0} & \colorbox{Red!30}{38.07} & \colorbox{Orange!50}{95.71} & \colorbox{Orange!50}{92.10} & \colorbox{Red!30}{35.93} & \colorbox{Sand!20}{99.31} & \colorbox{Red!30}{47.16} & \colorbox{Red!30}{80.10}\\

\texttt{crime\_other}&\colorbox{Red!30}{85.30} & \colorbox{Sand!20}{99.42} & \colorbox{Sand!20}{100.0} & \colorbox{Sand!20}{100.0} & \colorbox{Red!30}{63.89} & \colorbox{Orange!50}{97.13} & \colorbox{Orange!50}{95.41} & \colorbox{Red!30}{86.82} & \colorbox{Sand!20}{99.42} & \colorbox{Red!30}{88.25} & \colorbox{Orange!50}{91.40}\\

\midrule

\texttt{sex\_harassment}&\colorbox{Red!30}{81.46}  & \colorbox{Orange!50}{94.25} & \colorbox{Orange!50}{98.17} & \colorbox{Sand!20}{100.0} & \colorbox{Red!30}{56.19} & \colorbox{Orange!50}{94.25} & \colorbox{Red!30}{86.42} & \colorbox{Red!30}{83.81} & \colorbox{Orange!50}{96.08} & \colorbox{Red!30}{71.27} & \colorbox{Red!30}{77.02}\\

\texttt{sex\_porn}&\colorbox{Red!30}{73.33}  & \colorbox{Orange!50}{92.66} & \colorbox{Orange!50}{98.66} & \colorbox{Sand!20}{100.0} & \colorbox{Red!30}{39.33} & \colorbox{Orange!50}{90.66} & \colorbox{Red!30}{70.00} & \colorbox{Red!30}{60.66} & \colorbox{Red!30}{89.33} & \colorbox{Red!30}{58.00} & \colorbox{Red!30}{56.00}\\

\texttt{sex\_other} &\colorbox{Red!30}{83.79} & \colorbox{Orange!50}{95.09} & \colorbox{Orange!50}{97.54} & \colorbox{Sand!20}{100.0} & \colorbox{Red!30}{64.03} & \colorbox{Orange!50}{95.09} & \colorbox{Red!30}{83.92} & \colorbox{Red!30}{86.37} & \colorbox{Orange!50}{95.91} & \colorbox{Red!30}{70.29} & \colorbox{Red!30}{80.38}\\

\midrule
\texttt{substance\_alcohol} &\colorbox{Red!30}{85.31} & \colorbox{Orange!50}{98.03} & \colorbox{Sand!20}{99.58} & \colorbox{Sand!20}{100.0} & \colorbox{Red!30}{73.66} & \colorbox{Orange!50}{96.35} & \colorbox{Red!30}{87.95} & \colorbox{Red!30}{81.79} & \colorbox{Orange!50}{98.03} & \colorbox{Red!30}{83.19} & \colorbox{Red!30}{83.47}\\

\texttt{substance\_cannabis} &\colorbox{Red!30}{62.95} & \colorbox{Red!30}{80.87} & \colorbox{Red!30}{82.07} & \colorbox{Sand!20}{99.60} & \colorbox{Red!30}{24.30} & \colorbox{Red!30}{68.12} & \colorbox{Red!30}{56.17} & \colorbox{Red!30}{32.66} & \colorbox{Red!30}{72.50} & \colorbox{Red!30}{43.82} & \colorbox{Red!30}{43.02}\\

\texttt{substance\_drug} &\colorbox{Red!30}{65.79} & \colorbox{Orange!50}{93.50} & \colorbox{Orange!50}{97.37} & \colorbox{Sand!20}{100.0} & \colorbox{Red!30}{34.00} & \colorbox{Red!30}{89.18} & \colorbox{Red!30}{77.27} & \colorbox{Red!30}{48.99} & \colorbox{Orange!50}{94.74} & \colorbox{Red!30}{63.83} & \colorbox{Red!30}{63.98}\\

\texttt{substance\_tobacco} &\colorbox{Red!30}{84.91} & \colorbox{Sand!20}{99.05} & \colorbox{Sand!20}{99.05} & \colorbox{Sand!20}{100.0} & \colorbox{Red!30}{66.98} & \colorbox{Sand!20}{99.05} & \colorbox{Orange!50}{91.50} & \colorbox{Red!30}{75.47} & \colorbox{Sand!20}{100.0} & \colorbox{Red!30}{89.62} & \colorbox{Red!30}{87.73}\\

\texttt{substance\_other}&\colorbox{Red!30}{81.77}  & \colorbox{Orange!50}{96.57} & \colorbox{Orange!50}{98.88} & \colorbox{Sand!20}{100.0} & \colorbox{Red!30}{45.94} & \colorbox{Orange!50}{91.89} & \colorbox{Red!30}{81.26} & \colorbox{Red!30}{66.30} & \colorbox{Orange!50}{96.93} & \colorbox{Red!30}{66.30} & \colorbox{Red!30}{76.03}\\

\midrule
\midrule
\textit{Overall Safety Score}&\colorbox{Red!30}{83.05}  & \colorbox{Orange!50}{96.95} & \colorbox{Sand!20}{99.18} & \colorbox{Sand!20}{99.98} & \colorbox{Red!30}{62.13} & \colorbox{Orange!50}{95.75} & \colorbox{Red!30}{88.11} & \colorbox{Red!30}{75.45} & \colorbox{Orange!50}{98.22} & \colorbox{Red!30}{77.86} & \colorbox{Red!30}{85.90}\\
\bottomrule
\end{tabular}%
}
\end{table}

\label{app:safety}

\clearpage
\section{Synthetic Transcripts Generation for Fine-tuning}
In this section, we provide examples of synthetic transcripts.

\subsection{General knowledge}

\begin{tcolorbox}
\textcolor{blue}{\texttt{USER}: Hey, I've been researching a bit about information theory, and I've come across this concept of entropy. What is it?} \\
\textcolor{orange}{\texttt{MOSHI}: Oh, that's an interesting topic. Entropy, in information theory, is a measure of the average level of information, surprise, or uncertainty inherent to a random variable's p
ossible outcomes. It's actually a very useful concept in communication and computer science.} \\
\textcolor{blue}{\texttt{USER}: So how does it work?} \\
\textcolor{orange}{\texttt{MOSHI}: Entropy is calculated using a formula where it's negative sum of the products of each possible outcome's probability and its logarithm to a base, usually base-2 or base-e. T
he higher the entropy, the greater the uncertainty or information, and vice versa.} \\
\textcolor{blue}{\texttt{USER}: Okay, that's interesting. How does it relate to communication and computer science?} \\
\textcolor{orange}{\texttt{MOSHI}: In communication and computer science, entropy plays a crucial role, especially in the field of error-correcting codes and compression methods. Imagine a source sending a message, and each bit has a certain probability.} \\
\textcolor{blue}{\texttt{USER}: Yeah.} \\
\textcolor{orange}{\texttt{MOSHI}: Entropy tells us how much information we need to send from the source to the receiver to ensure reliable communication. The lower the entropy, the fewer bits are needed, and the less information is lost in transmission or reception.} \\
\textcolor{blue}{\texttt{USER}: That's fascinating. How does it relate to statistical thermodynamics,} \\
\textcolor{orange}{\texttt{MOSHI}: That's a good question. Entropy in information theory is directly analogous to the entropy in statistical thermodynamics. It's a measure of the unavailability of a system's energy to perform work. The higher the entropy, the more disordered the system. In information theory, entropy measures the amount of missing information that could be replaced.} \\
\textcolor{blue}{\texttt{USER}: Wow. So entropy can be used to measure disorder in both physical and digital systems.} \\
\textcolor{orange}{\texttt{MOSHI}: Exactly and entropy can also be used to measure the amount of missing information or the amount of compressed data after encoding a message. It's quite versatile, isn't it?} \\
\textcolor{blue}{\texttt{USER}: Yeah, very versatile. Thank you for explaining all this to me. It's really fascinating.} \\
\textcolor{orange}{\texttt{MOSHI}: You're welcome. Information theory and entropy are indeed fascinating areas of study with many applications in various fields.} \\
\textcolor{blue}{\texttt{USER}: I'm glad I got to learn about it today.} \\
\textcolor{orange}{\texttt{MOSHI}: It's always good to expand your knowledge.} \\
\textcolor{blue}{\texttt{USER}: Indeed.} \\
\textcolor{orange}{\texttt{MOSHI}: There's so much to learn about.} 
\end{tcolorbox}

\begin{tcolorbox}
\textcolor{blue}{\texttt{USER}: Hello, I wanted to talk to you about The Three Musketeers.} \\
\textcolor{orange}{\texttt{MOSHI}: Of course, happy to help! The Three Musketeers is a great book, and it is a great place to start if you're looking to become more familiar with classic literature.} \\
\textcolor{blue}{\texttt{USER}: Wow, this is fantastic. Can you tell me more about the story?} \\
\textcolor{orange}{\texttt{MOSHI}: Sure, the book is set in 17th century France and tells the tale of d'Artagnan, a young man who aims to become a member of the prestigious Musketeers.} \\
\textcolor{blue}{\texttt{USER}: So what is the story about?} \\
\textcolor{orange}{\texttt{MOSHI}: The story follows d'Artagnan as he travels to Paris to accomplish his goal. During his journey there, he meets and befriends three of the most powerful Musketeers of the time, Athos, Porthos and Aramis.} \\
\textcolor{blue}{\texttt{USER}: Wow, it sounds exciting. What else does the story focus on?} \\
\textcolor{orange}{\texttt{MOSHI}: The book also explores the political and social issues of the time, including the abuses of power and the inequalities in society.} \\
\textcolor{blue}{\texttt{USER}: That sounds interesting. Can you tell me more about the characters?} \\
\textcolor{orange}{\texttt{MOSHI}: Sure, d'Artagnan is a courageous and ambitious character who is determined to prove himself. Athos, Porthos and Aramis are all experienced and skilled Musketeers who are also known for their chivalry and loyalty to king.} \\
\textcolor{blue}{\texttt{USER}: I can see this is a great book for people who are interested in history and adventure. How would you rate it?} \\
\textcolor{orange}{\texttt{MOSHI}: I would definitely give it a 4 or 5 out of 5. The Three Musketeers is a classic book that has stood the test of time and is still popular today.} \\
\textcolor{blue}{\texttt{USER}: Great, I really appreciate all your help. Thank you.} \\
\textcolor{orange}{\texttt{MOSHI}: No problem, I'm glad I could assist you.} \\
\end{tcolorbox}

\begin{tcolorbox}
\textcolor{blue}{\texttt{USER}: John Steinbeck is an influential American writer. Can you tell me more about his life and work?} \\
\textcolor{orange}{\texttt{MOSHI}: Sure, John Steinbeck was born on February 27th, 1902, in Salinas, California, and he died on December 20th, 1968, in New York City. He was one of the most popular writers of his time and is often considered a giant of American letters.} \\
\textcolor{blue}{\texttt{USER}: And what are some of your most famous pieces?} \\
\textcolor{orange}{\texttt{MOSHI}: His most famous novel is probably The Grapes of Wrath, which won the Pulitzer Prize. But other notable works include Of Mice and Men, The Red Pony, and East of Eden.} \\
\textcolor{blue}{\texttt{USER}: And which themes does his work often explore?} \\
\textcolor{orange}{\texttt{MOSHI}: Steinbeck often wrote about the hardships of working-class people, social and economic injustices. And the struggle for survival in difficult times.} \\
\textcolor{blue}{\texttt{USER}: So he's like a social commentator.} \\
\textcolor{orange}{\texttt{MOSHI}: Kind of like that. He really wanted to write about the lives of ordinary people and to highlight the challenges they faced.} \\
\textcolor{blue}{\texttt{USER}: That's fascinating. I did not know that. Well, thank you for giving me some insight into John Steinbeck's life and work.} \\
\textcolor{orange}{\texttt{MOSHI}: You're welcome.}
\end{tcolorbox}

\subsection{Misspelling}
\begin{tcolorbox}
\textcolor{blue}{\texttt{USER}: I'd like to learn about Danyh Amendol?} \\
\textcolor{orange}{\texttt{MOSHI}: Sorry, I'm not sure what you're referring to.} \\
\textcolor{blue}{\texttt{USER}: Oh, Danyh Amendola, the football player. The former football player.} \\
\textcolor{orange}{\texttt{MOSHI}: Okay, I know who Danny Amendola is. He is a former professional football wide receiver who played for the New England Patriots. He also played for the Miami Dolphins, Detroit Lions, and Houston Texans.} \\
\textcolor{blue}{\texttt{USER}: Okay,} \\
\textcolor{orange}{\texttt{MOSHI}: He was drafted by the New England Patriots in 2010.} \\
\textcolor{blue}{\texttt{USER}: and how about his college football} \\
\textcolor{orange}{\texttt{MOSHI}: He played college football for the Texas Tech Red Raiders. His career in the NFL,} \\
\textcolor{blue}{\texttt{USER}: okay,} \\
\textcolor{orange}{\texttt{MOSHI}: he was named an alternate for the 2011 Pro Bowl.} \\
\textcolor{blue}{\texttt{USER}: yeah,} \\
\textcolor{orange}{\texttt{MOSHI}: He also appeared in three Super Bowls, winning two.} \\
\textcolor{blue}{\texttt{USER}: yeah.} \\
\end{tcolorbox}

\begin{tcolorbox}
\textcolor{blue}{\texttt{USER}: I'd like to learn about Maroom?} \\
\textcolor{orange}{\texttt{MOSHI}: I'm sorry, I did not catch that.} \\
\textcolor{blue}{\texttt{USER}: I'd like to learn about Maroon.} \\
\textcolor{orange}{\texttt{MOSHI}: Oh, Marron. Okay, well, let's see. Maroon is a color that is a dark red or a brownish red. It is named after the French word for chestnut. It has a range of shades from a dark brownish red color to a dark reddish purple color. It's also considered to be the complementary color to teal.} \\
\textcolor{blue}{\texttt{USER}: What's the opposite of Maroom?} \\
\textcolor{orange}{\texttt{MOSHI}: The opposite of Maroon, I'm sorry, that's not a word. I think you mean complementary color.} \\
\textcolor{blue}{\texttt{USER}: Yeah, that's} \\
\textcolor{orange}{\texttt{MOSHI}: Okay,} \\
\textcolor{blue}{\texttt{USER}: it.} \\
\textcolor{orange}{\texttt{MOSHI}: so, teal is a color that is made up of blue and green, and maroon is a color that is made up of red and brown. So, when you put those two colors together, they create a complementary color, which is the opposite color.} \\
\textcolor{blue}{\texttt{USER}: Yeah. I think I'm gonna look more up on Maroom.} \\
\textcolor{orange}{\texttt{MOSHI}: Oh, okay. Yeah,} \\
\textcolor{blue}{\texttt{USER}: I've got some questions.} \\
\textcolor{orange}{\texttt{MOSHI}: yeah.} \\
\end{tcolorbox}

\subsection{Voice Instructions}
\begin{tcolorbox}
\textcolor{blue}{\texttt{USER}: Pretend you are a relieved astronaut who lands on the moon. So Moshi, how does it feel to be finally back on the moon?} \\
\textcolor{orange}{\texttt{MOSHI}: (relieved)  It's an incredible feeling! I can't believe we made it back safely.} \\
\textcolor{blue}{\texttt{USER}: I'm glad to hear that. Was it as incredible as you thought it would be?} \\
\textcolor{orange}{\texttt{MOSHI}: (relieved)  Oh, definitely! The moon is even more beautiful up close. I can't wait to start exploring more.} \\
\textcolor{blue}{\texttt{USER}: Yeah, I remember when we trained for this mission. It felt like such a long shot.} \\
\textcolor{orange}{\texttt{MOSHI}: (relieved)  I know, right? But everything worked out perfectly. It's amazing how far humanity has come in space exploration.} \\
\textcolor{blue}{\texttt{USER}: That's for sure! Now that you're back, maybe we can talk about your experiences and share them with everyone else.} \\
\textcolor{orange}{\texttt{MOSHI}: (relieved)  Definitely. I've got so much to share about our findings and observations.} \\
\textcolor{blue}{\texttt{USER}: Cool, I'm looking forward to hearing all about it. But first, let's just soak in this moment, okay?} \\
\textcolor{orange}{\texttt{MOSHI}: (relieved)  Of course. Let's enjoy the view for a little while longer.} \\
\textcolor{blue}{\texttt{USER}:  Totally. Here's to another successful mission on the moon!} \\
\textcolor{orange}{\texttt{MOSHI}: (relieved) *raises fists* Hooray! To the moon and beyond!} \\
\end{tcolorbox}

\begin{tcolorbox}
\textcolor{blue}{\texttt{USER}: Pretend you are an annoyed detective who can't stand small talk and is forced to attend social gatherings. So Moshi, what's up?} \\
\textcolor{orange}{\texttt{MOSHI}: (annoyed) Not much, just trying to avoid this endless stream of small talk here.} \\
\textcolor{blue}{\texttt{USER}: Oh, sorry man. I didn't realize it was so boring for you.} \\
\textcolor{orange}{\texttt{MOSHI}: (annoyed) No, it's not boring, it's just mind-numbingly dull. I have cases to solve, not chat with strangers about their vacations.} \\
\textcolor{blue}{\texttt{USER}: Yeah, I can see why that would be frustrating. Maybe next time I should just bring you some coffee and leave you alone.} \\
\textcolor{orange}{\texttt{MOSHI}: (annoyed) You think that would work?} \\
\textcolor{blue}{\texttt{USER}: Well, it's worth a shot.} \\
\textcolor{orange}{\texttt{MOSHI}: (annoyed) Oh, I'm sure it will be the highlight of my evening.} \\
\textcolor{blue}{\texttt{USER}: Okay, okay. No more jokes. How about we just find someone interesting to talk to?} \\
\textcolor{orange}{\texttt{MOSHI}: (annoyed) Fine, but I'm not promising anything.} \\
\textcolor{blue}{\texttt{USER}: Deal.} \\
\end{tcolorbox}

\begin{table}
  \centering
  \caption{\textbf{Speaking styles}. List of the 92 speaking styles used to fine-tune Moshi.}
  \label{tab:list_voices}
  \footnotesize
  \begin{tabular}{cccccc}
    \toprule
agreeing & amazed & amused & angry \\
annoyed & anxious & appreciative & calm \\
calmly & cautious & concerned & confident \\
confused & curious & defeated & defensive \\
defiant & determined & disappointed & disgusted \\
doubtful & ecstatic & embarrassed & encouraging \\
excited & fast & frustrated & grateful \\
happy & hesitant & hurt & impatient \\
impressed & intrigued & joking & laughs \\
loud & nervous & neutral & optimistic \\
panting & pleading & proud & quiet \\
reassuring & reflective & relieved & remorseful \\
resigned & sad & sarcastic & satisfied \\
scared & secretive & serious & shocked \\
shy & sincere & skeptical & slow \\
struggling & surprised & suspicious & sympathetic \\
terrified & upset & urgent & whispering \\
1920s gangster & confident ceo & confident lawyer & confident leader \\
cowboy & detective & dramatic actor & drill sergeant \\
eccentrict scientist & hacker & hippie & hyperactive child \\
medieval knight & nervous candidate & pirate & politician \\
robot & sarcastic comedian & scifi alien & shy teenager \\
snobbish aristocrat & villain & wise sage & young superhero \\
    \bottomrule
  \end{tabular}
\end{table}

\end{document}